\documentclass{ws-ijbc}

\usepackage{graphicx}
\usepackage{url}

\newcommand{\be}{\begin{equation}}
\newcommand{\ee}{\end{equation}}

\newcommand{\ra}{\rightarrow}

\newcommand{\lbd}{\lambda}

\begin{document}

\catchline{}{}{}{}{} % Publisher's Area please ignore

\markboth{V.I. Yukalov, E.P. Yukalova, D. Sornette}{New Approach to
Modeling Symbiosis in Biological and Social Systems}

\title{NEW APPROACH TO MODELING SYMBIOSIS IN BIOLOGICAL AND SOCIAL SYSTEMS}

\author{V.I. YUKALOV}
\address{Department of Management, Technology and Economics, \\
ETH Z\"urich, Swiss Federal Institute of Technology,
Z\"urich CH-8092, Switzerland\\ and \\
Bogolubov Laboratory of Theoretical Physics, \\
Joint Institute for Nuclear Research, Dubna 141980, Russia \\
yukalov@theor.jinr.ru}

\author{E.P. YUKALOVA}
\address{Department of Management, Technology and Economics, \\
ETH Z\"urich, Swiss Federal Institute of Technology,
Z\"urich CH-8092, Switzerland\\ and \\
Laboratory of Information Technologies, \\
Joint Institute for Nuclear Research, Dubna 141980, Russia \\
lyukalov@ethz.ch}

\author{D. SORNETTE}
\address{Department of Management, Technology and Economics, \\
ETH Z\"urich, Swiss Federal Institute of Technology,
Z\"urich CH-8092, Switzerland \\ and \\
Swiss Finance Institute, c/o University of Geneva, \\
40 blvd. Du Pont d'Arve, CH 1211 Geneva 4, Switzerland \\
dsornette@ethz.ch}

\maketitle

\begin{history}
\received{(to be inserted by publisher)}
\end{history}

\vskip 2cm

\begin{abstract}
We suggest a novel approach to treating symbiotic relations between biological
species or social entities. The main idea is the characterisation of symbiotic
relations of coexisting species through their mutual influence on their
respective carrying capacities, taking into account that this influence can be
quite strong and requires a nonlinear functional framework. We distinguish
three variants of mutual influence, representing the main types of relations
between species: (i) passive symbiosis, when the mutual carrying capacities are
influenced by other species without their direct interactions; (ii) active
symbiosis, when the carrying capacities are transformed by interacting species;
and (iii) mixed symbiosis, when the carrying capacity of one species is influenced
by direct interactions, while that of the other species is not. The approach
allows us to describe all kinds of symbiosis, mutualism, commensalism, and
parasitism within a unified scheme. The case of two symbiotic species is analysed
in detail, demonstrating several dynamical regimes of coexistence, unbounded
growth of both populations, growth of one and elimination of the other population, convergence to evolutionary stable states, and everlasting population oscillations.
The change of the dynamical regimes occurs by varying the system parameters
characterising the destruction or creation of the mutual carrying capacities.
The regime changes are associated with several dynamical system bifurcations.
\end{abstract}


\newpage

\section{Introduction}

Symbiosis is one of the most widespread phenomenon in nature. Actually,
practically all existing species enjoy some kind of symbiotic relations.
As examples, we can start from ourselves, as human beings, whose bodies are
in symbiotic relations with ten thousand or more bacterial species
(National Institute of Health, USA,
\url{http://www.nih.gov/news/health/jun2012/nhgri-13.htm}) and around $10^{14}$
different microorganisms. Other well known examples include the association
between plant roots and fungi, between coral organisms and various types
of algae, between cattle egrets and cattle, between alive beings and
different bacteria and viruses. Numerous other examples can be found in the
literature \cite{Boucher_1,Douglas_2,Sapp_3,Ahmadjian_4,Townsend_5,Yukalov_6}.

Symbiotic relations are not restricted only to biological species, but may
be noticed in a variety of economic, financial, and other relations in social
systems. For example, as a variant of symbiosis one can treat the
interconnections between economics and arts, between basic and applied science,
between culture and language, between banks and firms, between different
enterprises, between firm owners and hired personal, and so on. Many more
examples are discussed in the following publications
\cite{Hippel_7,Grossman_8,Richard_9,Graedel_10,Yukalov_6}.

Coexistence of different species, such as predators and preys, is usually
described by the evolution equations of the Lotka-Volterra type \cite{Lotka_11},
where the species directly interact with each other. Such equations are not
valid for describing symbiosis since, in symbiotic relations, the species
do not eat or kill each other, but influence their mutual livelihoods
\cite{Boucher_1,Douglas_2,Sapp_3,Ahmadjian_4,Townsend_5,Yukalov_6,Yukalov_12}.

From the mathematical point of view, the replicator equations characterizing
trait groups, e.g., cooperators, defectors, and punishers, are also of the
Lotka-Volterra type, as they consider direct interactions. This also concerns
the public-goods games for the spatially structured trait-group coexistence on
networks \cite{Perc_13,Perc_14,Szolnoki_15,Szolnoki_16,Perc_17}, as well as
the utility-rate equations \cite{Yukalov_18}. Such equations do not seem to be
applicable for describing symbiotic relations.

The special feature of symbiotic relations is that the coexisting species
interact with each other mainly through changing the carrying capacities of
each other \cite{Boucher_1,Douglas_2,Sapp_3,Ahmadjian_4,Townsend_5,Yukalov_6}.
Therefore, for the correct description of symbiosis, it is necessary to consider
the species carrying capacities as functionals of the species population
fractions. In the previous publications \cite{Yukalov_6,Yukalov_12}, the species
carrying capacity functionals were considered as polynomial expansions in powers
of population fractions, with taking into account only the lowest orders of these
expansions, which resulted in linear or bilinear approximations, depending on the
symbiosis type. Such lowest-order approximations presuppose that the symbiotic
interactions are sufficiently weak, only slightly varying the mutual carrying
capacities. The formal use of the linear or bilinear approximations in some
cases leads to artificial finite-time singularity or finite-time population
death. This defect could be avoided by employing a nonlinear form for the
carrying capacities.

It is the aim of the present paper to formulate the evolution equations for
coexisting symbiotic species interacting through the influence on the carrying
capacities of each other, keeping in mind that this influence can be arbitrarily
strong. This implies the use of a nonlinear form for the carrying capacity
functional. We follow the idea used for the generalization of the evolution
equation for a single population \cite{Yukalov_19}, where some artificial
singularities could be eliminated by a nonlinear carrying-capacity functional.
But now, we consider several symbiotic coexisting species, which makes the
principal difference from the single-species evolution.

In Sec. 2, we formulate the main idea of the approach for describing symbiosis,
not as a direct interaction of different species as in the Lotka-Volterra
equation but, through the mutual influence on the carrying capacities of each
other. We distinguish three types of the mutual influence characterizing
passive symbiosis, active symbiosis, and mixed-type symbiosis. The origin of
the problems spoiling the linear or bilinear approximations is explained. In
Sec. 3, we introduce the generalized formulation for the symbiotic equations
with nonlinear carrying-capacity functionals. We stress that this formulation
allows one to treat symbiotic relations of arbitrary strength and of any type,
such as mutualism, commensalism, and parasitism. In the following sections,
we give a detailed analysis of the dynamics and of the evolutionary stable
states for each kind of the symbiotic relations characterizing passive
symbiosis (Sec. 4), active symbiosis (Sec. 5), and mixed symbiosis (Sec. 6)
for the cases of mutualism and parasitism. All possible dynamical regimes
are investigated, occurring as a result of bifurcations. Of special interest
is a supercritical Hopf bifurcation happening in the case of mixed symbiosis.
Section 7 summarizes the results for commensalism that is a marginal case
between mutualism and parasitism. Section 8 concludes.

\section{Modeling Symbiosis by Mutual Influence on Carrying Capacities}

\subsection{General structure of equations}

Suppose several species, enumerated as $i = 1, 2, \ldots, S$, coexist, being
in mutual symbiosis with each other. It is always admissible to reduce the
consideration to dimensionless units, as has been thoroughly explained in
the previous articles \cite{Yukalov_6,Yukalov_12}. So, in what follows, we shall
deal with dimensionless quantities, such as the species fractions $x_i = x_i(t)$
depending on dimensionless time $t\geq 0$. The species can proliferate, with the
inverse of the proliferation rates characterizing their typical lifetimes.
Generally, in applications of dynamical theory, the time scales of the connected
equations could be different \cite{Desroches_36}. However, in the case of symbiosis,
one has to keep in mind that the relations between the species, by definition,
are assumed to last sufficiently long to be treated as genuinely symbiotic
\cite{Boucher_1,Douglas_2,Sapp_3,Ahmadjian_4,Townsend_5,Yukalov_6}.
While this does not cover all symbiotic relationships, such as those
between long-lived humans and very short-lived bacterias for instance,
we will restrict our considerations to the class of symbiotic relationships in
which the lifetimes of the involved species are comparable. This corresponds
to the situations where these species compete for the corresponding
carrying capacities $y_i$. Since the meaning of symbiotic relations is the
variation of the mutual carrying capacities, the latter have to be functionals
of the population fractions:
\be
\label{1}
 y_i = y_i(x_1,x_2,\ldots,x_S) \;  .
\ee
We consider the society, composed of the symbiotic species, as being closed,
so that the carrying capacities are not subject to variations caused by
external forces, but are only influenced by the mutual interactions of the
coexisting species. Therefore the population dynamics is completely defined
by the type of symbiotic relations between the species. In the other case,
when external forces would be present, the society dynamics would also be
governed by such external forces and would strongly depend on the kind of
chosen forces \cite{Leonov_37,Pong_38}.

The evolution equation for an $i$-th species takes the form
\be
\label{2}
  \frac{dx_i}{dt} = x_i - \; \frac{x_i^2}{y_i} \; ,
\ee
where $x_i$ and $y_i$ are non-negative. In particular, for the case of two
species, $x_1 \equiv x$ and $x_2 \equiv z$, we have
\be
\label{3}
\frac{dx(t)}{dt} = x(t) - \; \frac{x^2(t)}{y_1(x,z)} \;  , \qquad
\frac{dz(t)}{dt} = z(t) - \; \frac{z^2(t)}{y_2(x,z)} \; .
\ee
To proceed further, it is necessary to specify the expressions for $y_i$.
It is possible to distinguish three types of symbiosis, depending on the
process controlling the variations of the mutual carrying capacities.

\subsection{Passive weak symbiosis}

To clearly characterize the difference between the symbiosis types, we start
with the case of weak symbiosis, when the mutual influence on the carrying
capacities is not strong, so that the carrying-capacity functionals can be
approximated by power-law expansions over the species fractions, limiting
ourselves by the lowest orders of such expansions. In what follows, we consider
the case of two symbiotic species.

Probably the most often met type of symbiosis is when the species influence
the mutual carrying capacities just by the existence of the species themselves,
resulting in the change on the mutual livelihoods caused by the species vital
activity. For instance, most land ecosystems rely on symbiosis between the
plants that extract carbon from the air and mycorrhizal fungi extracting
minerals from the ground. If the mutual influence is weak, the corresponding
carrying capacities can be modeled by the linear approximation
\be
\label{4}
  y_1(x,z) \simeq 1 + b_1 z \; , \qquad y_2(x,z) \simeq 1 + g_1 x \; ,
\ee
in which $b_1$ and $g_1$ are the parameters characterizing the productive or
destructive influence of the related species on their counterparts. This type
of symbiosis can be termed passive, since the species do not directly interact
in the process of varying the livelihoods of their neighbors, in the sense
that the impacts of species abundance on carrying capacities are linear.

\subsection{Active weak symbiosis}

If in the process of influencing the livelihoods of each other the species
directly interact, then their carrying capacities are approximated by the
bilinear expressions
\be
\label{5}
  y_1(x,z) \simeq 1 + b_2 xz \; , \qquad y_2(x,z) \simeq 1 + g_2 xz \;  ,
\ee
assuming that their interactions are sufficiently weak. The parameters $b_2$
and $g_2$ again can be interpreted as production or destruction coefficients,
according to the production or destruction affecting the related carrying
capacities. Examples of this type of symbiosis could be the relations between
different firms producing goods in close collaboration with each other, or the
relation between basic and applied sciences.

\subsection{Mixed weak symbiosis}

The intermediate type of symbiosis is when one of the species, in the process
of influencing the carrying capacity, interacts with the other species, while
the latter acts on the carrying capacity of the neighbor just by means of its
vital activity, without explicit interactions. This type of symbiosis is common
for many biological and social systems, when one of them is a subsystem of the
larger one. Thus, such relations exist between the country gross domestic
product and the level of science, or between culture and language. In the case
of weak influence of this type, the carrying capacities are represented as the
expansions
\be
\label{6}
  y_1(x,z) \simeq 1 + b_2 xz \; , \qquad y_2(x,z) \simeq 1 + g_1 x \;  .
\ee

\section{Arbitrarily Strong Influence on Livelihoods of Symbiotic Species}

When one or both of the species are parasites, the corresponding parameters
$b_i$ or $g_i$ are negative. Hence the effective carrying capacity $y_i$ can
become zero or negative. Then, with the approximations (4), (5) or (6), the
evolution equations (3) can result in finite-time singularities of the population
fraction or in the disappearance of the solutions, as is found in the previous
publications \cite{Yukalov_6,Yukalov_12}. Such a situation looks artificial,
being just the result of the linear or bilinear approximations involved. In
order to avoid that carrying capacities can become zero and negative, it is
necessary to employ a more elaborated expression for the carrying-capacity
functionals.

\subsection{Passive nonlinear symbiosis}

In order to be able to consider mutual symbiotic relations of arbitrary
strength, it is necessary to generalize the carrying capacities. The latter
can be treated as expansions in powers of the species fractions. In the case
of weak symbiosis, it has been possible to limit oneself to the first terms
of such expansions. However, for generalizing the applicability of the
approach, it is necessary to define the effective sums of such expansions.
The effective summation of power series can be done by means of the
self-similar approximation theory
\cite{Yukalov_20,Yukalov_21,Yukalov_22,Yukalov_23,Gluzman_24}.
Under the condition of obtaining an effective sum that does not change its
sign for the variables defined on the whole real axis, one needs to resort
to the exponential approximants \cite{Yukalov_25,Gluzman_26,Gluzman_27,Gluzman_28,
Yukalov_29,Gluzman_30}. Limiting ourselves by the simplest form of such an
exponential approximation, for the case of expansion (4), we have the
effective sums
\be
\label{7}
 y_1(x,z) = e^{bz} \; , \qquad y_2(x,z) = e^{gx} \;  .
\ee
The parameters $b$ and $g$, depending on their signs, characterize the
creative or destructive influence of the species on the carrying capacities of
their coexisting neighbors.

\subsection{Active nonlinear symbiosis}

Similarly involving the method of self-similar exponential approximants to
perform the summation for the previous case of passive symbiosis,
starting from the terms (5), we obtain the nonlinear carrying capacities
\be
\label{8}
 y_1(x,z) = e^{bxz} \; , \qquad y_2(x,z) = e^{gxz} \;  ,
\ee
with the same interpretation of the symbiotic parameters $b$ and $g$.

\subsection{Mixed nonlinear symbiosis}

In the case of the mixed symbiosis, with the carrying capacities starting from
terms (6), we get the effective summation resulting in the nonlinear expressions
\be
\label{9}
y_1(x,z) = e^{bxz} \; , \qquad y_2(x,z) = e^{gx} \;   .
\ee

In this way, we obtain the generalizations for the carrying capacities that are
applicable to the mutual symbiotic relations of arbitrary strength.

\subsection{Variants of symbiotic coexistence}

It is worth stressing that the suggested approach allows us to treat all known
variants of symbiosis, which is regulated by the values of the symbiotic
parameters $b$ and $g$. Thus, when both these parameters are positive, this
corresponds to mutualism, which is a relationship between different species where
both of them derive mutual benefit:
\be
\label{10}
 b > 0 \; , \qquad g > 0 \qquad (mutualism) \;  .
\ee

When one of the species benefits from the coexistence with the other species,
while the other one is neutral, getting neither profit nor harm, this
relationship corresponds to commensalism, defined by one of the conditions
\begin{eqnarray}
\label{11}
\begin{array}{lll}
b > 0 \; , & ~~~ g = 0 \; ; & \\
b = 0 \; , & ~~~ g > 0      & ~~~ (commensalism) \;  .
\end{array}
\end{eqnarray}

Finally, if at least one of the coexisting species is harmful to the other
one, this is typical of parasitism that is characterized by the validity of
one of the conditions
\begin{eqnarray}
\label{12}
\begin{array}{lll}
b > 0 \; , & ~~~ g < 0 \; ; & \\
b = 0 \; , & ~~~ g < 0 \; ; & \\
b < 0 \; , & ~~~ g < 0 \; ; & \\
b < 0 \; , & ~~~ g = 0 \; ; & \\
b < 0 \; , & ~~~ g > 0      & ~~~(parasitism) \;  .
\end{array}
\end{eqnarray}

Varying the symbiotic parameters $b$ and $g$ results in a variety of
bifurcations between different dynamical regimes. Below, we study these
bifurcations employing the general methods of dynamical theory
\cite{Thompson_31,Kuznetsov_32,Chen_33,Leonov_34}.

\section{Passive Nonlinear Symbiosis}

Equations (3), with the carrying capacities (7) acquire the form
\be
\label{eq1}
\frac{dx}{dt} = x - x^2 e^{-bz} \; , \qquad
\frac{dz}{dt} = z -z^2 e^{-gx} \; ,
\ee
with the parameters $g \in (-\infty, \infty)$ and $b \in (-\infty, \infty)$,
and with the initial conditions $x_0 = x(0)$, $z_0 = z(0)$. We are looking
for non-negative solutions $x(t) \geq 0$ and $z(t) \geq 0$ to the system of
these differential equations (\ref{eq1}).

We may note that system (\ref{eq1}) is symmetric with respect to the change
$b \longleftrightarrow g$, $x_0 \longleftrightarrow z_0$, and
$x(t) \longleftrightarrow z(t)$.

The overall phase portrait for the case of the passive symbiosis will be
displayed below.

\subsection{Existence of evolutionary stable states}

Depending on the signs of the parameters $g$ and $b$, the system of
equations (\ref{eq1}) possesses different numbers of stationary states
(fixed points) $\{x^*,z^*\}$. For any values of the parameters, there always
exist three trivial fixed points, $\{ x^* = 0, z^* = 0\}$,
$\{ x^* = 1, z^* = 0\}$, and $\{ x^* = 0, z^* = 1 \}$, which are unstable
for all $g$ and $b$.

Nontrivial fixed points $\{x^* \neq 0, z^* \neq 0\}$ are the solutions to
the equations:
\be
\label{eq2}
x^* = e^{bz^*} \; , \qquad z^* = e^{gx^*} \; .
\ee

The characteristic exponents, defining the stability of the stationary
solutions, are given by the expression
\be
\label{eq3}
\lbd_{1,2} = - 1 \pm \sqrt{bgx^*z^*} \; ,
\ee
where $x^*$ and $z^*$ are the fixed points defined by Eqs. (\ref{eq2}),
which can also be represented as
$$
x^* = \exp(b e^{gx^*}) \; , \qquad z^* = \exp(g e^{bz^*}) \;  .
$$

Analysing the existence and stability of the fixed points, we find that,
when $b \in (0, \infty)$ and $g \in (0, \infty)$, there exists a line
$g = g_c(b)$, with respect to which two possibilities can occur:

\begin{itemize}

\item
If $0 < g < g_c(b)$, then Eqs. (\ref{eq2}) have two solutions
$\{x_1^*, z_1^*\}$ and $\{x_2^*, z_2^*\}$, such that $1 < x_1^* < x_2^*$
and $1 < z_1^* <z_2^*$. Numerical investigation of (\ref{eq2}) and (\ref{eq3})
shows that the lower fixed point $\{x_1^*,z_1^*\}$ is stable, while the higher
fixed point $\{x_2^*,z_2^*\}$ is unstable.

\item
If $g > g_c(b)$, then Eqs. (\ref{eq2}) do not have solutions, hence
Eqs. (\ref{eq1}) do not possess stationary states. When $b = 1/e$, then
$g_c(b) = 1/e$ and $x^* = z^* = e$.

\end{itemize}

\vskip 2mm

If either $b \leq 0$ and $ g > 0$, or $b > 0$ and $g \leq 0$, then there
exists only one stationary solution $\{x^*,z^*\}$, which is a stable fixed
point characterized by the corresponding characteistic exponents (\ref{eq3})
with ${\rm Re} \lbd_{1,2} < 0$, which follows from the inequalities
$\lbd_1 \lbd_2 = 1 - bgx^*z^* > 0$ and $-(\lbd_1 + \lbd_2) = 2 > 0$.

Note that if $b \leq 0$ and $g > 0$, then $x^* < 1$ and $z^* > 1$, while if
$b > 0$ and $g \leq 0$, then $x^* > 1$ and $z^* < 1$.

When $b < 0$ and $g < 0$, there can exist up to three fixed points, such that
$x^* < 1$ and $z^* < 1$. For $b \leq -e$, there exist two lines $g_1=g_1(b)<0$
and $g_2 = g_2(b) < 0$, for which $g_1(-e) = g_2(-e) = -e$. The following two
possibilities can  happen:

\begin{itemize}

\item
If $b < -e$ and $g_1(b) < g < g_2(b)$, then Eqs. (\ref{eq2}) have 3 solutions.
Numerical analysis shows that two fixed points, $\{x_1^*,z_1^*\}$ and
$\{x_2^*,z_2^*\}$, are stable, while the third fixed point, $\{x_3^*,z_3^*\}$,
is unstable.

\item
If either $-e \leq b < 0$ and $g < 0$, or $b < -e$ and $g < g_1(b)$, or
$b < -e$ and $g_2(b) < g < 0$, then Eqs. (\ref{eq2}) have only one solution,
which is a stable fixed point.

\end{itemize}

\vskip 2mm

In the limiting cases, if $g \ra 0$, then $z^* \ra 1$ and $x^* \ra e^b$.
If $b \ra 0$, then $x^* \ra 1$ and $z^* \ra e^g$.

If $b \ra -\infty$ and $g > 0$ is fixed, then $x^* \ra 0$ and $z^* \ra 1$.
When $g \ra -\infty$ and $b > 0$ is fixed, then $z^* \ra 0$ and $x^* \ra 1$.

If $b \ll -1$, and $g \ll -1$, then, depending on the relation between the
parameters, it may be that either $x^* \ra 0$ and $z^* \ra 1$, or $x^* \ra 1$
and $z^* \ra 0$, or $x^* \ra 0$ and $z^* \ra 0$, but at finite parameter
values, one always has finite $x^* \neq 0$ and $z^* \neq 0$.

Figure 1a clarifies the regions of the stable fixed-point existence in the
plane $b - g$. The region of existence for $b, g < 0$ is detailed in Fig. 1b.

\subsection{Dynamics of populations under mutualism ($b > 0, g > 0$)}

There exist two types of dynamical behavior: (i) {\it Unbounded growth of
populations} or (ii) {\it convergence to a stationary state}.

\vskip 2mm

(i) {\it Unbounded growth of populations}

\vskip 2mm

When $0 < b < \infty$ and $g > g_c(b) > 0$, then Eqs. (\ref{eq2}) do not have
solutions, hence Eqs. (\ref{eq1}) do not have stationary states. For any
choice of the initial conditions $\{x_0,z_0\}$, solutions
$x(t), z(t) \ra \infty$, as $t \ra \infty$. The corresponding behavior
of $x(t)$ and $z(t)$ is shown in Fig.2a.

\vskip 2mm

(ii) {\it Convergence to stationary states}

\vskip 2mm

If $0 < b < \infty$ and $0 < g < g_c(b)$, there exist two solutions to
Eqs. (\ref{eq2}), such that $1 < x_1^* < x_2^*$ and $1 < z_1^* < z_2^*$.
Numerical analysis shows that the lower solution $\{x_1^*,z_1^*\}$ is stable,
while $\{x_2^*,z_2^*\}$ is unstable. Around the stable fixed point, there
exists a finite basin of attraction, so that if the initial conditions
$\{x_0,z_0\}$ belong to it, then $x(t) \ra x_1^*$, and $z(t) \ra z_1^*$,
as $t \ra \infty$. But if the initial conditions do not belong to the basin
of attraction, the species fractions rise to infinity for $t \ra \infty$.

Figure 2b demonstrates the behavior or solutions for $g$ slightly above the
boundary ($g > g_c(b)$) and slightly below it ($g < g_c(b))$. Below the
boundary $g_c = g_c(b)$, there exists a region of parameters $b,g$, such
that $x(t)$ and $z(t)$ converge to a stationary state, provided that the
initial conditions are inside the basin of attraction. Figures 2c and 2d
illustrate the behaviour of $x(t)$ and $z(t)$ for the same choice of the
parameters $b,g$, but for different initial conditions.

\subsection{Dynamics of populations with one parasitic species
(either $b < 0$, $g> 0$ or $b > 0$, $g < 0$)}

Only one regime exists, when the {\it populations tend to their stationary
states}, with the attraction basin being the whole region of $x_0, z_0$.

Recall that there is a symmetry in Eqs. (\ref{eq1}), such that under the
replacement $b \longleftrightarrow g$ and $x_0 \longleftrightarrow z_0$,
we have $x(t) \longleftrightarrow z(t)$ and $x^* \longleftrightarrow z^*$.
Therefore, it is sufficient to consider only one case, say, when
$b < 0$ and $g > 0$.

When $b < 0$ and $g > 0$, Eqs. (\ref{eq2}) have a single solution
$\{x^* < 1, z^* > 1 \}$, which is a stable fixed point. Conversely, when
$b > 0$ and $g < 0$, then $x^* > 1$ and $z^* < 1$. This tells us that the
species population coexisting with a parasite is suppressed.

The populations converge to their stationary points irrespectively of the
choice of the initial conditions. The convergence can be either monotonic
or non-monotonic, as is demonstrated in Figs. 3 and 4.

\subsection{Dynamics of populations with two parasitic species
($b < 0$, $g < 0$)}

Depending on the symbiotic parameters, there can exist either
(i) {\it a single stationary state} or (ii) {\it bistability with two
stationary states}.

When $b < 0$ and $g < 0$, then Eqs. (\ref{eq2}) have either a single solution
$\{x^*,z^*\}$, which is a stable fixed point, or three solutions, among which
two solutions, $\{x_{1,2}^*,z_{1,2}^*\}$, are stable fixed points and the third,
$\{x_3^*,z_3^*\}$, is a saddle. In all the cases, $x^* < 1$ and $z^* < 1$,
which means that two parasitic species cannot develop large populations.

\vskip 2mm

(i) {\it Single stationary state}

For $b,g \leq -e$, there exist two lines, $g_1(b)$ and $g_2(b)$, such that
$g_1(-e) = g_2(-e) = -e$ and $g_1(b) < g_2(b) < 0$. If either $-e < b < 0$ and
$g < 0$, or $b < -e$ and $g_2(b) < g < 0$, or $b < -e$ and $g < g_1(b)$, then
there exists a single solution to Eqs. (\ref{eq2}), which is a stable fixed
point.

\vskip 2mm

(ii) {\it Bistability with two stationary states}.

\vskip 2mm

If $b < -e$ and $g_1(b) < g < g_2(b)$, there exist three solutions to
Eqs. (\ref{eq2}). Two solutions are stable fixed points and the third solution
is a saddle. The bifurcation point corresponds to $b = g = -e$. At this point,
$x^* = z^* = 1/e$.

\vskip 2mm

The populations $x(t)$ and $z(t)$ tend monotonically or non-monotonically to
their stable stationary states, from above or from below, as is shown in
Figs. 5 and 6. In the region of the symbiotic parameters $b,g < 0$, where
a single stable fixed point exists, $x(t) \ra x^*$, $z(t)\ra z^*$, as
$t \ra \infty$, irrespectively of the initial conditions.

In the region of the parameters $b$ and $g$, where two stable fixed point
exist, the population convergence depends on the choice of initial conditions.
For $t \ra \infty$, the population $\{x(t), z(t)\}$ tend either to
$\{x^*_1, z^*_1\}$ or to $\{x^*_2, z^*_2\}$, depending on the initial
conditions $\{x_0, z_0\}$ being in the attraction basin of the related fixed
point.

On the line $-e < b = g < 0$, there exists a single stationary state
$\{x^* = z^*\}$ that is a stable fixed point.

On the line $b = g < -e$, there are three fixed points. Two fixed points
are stable, so that $x_1^* = z_2^*$ and $z_1^* = x_2^*$. The third fixed
point, $\{x_3^* = z_3^*\}$ is unstable. For $b = g \ra -\infty$, we have
$x_1^* \ra 0$, $z_1^* \ra 1$, $x_2^* \ra 1$, $z_2^* \ra 0$, and
$x_3^* = z_3^* \ra 0$.

\subsection{Phase portrait for passive nonlinear symbiosis}

Fig. 7 provides an overview of the different regimes analysed in this
section concerned with the analysis of equation (\ref{eq1}).

Panels (a) and (b) illustrate the regime of mutualism, in which
the two species benefit from each other. Panel (a) represents the
situation of sufficiently symmetric mutualism, in which the long-term
behaviour is characterised by an exponential growth for both species.
Note that asymmetric initial conditions lead to a non-monotonous behaviour
of the species population that is initially too large, which has to shrink
first before growing again in synergy with the other species.
Panel (b) corresponds to the situation of large mutualism asymmetries.
In this case, the two population need to be above a certain threshold
in order to reach the regime of unbounded growth. Otherwise, their
populations are trapped and converge to a steady state.

Panel (c) illustrates the dynamics of a very asymmetric situation where
one species augments the carrying capacity of the other, while the later
has a negative effect on the carrying capacity of the former. In this case,
the dynamics of the two species population converges to a stable fixed point,
a steady state characterising a compromise in the destructive interactions
between the two species.

In the situation shown in panel (d) in which both species tend to destroy the
carrying capacity of the other, a saddle node separates two symmetrical stable
fixed points, where one species in general profits much more than the other.
This is an example of a spontaneous symmetry breaking, in which the two stable
fixed points are exactly symmetric under the transformation
$x \longleftrightarrow z$. The selection of one or the other stable fixed points
depends on the initial conditions, as usual in spontaneous symmetry breaking.
A slight initial advantage of $x$ above $z$ is sufficient to consolidate into
a very large asymptotic population difference. Such behaviours are reminiscent
to many real-life situations in which a slight initial favourable situation
becomes entrenched in a very strong prominent role. Our analysis shows that such
a situation occurs generically in the case of passive symbiosis when the two
species are competing destructively. It is tempting to interpret real-life
geopolitical and economic situations in particular as embodiment of such a
scenario.

\section{Active Nonlinear Symbiosis}

This case is described by the system of equations
\be
\label{eqq1}
\frac{dx}{dt} = x - x^2 e^{-bxz} \; , \qquad
\frac{dz}{dt} = z -z^2 e^{-gxz} \; ,
\ee
with the symbiotic parameters $g \in (-\infty, \infty)$,
$b \in (-\infty, \infty)$, and the initial conditions $x_0 = x(0)$,
$z_0 = z(0)$. Again, only non-negative solutions
$x(t) \geq 0$ and $z(t) \geq 0$ are of interest. Note that the system (\ref{eqq1})
is symmetric with respect to the replacement
$b \longleftrightarrow g$, $x_0 \longleftrightarrow z_0$, and
$x(t) \longleftrightarrow z(t)$.

\subsection{Existence of evolutionary stable states}

Similarly to the previous case, there always exist three trivial fixed
points, $\{ x^* = 0, z^* = 0 \}$, $\{ x^* = 1, z^* = 0 \}$, and
$\{ x^* = 0, z^* = 1 \}$, which are unstable for any $g$ and $b$.

Nontrivial fixed points $\{ x^*\neq 0, z^*\neq 0 \}$ are the solutions to the
equations:
\be
\label{eqq2}
x^* = e^{bx^*z^*} \; , \qquad z^* = e^{gx^*z^*} \; .
\ee
These Eqs. (\ref{eqq2}) can be represented as
\be
\label{eqq3}
x^* = \exp(b(x^*)^{1+g/b}) \;  , \qquad z^* = \exp(g(z^*)^{1+b/g}) \; .
\ee

The characteristic exponents are given by the equations
\be
\label{eqq4}
\lbd_1 = -1 \; , \qquad \lbd_2 = -1 +(b+g)x^*z^* \; ,
\ee
where $x^*$ and $z^*$ are the solutions to Eqs. (\ref{eqq2}). From here,
it follows that the fixed points are stable for all $b + g < 0$.

The following situations can occur:

\begin{itemize}

\item
If $b + g > 1/e$, then Eqs. (\ref{eqq2}), or (\ref{eqq3}), do not have
solutions.

\item
If $0 < b + g < 1/e$, then Eqs. (\ref{eqq2}) or (\ref{eqq3}) have two
solutions $\{x_1^*, z_1^*\}$ and $\{x_2^*, z_2^*\}$.

Here, when $b > 0$, $g > 0$, then $1 < x_1^* < x_2^*$ and $1 < z_1^* < z_2^*$.
When $b < 0$ and $g > 0$, then $1 > x_1^* > x_2^*$ and $1 < z_1^* < z_2^*$.
If $b > 0$ and $g < 0$, then $1 < x_1^* < x_2^*$, but $1 > z_1^* > z_2^*$.
Numerical investigation shows that the fixed point $\{x_1^*, z_1^*\}$ is stable,
while the second fixed point $\{x_2^*, z_2^*\}$ is a saddle.

\item
If $b + g < 0$, then there exists only one solution to Eqs. (\ref{eqq2})
or (\ref{eqq3}) that is a stable fixed point.

\end{itemize}

\vskip 2mm

The region of the existence of the stable fixed-point in the plane $b-g$ is
presented in Fig. 8.

When $0 < b + g < 1/e$, the stable fixed point possesses a finite basin
of attraction. If $b + g \leq 0$, then the basin of attraction is the whole
region $\{x_0 \geq 0, \; z_0 \geq 0 \}$.

\subsection{Population dynamics under $b + g > 1/e$}

In this case, Eqs. (\ref{eqq2}) or (\ref{eqq3}) do not have solutions,
which means that there are no fixed points. There can happen two types
of solutions: (i) {\it Unbounded growth of mutualistic populations} or
(ii) {\it Unbounded growth of a parasitic population and dying out of the
host population}.

\vskip 2mm

(i) {\it Unbounded growth of mutualistic populations}

\vskip 2mm

In the case of mutualism, when $b > 0$ and $g > 0$, both populations
grow so that $x(t) \ra \infty$ and $z(t) \ra \infty$, as $t \ra \infty$.

\vskip 2mm

(ii) {\it Unbounded growth of parasitic population and dying out of
host population}

When one of the species is parasitic, the host population dies out.
If $b < 0$, but $g > 0$, then $x(t) \ra 0$ and $z(t) \ra \infty$, as
$t \ra \infty$. Conversely, if $b > 0$, but $g < 0$, then $x(t) \ra \infty$,
while $z(t) \ra 0$, as $t \ra \infty$.

\vskip 2mm

The population dynamics for the case $b + g > 1/e$ is presented in Fig. 9.
Because of the symmetry of Eqs. (\ref{eqq1}), it is sufficient to consider
only the cases when $b > 0$, $g > 0$ and $b < 0$, $g > 0$.

\subsection{Population dynamics under $b + g < 1/e$}

There exists only one stable fixed point, so that the population dynamics
depends on whether the initial conditions are inside the attraction basin
or not. Respectively, there can happen three kinds of behavior:
(i) {\it Convergence to a stationary state}; (ii) {\it Unbounded growth of
both populations}; (iii) {\it Unbounded growth of parasitic population and
dying out of host population}.

\vskip 2mm

(i) {\it Convergence to stationary state}

\vskip 2mm

When $0 < b + g < 1/e$ and the initial conditions $\{x_0, z_0\}$ are in the
basin of attraction of the fixed point $\{x^*, z^*\}$, then the populations
$x(t) \ra x^*$ and $z(t) \ra z^*$, as $t \ra \infty$.

If $b + g \leq 0$, then the basin of attraction is the whole region of
$x_0 \geq 0$ and $z_0 \geq 0$, that is, for any choice of the initial
conditions $\{x_0, z_0\}$ the solutions $\{x(t), z(t)\} \ra \{x^*, z^*\}$,
as $t \ra \infty$.

\vskip 2mm

(ii) {\it Unbounded population growth}

\vskip 2mm

When $0 < b + g < 1/e$ and $b,g>0$, there exists a finite basin of attraction,
so that if the initial conditions $\{x_0, z_0\}$ are outside of the attraction
basin, then the populations $\{x(t), z(t)\}$ unboundedly grow with increasing
time $t \ra \infty$.

\vskip 2mm

(iii) {\it Unbounded growth of parasitic population and
dying out of host population}

When $0 < b + g < 1/e$, with $b < 0$ but $g > 0$, and initial conditions
$\{x_0, z_0\}$ are outside of the attraction basin, then for $t\ra\infty$, one
of the species is dying out, $x(t)\ra 0$, while another one experiences unbounded
growth $z(t)\ra \infty$. Conversely, if $b>0$, but $g < 0$, then $x(t) \ra \infty$,
while $z(t) \ra 0$, as $t \ra \infty$.

Figure 10 demonstrates different types of the population convergence to a
stable stationary state $\{x^*, z^*\}$.

\subsection{Phase portrait for active nonlinear symbiosis}

Figure 11 provides an overview of the different regimes analysed in this
section concerned with the analysis of equation (\ref{eqq1}).
Qualitatively, the nature of the different regimes are similar to those
discussed with respect to the phase portrait shown in Fig. 7
summarising the different regime for passive symbiosis described by
equation (\ref{eq1}).

\section{Mixed Nonlinear Symbiosis}

The system of equations describing this case is
\be
\label{eqqq1}
\frac{dx}{dt} = x - x^2 e^{-bxz} \; , \qquad
\frac{dz}{dt} = z -z^2 e^{-gx} \; ,
\ee
with the symbiotic parameters $g \in(-\infty, \infty)$,
$b \in(-\infty, \infty)$, and the initial conditions $x_0 = x(0)$, $z_0 = z(0)$.
As usual, only non-negative solutions $x(t) \geq 0$ and $z(t) \geq 0$ are of
interest.

Note that contrary to Eqs. (\ref{eq1}) and (\ref{eqq1}), the system of
equations (\ref{eqqq1}) is not symmetric. The carrying capacity of the
population $x(t)$ is formed in the process of interactions between the coexisting
species. In contrast, the carrying capacity of the population $z(t)$ is influenced
only by the population $x(t)$. Therefore, we may call the population $x(t)$
{\it active}, but the population $z(t)$ {\it passive}.

\subsection{Existence of evolutionary stable states}

Similarly to the previous cases, there always exist 3 trivial fixed points,
$\{ x^* = 0, z^* = 0\}$, $\{ x^* = 1, z^* = 0\}$, and $\{ x^* = 0, z^* = 1 \}$,
which are unstable for all $b, g \in (-\infty, \infty)$.

Nontrivial fixed points $\{x^*\neq 0, z^*\neq 0\}$ are the solutions to the
equations:
\be
\label{eqqq2}
x^* = e^{bx^*z^*} \; , \qquad z^* = e^{gx^*} \; ,
\ee
which can be transformed into
\be
\label{eqqq3}
x^* = \exp(bx^*e^{gx^*}) \; , \qquad z^* = \exp(gz^{bz^*/g}) \; .
\ee

The characteristic exponents, defining the stability of the stationary states,
are given by the equations
\be
\label{eqqq4}
\lbd_{1,2} = \frac{1}{2} \left [ bx^*z^*-2 \pm x^* \sqrt{b z^* (4g + bz^*) } \right ] \; ,
\ee

The following cases can occur:

\begin{itemize}

\item

When $b > 0$, there exists a boundary $g = g_c(b)$, such that if $g < g_c(b)$,
then there can exist up to three fixed points, but only one of them being
stable.

\item

If either $0 < b < 1/e$ and $g > g_c(b) > 0$, or $b > 1/e$ and $g \geq 0$, then
Eqs. (\ref{eqqq2}) do not have solutions, hence, there are no fixed points.

\item

For $b > 1/e$ and $g_c(b) < g < 0$, there exists only one unstable fixed
point.

\item

If $0 < b < 1/e$ and $0 \leq g < g_c(b)$, there is one stable and one unstable
fixed points.

\item

When $b \leq 0$ and $g \in (-\infty, \infty)$, there exists a single fixed
point that is stable. For $b < 0$ and $g > 0$, we have $x^* < 1, z^* > 1$,
while when $b < 0$ and $g < 0$, then $x^* < 1, z^* < 1$.

\item

For $0 \leq b \leq b_0 \approx 0.47$, there exists an additional
line $g_0(b)$, such that $g_0(0) = 0$ and $g_0(b_0) = g_c(b_0)\approx -0.075706$.
If either $0 < b < b_0$ and $g < g_0(b)$ or $b \geq b_0$ and $g < g_c(b)$,
there is a single fixed point that is stable.

\item

When either $0 < b < 1/e$ and $g_0(b) < g < 0$, or $1/e \leq b < b_0$ and
$g_0(b) < g < g_c(b)$, then Eqs. (\ref{eqqq2}) have three solutions, but only
one of them is a stable fixed point.

\end{itemize}

\vskip 2mm

In the limiting case, we have $g_c(1/e) = 0$. When $b \ra +0$, then
$g_c(b) \ra \infty$. If $b \ra \infty$, then $g_c(b) \ra -\infty$. On the
boundary $\{b \in (0, b_0), g = g_0(b)\}$, two solutions, corresponding to
unstable fixed points, coincide and disappear as soon as $g \leq g_0(b)$.

\vskip 2mm

The region of the existence of the fixed-point in the plane $b-g$ is presented
in Fig. 12.

The dynamics of the symbiotic populations strongly depends on whether the influence
of the passive species $z(t)$ on the active species $x(t)$ is mutualistic
or parasitic, which is described by their interaction symbiotic parameter $b$.
Therefore, these two cases will be treated separately.

\subsection{Dynamics of populations with mutualistic passive species ($b > 0$)}

Depending on the symbiotic parameters $b$ and $g$, there are three types of
dynamic behavior: (i) {\it Unbounded growth of populations}; (ii) {\it everlasting
oscillations}, and (iii) {\it convergence to a stationary state}.

\vskip 2mm

(i) {\it Unbounded growth of populations}

\vskip 2mm

When either $0 < b < 1/e$ and $g > g_c(b) > 0$, or $b > 1/e$ and $g \geq 0$,
then Eqs. (\ref{eqqq1}) do not have fixed points. Populations $x(t) \ra \infty$
and $z(t) \ra \infty$, either monotonically or non-monotonically, as
$t \ra \infty$. The logarithmic behaviour of diverging solutions $x(t)$ and
$z(t)$ is shown in Figs. 13a and 13b.

When $0 < b < 1/e$ and $0 \leq g < g_c(b)$, there are two fixed points, but
only one of them is stable. In this case, there exists a finite basin of
attraction for the stable fixed point. If the initial conditions $\{x_0, z_0\}$
are outside of the attraction basin, the populations $x(t)$ and $z(t)$ diverge,
as $t \ra \infty$.

\vskip 2mm

(ii) {\it Everlasting oscillations}

\vskip 2mm

When $b > 1/e$ and $g_c(b) < g < 0$, then Eqs. (\ref{eqqq2}) have only one
solution $\{x^*,z^*\}$, which is an unstable focus, with complex
characteristic exponents, such that $\lbd_2 = \lbd_1^*$ and
${\rm Re}\lbd_{1,2} > 0$. In this region of the parameters $b$ and $g$,
there exists a limit cycle, and the populations $x(t)$ and $z(t)$ oscillate
without convergence for all $t > 0$.

When $g \ra -0$, then the amplitude of oscillations drastically increases,
but remains finite. On the half line, $b > 1/e$ and $g = 0$, solutions $x(t)$
and $z(t)$ diverge, as $t \ra \infty$.

\vskip 2mm

The dynamics of populations $x(t)$ and $z(t)$ are shown in Figs. 13c,d, 14a,b,
15c,d, and 16c,d. Figure 17 presents the logarithmic behavior of the
populations $x(t)$, $z(t)$ for the fixed $b = 0.4$ and varying $g$, such
that either $g_c(b) < g < 0$ or $g_0(b) < g < g_c(b)$.

\vskip 2mm

(iii) {\it Convergence to stationary states}

When $b > 0$ and $g < g_c(b)$, there exists one stable fixed point
$\{x^*, z^*\}$.

When $0 < b < 1/e$ and $0 < g < g_c(b)$, then there exists a finite basin
of attraction for the fixed point. If the initial conditions $\{x_0, z_0\}$
are inside the basin of attraction, then $x(t) \ra x^*$ and $z(t) \ra z^*$,
as $t \ra \infty$.

If either $0 < b < 1/e$ and $g \leq 0$, or $b > 1/e$ and $g < g_c(b) < 0$,
then the basin of attraction is the whole region of $\{x_0 \geq 0, z_0 \geq 0\}$.
In that case, for any initial conditions, $x(t) \ra x^*$ and $z(t) \ra z^*$,
as $t \ra \infty$.

There are two possible ways of convergence to a stationary state, either
with oscillations, when the fixed point is a focus, or without oscillations,
when the fixed point is a node.

When $b > b_0$ and $g < g_c(b)$, and $g$ is in the vicinity of $g = g_c(b)$,
then $x(t) \ra x^*$ and $z(t) \ra z^*$ with oscillations. The
corresponding behaviour is shown in Figs. 14c,d, 15c,d (solid lines), and
Figs. 16c,d (dashed lines).

When $0 < b < 1/e$ and $0 < g < g_c(b)$ and the initial conditions
$\{x_0, z_0\}$ are inside the attraction basin of the fixed point, then
$x(t) \ra x^*$, $z(t) \ra z^*$ without explicit oscillations, though the
convergence can be either monotonic or non-monotonic.

If either $0 < b < 1/e$ and $g \leq 0$, or $1/e < b < b_0$ and $g < g_c(b)$,
or $b > b_0$ and $g \ll g_c(b)$, then $x(t) \ra x^*$ and $z(t) \ra z^*$
without oscillations.

Examples of convergence without oscillations are given in Figs. 15 and 16.

\vskip 2mm

The bifurcation line $g = g_c(b) < 0$ for $b > b_0$ is the line of a
supercritical Hopf bifurcation. It separates the regions where the real parts
of the complex characteristic exponents have different signs. When $b > b_0$
and $g < g_c(b) < 0$, then the characteristic exponents of the stable fixed
point are complex, with the negative Lyapunov exponents ${\rm Re}\lbd_{1,2} < 0$.
When $b > b_0$ and $g_c(b) < g < 0$, then the stable focus transforms into
an unstable focus, with positive Lyapunov exponents ${\rm Re}\lbd_{1,2} > 0$.
At the same time, there appears a stable limit cycle. On the bifurcation line,
where $b > b_0$ and $g = g_c(b)$ the Lyapunov exponents of the focus are zero,
${\rm Re}\lbd_{1,2} = 0$, as it should be under a Hopf bifurcation
\cite{Kuznetsov_32,Chen_33,Cobiaga_35}.

A very interesting behaviour is associated with the transition across the line
$g_c(b)$, when $1/e < b < b_0$. In the region where either $0 < b < 1/e$ and
$g_0 < g < 0$, or $1/e < b < b_0$ and $g_0(b) < g < g_c(b)$, there are three
fixed points, a stable node, a saddle, and an unstable focus. In the region
$b > 1/e$ and $g > g_c(b)$, there is one unstable focus and a limit cycle.
When approaching the boundary $g_c(b)$, moving from one region to the other,
the stable node and saddle become close to each other and on the boundary they
coincide, while disappearing to the right from this line $g_c(b)$. The unstable
focus safely moves through the boundary to the region where a limit cycle appears.

\subsection{Dynamics of populations with parasitic passive species ($b < 0$)}

There is only one regime of {\it convergence to stationary states}.

When $b \leq 0$ and $g \in (-\infty, \infty)$, there exists just one
solution to Eqs. (\ref{eqqq2}), or (\ref{eqqq3}), and it is a stable
fixed point. The attraction basin is the whole region of the initial
conditions $\{x_0, z_0\}$, so that, for any initial condition, populations
$x(t) \ra x^*$ and $z(t) \ra z^*$, as $t \ra \infty$.

The corresponding behaviour of the populations is presented in Fig. 16.

\subsection{Phase portrait for mixed nonlinear symbiosis}

The phase portraits for the case of the mixed symbiosis described by
equation (\ref{eqqq1}), under different symbiotic parameters, are presented
in Figs. 18 and 19.

Fig. 18 for $g > 0$ shows three different regimes that are qualitatively
similar to those shown for the passive symbiosis cases in panels (a)-(c)
of Fig. 7.

Fig. 19 corresponds to the situation where $b > 0$ and $g < 0$, so that
the species $x$ tends to destroy the carrying capacity of the species $z$,
while the latter tends to augment nonlinearly the carrying capacity of the
former. As a consequence of the asymmetry of the equations and of this pair
of parameters, either convergent oscillations to a stable fixed point
or oscillatory solutions occur.

\section{Commensalism as Marginal Type of Symbiosis (either $b = 0$, or $g = 0$)}

\subsection{Independent species ($b = 0$ and $g = 0$)}

This is the extreme case, when the species are actually independent of each other.
If $b = 0$ and $g = 0$, then the systems of the evolution equations turn into two
independent logistic equations
\be
\label{d}
\frac{dx}{dt}= x - x^2 \; , \qquad  \frac{dz}{dt}= z - z^2 \; ,
\ee
with the initial conditions $x_0 = x(0)$ and $z_0 = z(0)$.

Solutions to these equations are known:
\be
\label{d1}
x(t)= \frac{x_0}{x_0-(x_0-1)e^{-t}} \; , \qquad
z(t)= \frac{z_0}{z_0-(z_0-1)e^{-t}} \; .
\ee
That is, $x(t) \ra x^* = 1$ and $z(t) \ra z^* = 1$, as $t \ra \infty$.

\subsection{Commensalism under $b = 0$, $g \neq 0$}

When $b = 0$ and $g \neq 0$ in equation (\ref{eqqq1}), this can be treated
as a limiting case of either passive or mixed symbiosis. Then the system of
equations (\ref{eq1}), or (\ref{eqqq1}), turns into
\be
\label{b}
\frac{dx}{dt}= x - x^2 \; , \qquad  \frac{dz}{dt}= z - z^2 e^{-gx(t)}  \; .
\ee
The solution $x(t)$ to the first equation of (\ref{b}) is given by the first
equation in (\ref{d1}). The stable set of fixed points is defined by the
expressions
$$
x^* = 1 \; , \qquad z^* = e^g \; ,
$$
which exist for all $g \in (-\infty, \infty)$. Populations
$\{x(t), z(t)\} \ra \{x^*, z^*\}$ monotonically either from above or from
below, depending on the given initial conditions.

\subsection{Commensalism under $b \neq 0$, $g = 0$}

When $b \neq 0$ and $g = 0$ in equation (\ref{eqqq1}), this is the limiting
case of either active or mixed symbiosis. The system of equations (\ref{eqq1}),
or (\ref{eqqq1}), becomes
\be
\label{c}
\frac{dx}{dt}= x - x^2  e^{-bxz}\; , \qquad  \frac{dz}{dt}= z - z^2  \; .
\ee
The solution $z(t)$ to the second equation of (\ref{c}) is given by the
second expression in (\ref{d1}). The fixed point $z^* = 1$ is stable, while
the fixed point $z^* = 0$ is unstable.

For the first equation in (\ref{c}), the trivial fixed point $x^* = 0$ is
also unstable. Nontrivial fixed points $x^* \neq 0$ are the solutions to
the equation
\be
\label{c1}
x^* = e^{bx^*}  \; .
\ee

When $b \leq 0$ there exists a single fixed point $x^* \leq 1$ that is
stable. Then, for any initial conditions, $x(t) \ra x^*$ and $z(t) \ra z^* = 1$,
as $t \ra \infty$.

If $0 < b < 1/e$, there exist two solutions, but only one solution, such that
$1 < x^* < e$, is stable, possessing a finite basin of attraction.

When $b > 1/e$, there are no fixed points for $x(t)$, hence $x(t) \ra \infty$,
though $z(t) \ra 1$, as $t \ra \infty$.

\vskip 2mm

Dynamics of the populations $x(t)$ and $z(t)$ in the case of commensalism,
described by Eqs. (\ref{d}), (\ref{b}), and (\ref{c}), is demonstrated
in Fig. 20.

\section{Summary of Main Symbiotic Behaviors}

For the convenience of the reader, we summarize here the main types of symbiotic
behaviors for the considered cases of symbiosis, classifying the population
dynamics with respect to the related symbiotic parameters.

\subsection{Dynamics under passive symbiosis}

The following types of population behavior can happen:

\begin{itemize}

\item
{\it Unbounded growth} for
\be
\label{S1}
0 < b < \infty \; , \qquad g > g_c(b) > 0 \;   .
\ee

\item
{\it Convergence to stationary states}, if the initial conditions are
inside the attraction basin, and {\it unbounded growth}, when the initial
conditions are outside the attraction basin, which occurs for
\be
\label{S2}
0 < b < \infty \; , \qquad 0 < g < g_c(b)  \;   .
\ee

\item
{\it Convergence to stationary states} for arbitrary initial conditions,
when either
\be
\label{S3}
 b <  0 \; , \qquad g > 0 \;  ,
\ee
or
\be
\label{S4}
  b > 0 \; , \qquad g < 0 \;   .
\ee

\item
{\it Convergence to stationary states} in the presence of {\it bistability},
when
\be
\label{S5}
  b <  0 \; , \qquad g < 0 \;   .
\ee

\end{itemize}

\subsection{Dynamics under active symbiosis}

The following regimes of population dynamics exist:

\begin{itemize}

\item
{\it Unbounded growth} if either
\be
\label{S6}
  b + g > \frac{1}{e} \;   , \qquad b > 0 \; , \qquad g > 0 \; ,
\ee
or
\be
\label{S7}
b + g <  \frac{1}{e} \; , \qquad b  > 0 \; , \qquad g > 0 \;  ,
\ee
and initial conditions are outside of the attraction basin.

\vskip 2mm

\item
{\it Dying out of the host species and unbounded proliferation of parasitic
species}, when either
\be
\label{S8}
  b <  0 \; , \qquad g > \frac{1}{e} + | b| \;  ,
\ee
or
\be
\label{S9}
b > \frac{1}{e} + | g | \; , \qquad g < 0 \;  ,
\ee
or
\be
\label{S91}
b  < 0 \; , \qquad | b | < g < \frac{1}{e} + | b| \; ,
\ee
and initial conditions are outside of the attraction basin,
or
\be
\label{S92}
g  < 0 \; , \qquad | g | < b < \frac{1}{e} + | g| \; ,
\ee
and initial conditions are outside of the attraction basin.

\vskip 2mm

\item
{\it Convergence to stationary states} for the initial conditions inside
the attraction basin if
\be
\label{S10}
 b + g \leq \frac{1}{e} \;  .
\ee

\end{itemize}

\subsection{Dynamics under mixed symbiosis}

There can exist the following regimes of population dynamics:

\begin{itemize}
\item
{\it Unbounded growth} when either
\be
\label{S11}
 0 < b < \frac{1}{e} \; , \qquad g > g_c(b) > 0 \;  ,
\ee
or
\be
\label{S12}
 b > \frac{1}{e} \; , \qquad g \geq 0 \;  .
\ee

\item
{\it Convergence to stationary states} for the initial conditions inside
the attraction basin and {\it unbounded growth} for the initial conditions
outside the attraction basin, when
\be
\label{S13}
 0 < b < \frac{1}{e} \; , \qquad 0 \leq g < g_c(b) \;  .
\ee

\item
{\it Convergence to stationary states} for all initial conditions, if either
\be
\label{S14}
  b > \frac{1}{e} \; , \qquad  g < g_c(b) < 0 \;  ,
\ee
or
\be
\label{S15}
0 < b < \frac{1}{e} \; , \qquad g \leq 0 \;   ,
\ee
or
\be
\label{S16}
 b \leq 0 \; , \qquad -\infty < g < \infty \;  .
\ee

\item
{\it Everlasting oscillations} under the condition
\be
\label{S17}
 b > \frac{1}{e} \; , \qquad g_c(b) < g < 0 \;  .
\ee

\end{itemize}

\section{Conclusion}

A novel approach to treating symbiotic relations between biological or social
species has been suggested. The principal idea of the approach is the
characterization of symbiotic relations of coexisting species through the
carrying capacities of each other. Taking into account that the mutual influence
can be quite strong, the carrying capacities are modeled by nonlinear
functionals. We opt for the exponential form of this functional, since such
a form naturally appears as an effective sum of a series, summed in a
self-similar way, under the condition of semi-positivity.

We distinguish three variants of mutual influence, depending on the type of
relations between the species. In the case of {\it passive symbiosis}, the
mutual carrying capacities are influenced by other species without their direct
interactions. In {\it active symbiosis}, the carrying capacities are influenced
by interacting species. And {\it mixed symbiosis} describes the situation when
the carrying capacity of one species is influenced by direct interactions, while
that of the other species is not.

The approach allows us to describe all kinds of symbiosis, that is,
{\it mutualism, commensalism, and parasitism}. The case of two symbiotic species
is analyzed in detail. Depending on the symbiotic parameters, characterizing the
destruction or creation of the mutual carrying capacities, that is, describing
mutualistic or parasitic species, several dynamical regimes of coexistence
are possible: unbounded growth of both populations, growth of one and the
elimination of the other population, convergence to evolutionary stable states,
and everlasting population oscillations.

The main difference of the described symbiosis with nonlinear functional
carrying capacities, compared with the model of \cite{Yukalov_6,Yukalov_12}
considered earlier of symbiosis with carrying capacities in the linear or
bilinear approximations, is the absence of finite-time singularities and of
an abrupt death of populations. The latter have appeared in the previous works
\cite{Yukalov_6,Yukalov_12} owing to the artificial change of sign of the
carrying capacities, when they were crossing zero. The nonlinear carrying
capacities, employed in the present paper, are semi-positive defined, which
ensures that the crossing-zero problem is avoided.

Among various bifurcations, when the dynamic regimes of the population
evolution qualitatively changes, the most interesting ones are the two
bifurcations occurring for the case of mixed symbiosis, with the coexistence
of two parasitic species, when crossing the bifurcation line $g_c(b)$. One
is the supercritical Hopf bifurcation, when a stable focus becomes unstable
and a limit cycle arises. The other is the coalescence of a stable node with
a saddle, with their disappearance and the appearance of a limit cycle.

\newpage

%Figure 1
\begin{figure} [ht]
\centering
\begin{tabular}{lr}
\epsfig{file=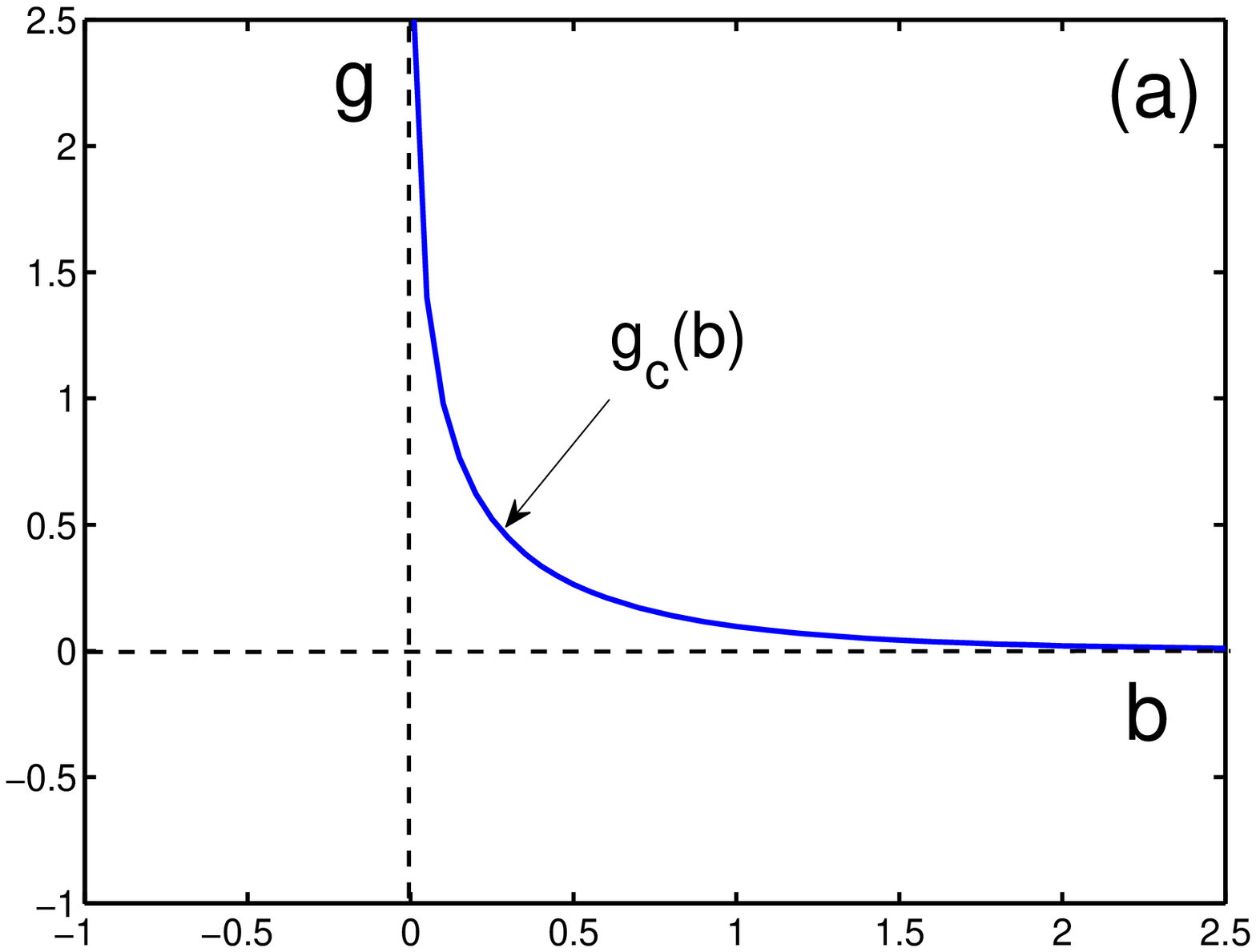,width=8cm} &
\epsfig{file=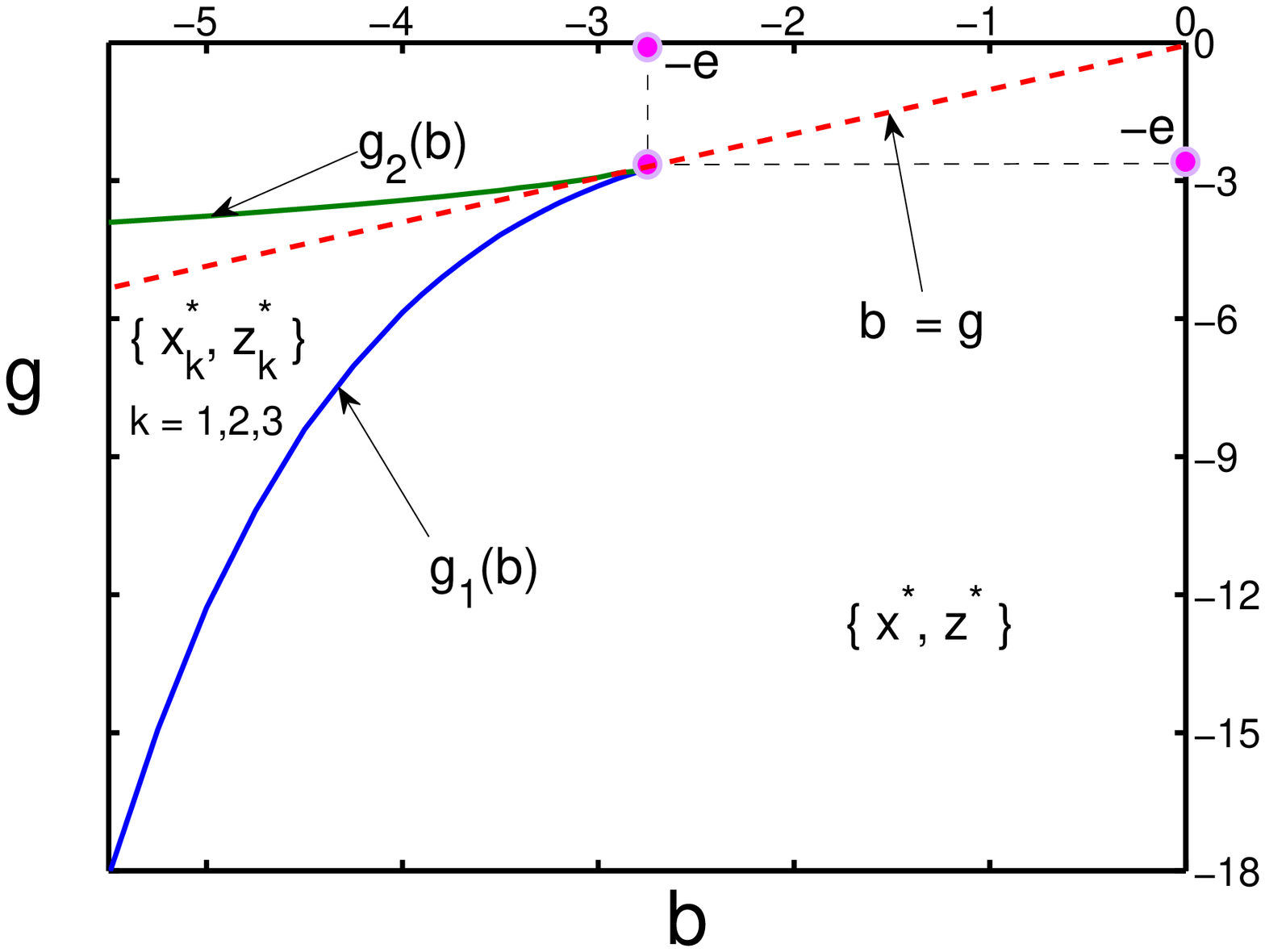,width=8cm}
\end{tabular}
\caption{
(a)
The regions of existence of stationary states for passive symbiosis in the
plane $b - g$. When $b > 0$ and $0 < g < g_c(b)$, there exist two fixed points,
but only one is stable. If $b \leq 0$ and $g > 0$, or $b > 0$ and $g \leq 0$,
there exists a single fixed point that is stable. If $b,g < 0$, there exist
up to three fixed points, but not more then two of them are stable.
(b)
The region of fixed-point existence under mutual parasitism, when $b,g < 0$.
For either $-e < b < 0$ and $g < 0$, or for $b < -e$ and $g < g_1(b)$, or for
$b < -e$ and $g_2(b) < g < 0$, there exists only one fixed point that is stable.
For $b < -e$ and $g_1(b) < g < g_2(b)$, there exist three fixed points, but only
two of them are stable.
}
\label{fig:Fig.1}
\end{figure}

\newpage

%Figure 2
\begin{figure} [ht]
\centering
\begin{tabular}{lr}
\epsfig{file=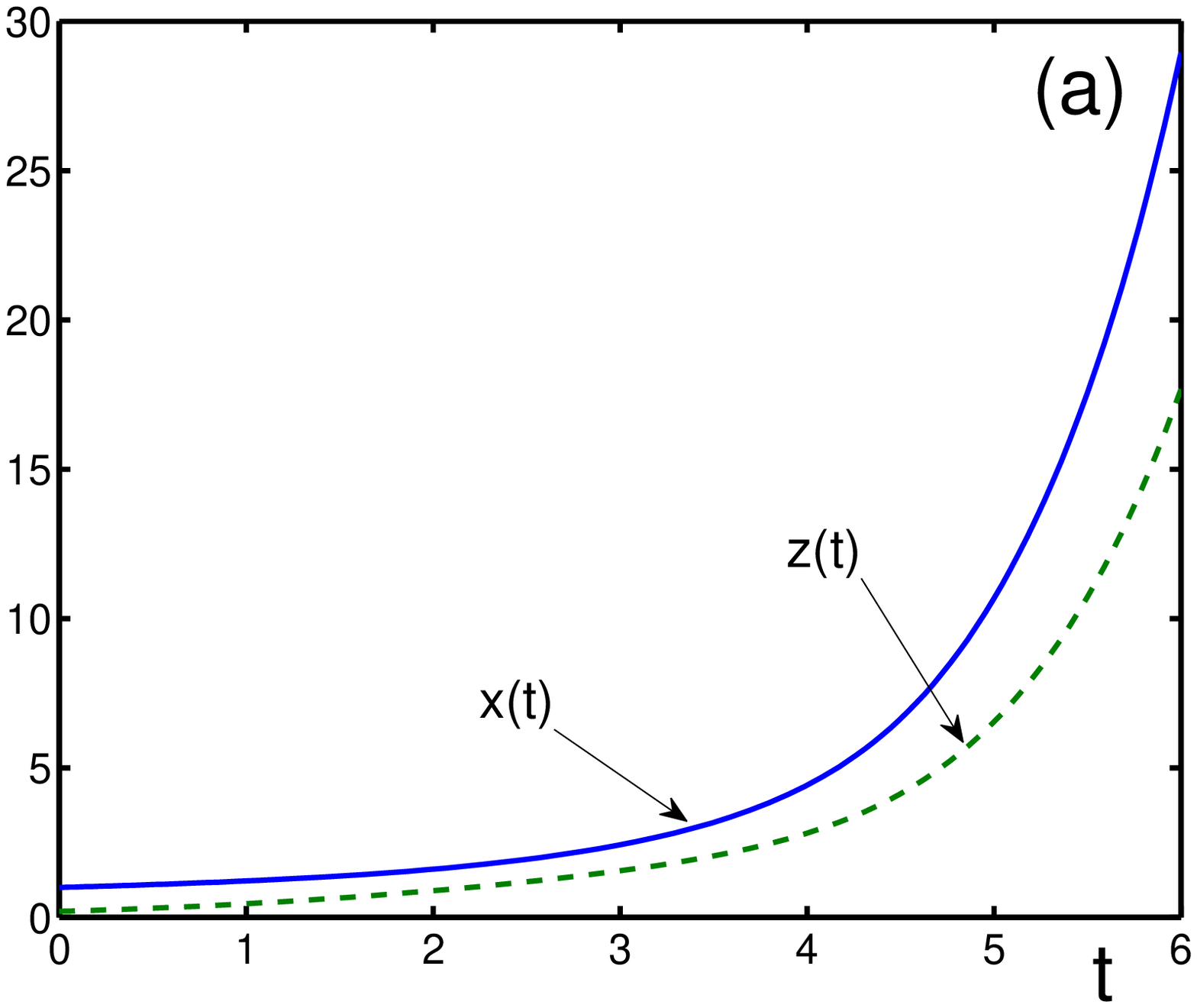,width=8cm}  &
\epsfig{file=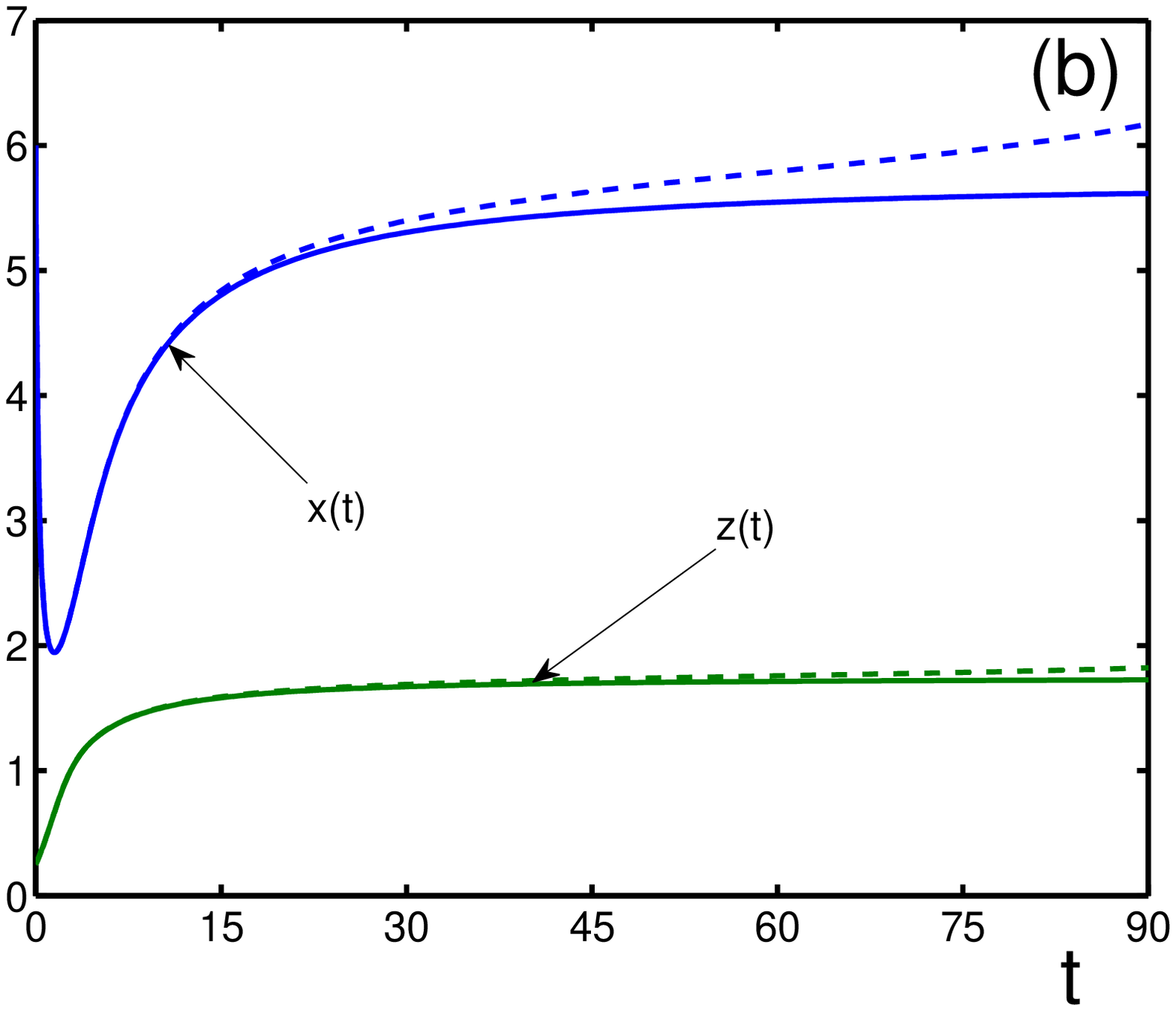,width=8cm} \\
\epsfig{file=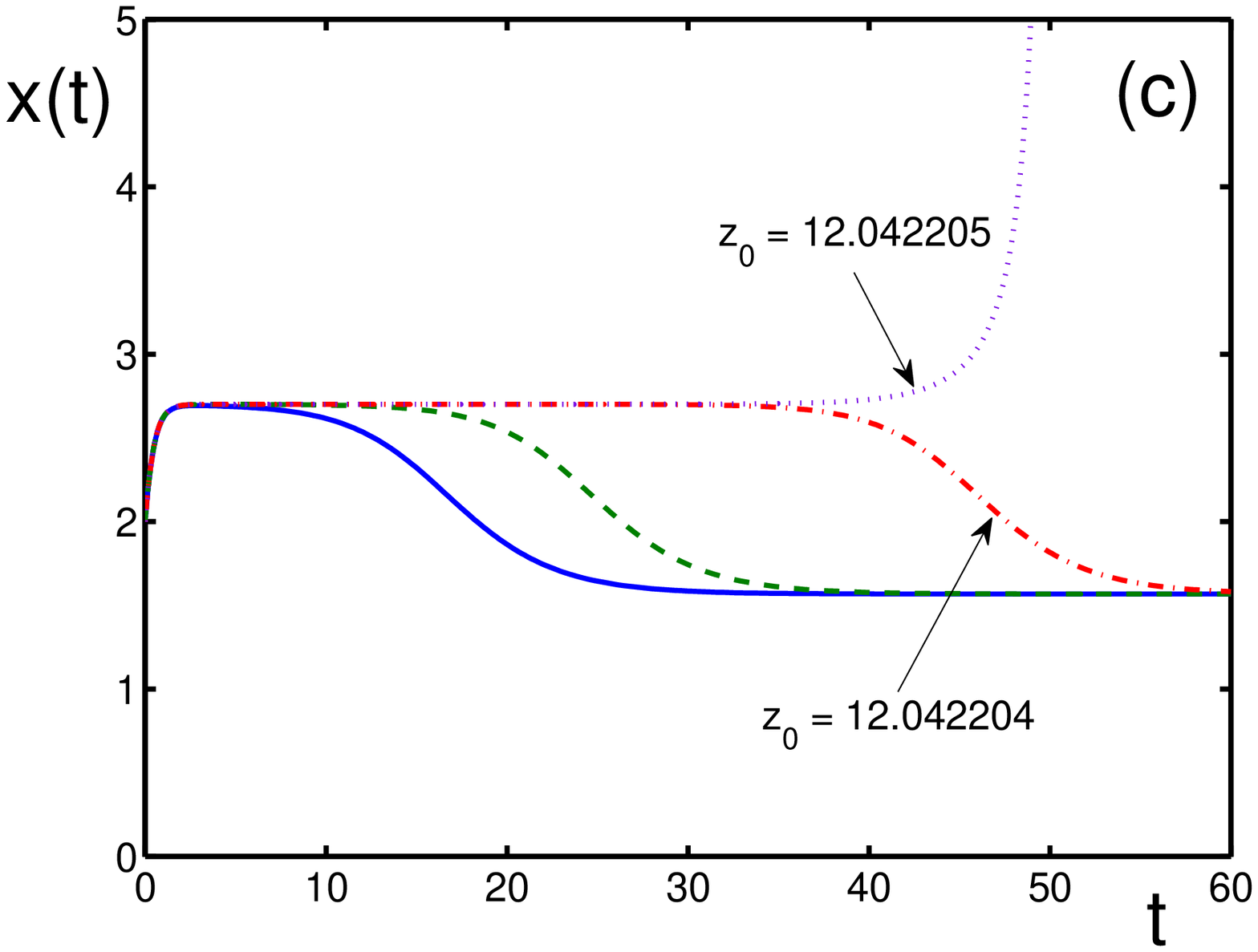,width=8cm} &
\epsfig{file=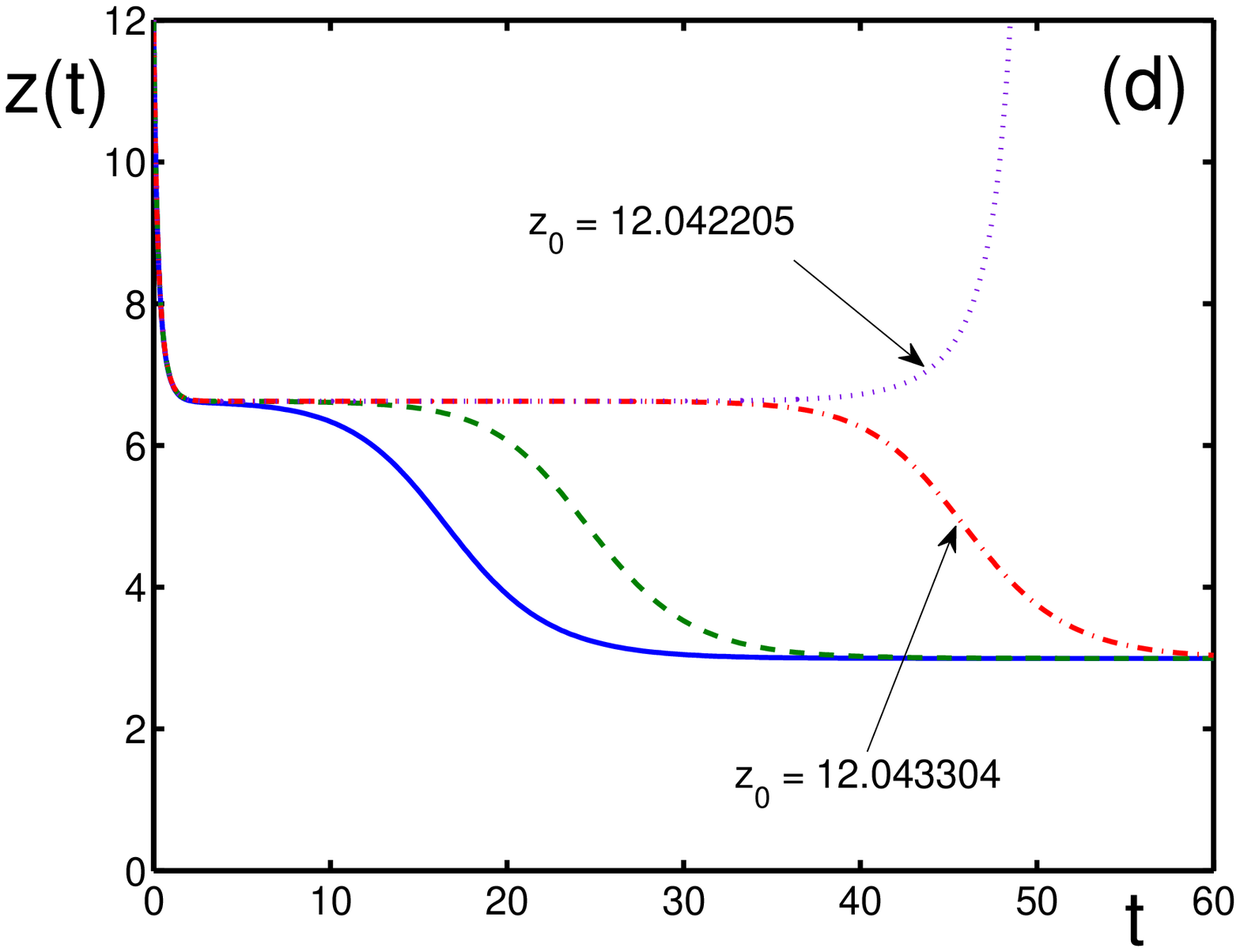,width=8cm}
\end{tabular}
\caption{Dynamics of populations in the case of mutualistic passive symbiosis
for different symbiotic parameters $b > 0$ and $g > 0$, and different initial
conditions:
(a)
For $b = 1$, $g = 0.5 > g_c(b)\approx 0.0973$, and $ x_0 = 1$, $z_0 = 0.2$,
populations $x(t)$ (solid line) and $z(t)$ (dashed line) monotonically increase,
as $t\ra \infty$;
(b)
For the same $b = 1$, as in Fig. 2a, with the initial conditions $x_0 = 6$
and $z_0 = 0.25$, for different $g$ taken slightly below and above the critical
line $g = g_c(b)$. Below the critical line, when $g = 0.0972 < g_c = 0.0973$,
the population $x(t)$ tends to its stable fixed point $x^* = 5.672$ and the
population $z(t)$ tends monotonically from below to $z^* = 1.735$. Both
populations are shown by solid lines. Above the critical line, when
$g = 0.0976 > g_c$, both populations, shown by dashed lines, tend to infinity,
as $t \ra \infty$;
(c)
Dynamics of the population $x(t)$ for the symbiotic parameters $b = 0.15$
and $g = 0.7 < g_c = 0.765$, with the initial condition $x_0 = 2$, but for
different initial conditions $z_0$. The population $x(t)$ tends non-monotonically
to its stable fixed point $x^* = 1.567$, when $z_0$ is taken from the basin
of attraction of this point: $x(t) \ra x^*$ for $z_0 = 12$ (solid line),
$z_0 = 12.04$ (dashed line), $z_0 = 12.042204$ (dashed-dotted line), and
$x(t) \ra \infty$, as $t \ra \infty$, when $z_0 = 12.042205$ (dotted line);
(d)
Dynamics of the population $z(t)$ for the same parameters $b,g$ and the same
initial conditions, as in Fig. 2c. The population converges to a stationary
state, when the initial conditions are inside the attraction basin, and diverges,
if the initial conditions are outside of this basin.}
\label{fig:Fig.2}
\end{figure}

\newpage

%Figure 3
\begin{figure} [ht]
\centering
\begin{tabular}{lr}
\epsfig{file=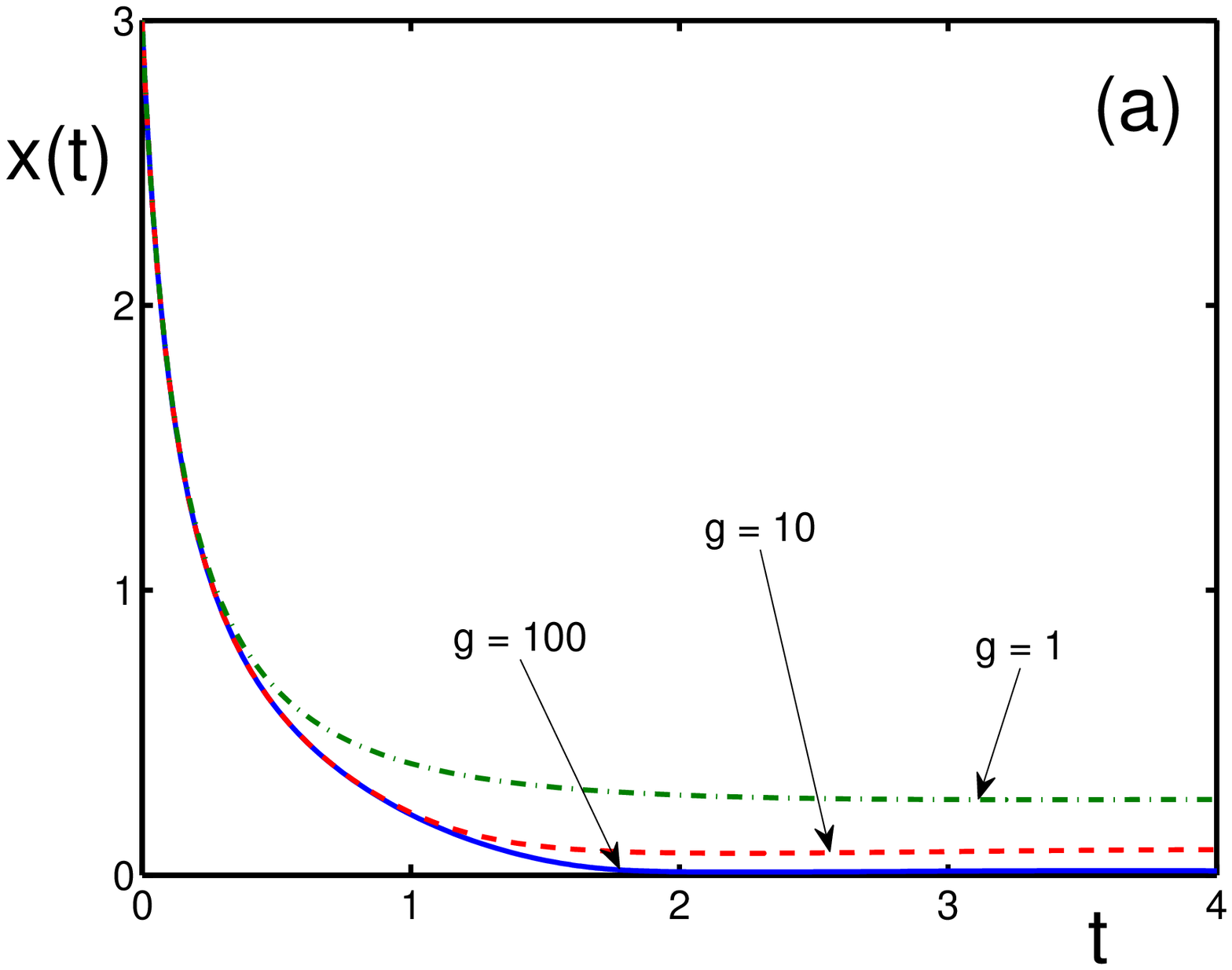,width=8cm}  &
\epsfig{file=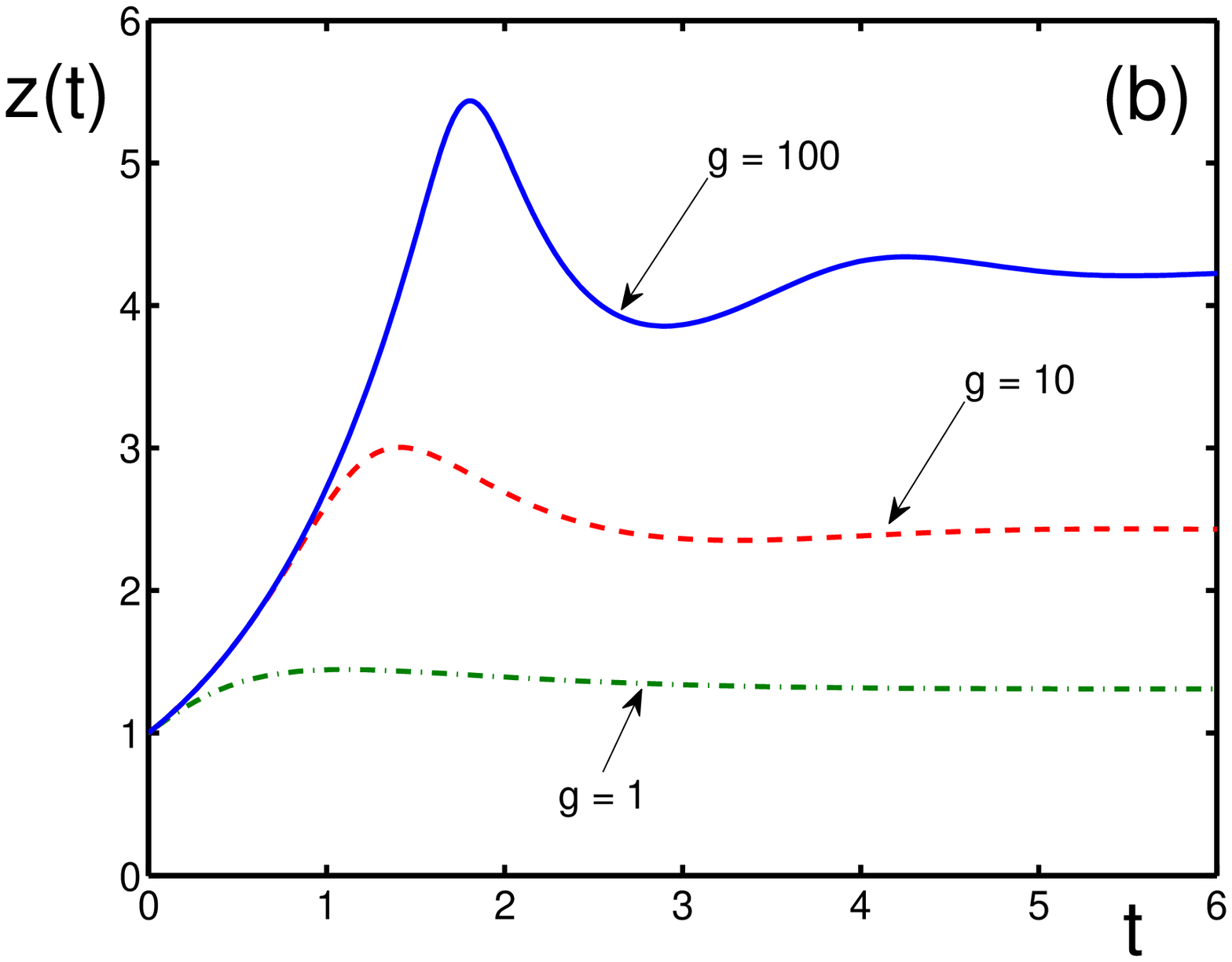,width=8cm}
\end{tabular}
\caption{Dynamics of populations under passive symbiosis with one parasitic
species. Influence of the symbiotic parameter $g > 0$ on the behaviour of
the populations $x(t),z(t)$, for the fixed $b = -1$, and the initial
conditions $x_0 = 3$ and $z_0 = 1$. Different lines correspond to the varying
parameter $g$: For $g = 1$ (dashed-dotted line) the stable fixed point is
$\{x^* = 0.270, z^* = 1.31\}$; for $g = 10$ (dashed line), the fixed point
is $\{x^* = 0.089, z^* = 2.42\}$; and for $g = 100$ (solid line), the
stationary point is $\{x^* = 0.014, z^* = 4.24\}$.
(a)
Population $x(t)$ tends monotonically from above to the corresponding stable
stationary state, as $t\ra\infty$;
(b)
Population $z(t)$ tends non-monotonically from below to the corresponding stable
stationary state, as $t\ra\infty$.
}
\label{fig:Fig.3}
\end{figure}

\newpage

%Figure 4
\begin{figure} [ht]
\centering
\begin{tabular}{lr}
\epsfig{file=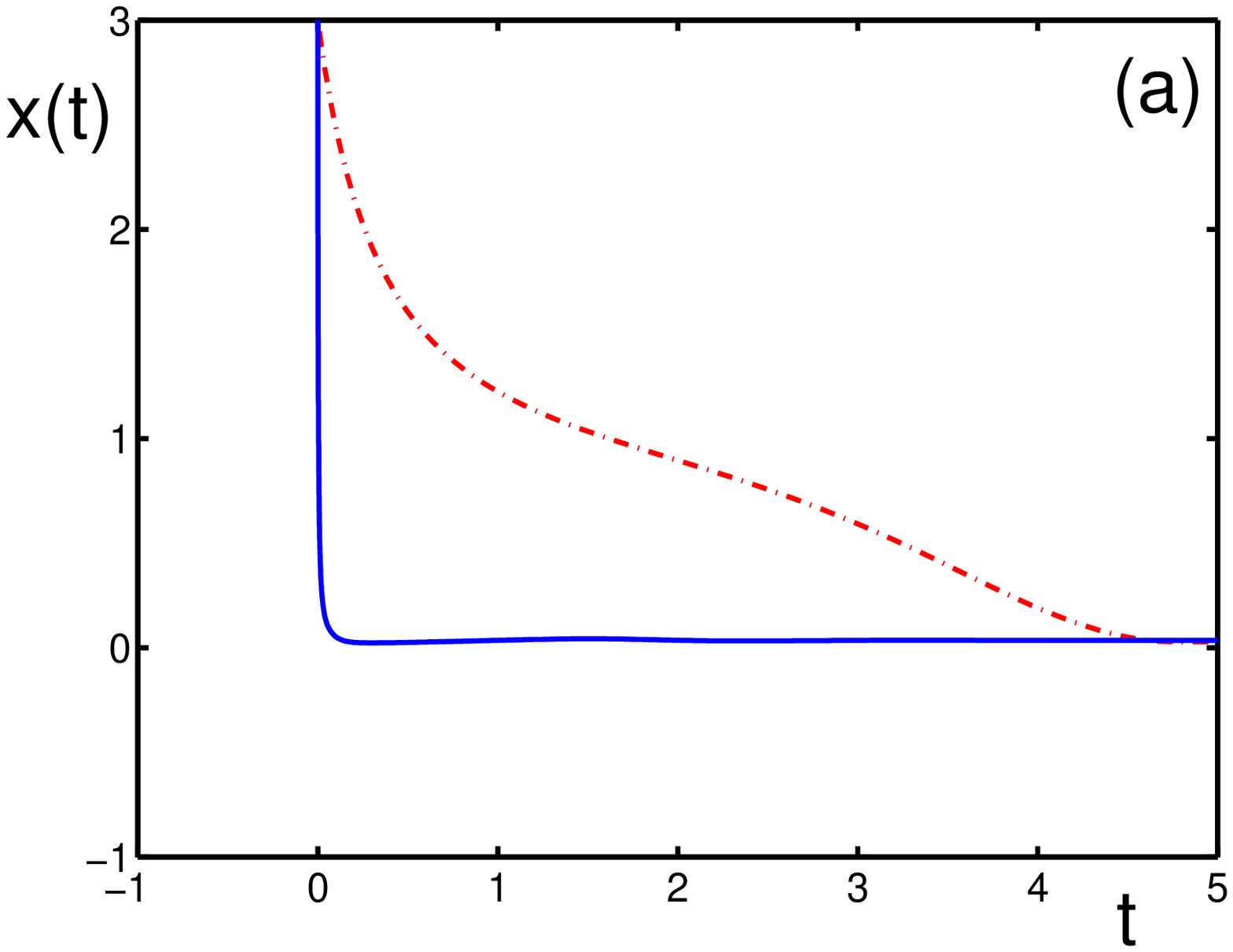,width=8cm}  &
\epsfig{file=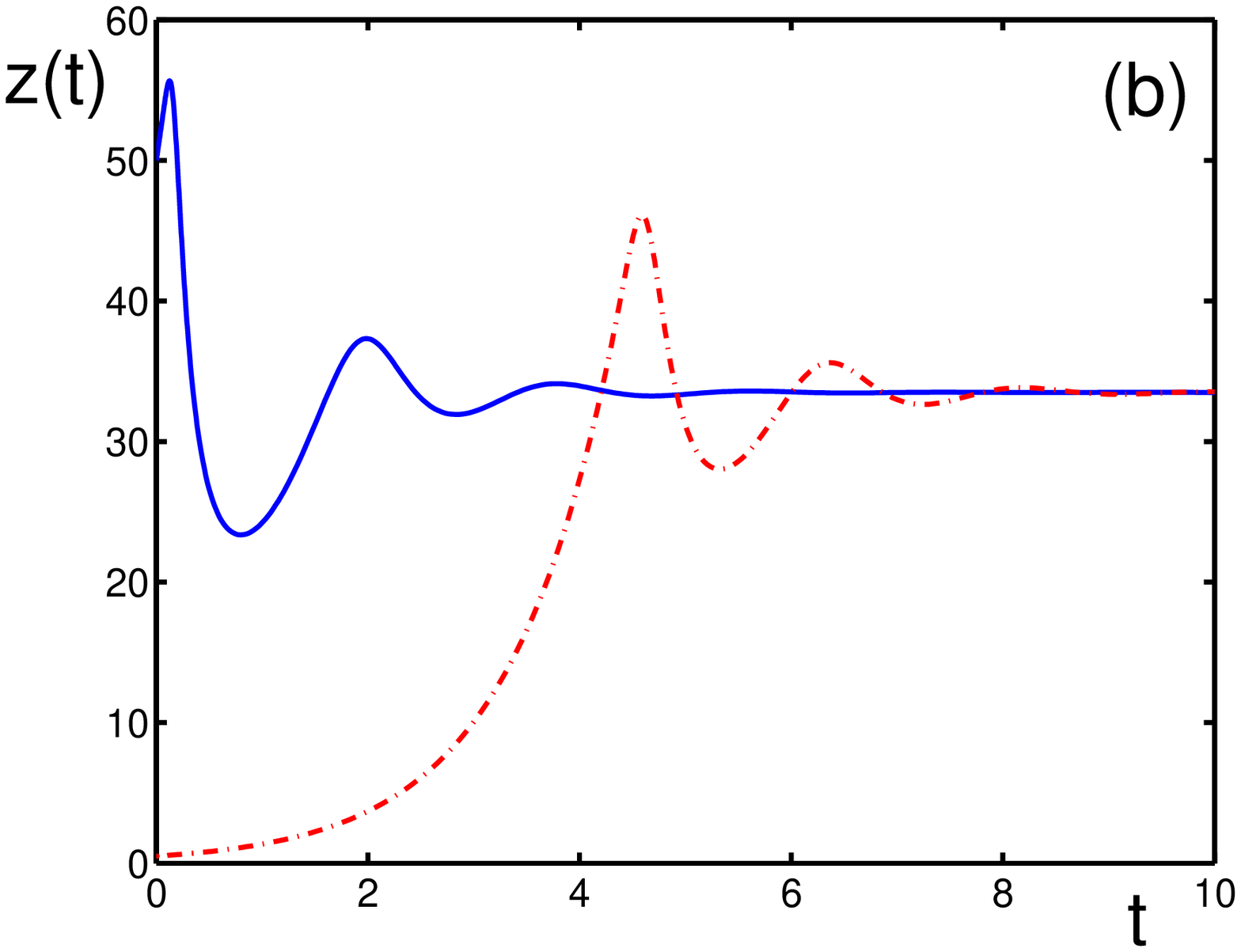,width=8cm}
\end{tabular}
\caption{Dynamics of populations under passive symbiosis with one parasitic
species. Influence of the initial condition $z_0$ on the behavior of solutions
for the fixed $b = -0.1$, $g = 100$, and $x_0 = 3$. The corresponding
stable fixed point is $\{x^* = 0.0335, z^* = 33.49\}$.
(a)
Population $x(t)$ tends monotonically, as $t \ra \infty$, from above to the
stable stationary state $x^*$ for $z_0 = 50$ (solid line) and for $z_0 = 0.5$
(dashed line);
(b)
Population $z(t)$ tends, as $t\ra\infty$, non-monotonically from above to the
stable stationary state $z^*$ for $z_0 = 50$ (solid line) and non-monotonically
from below to $z^*$ for $z_0 = 0.5$ (dashed line).
}
\label{fig:Fig.4}
\end{figure}

\newpage

%Figure 5
\begin{figure} [ht]
\centering
\begin{tabular}{lr}
\epsfig{file=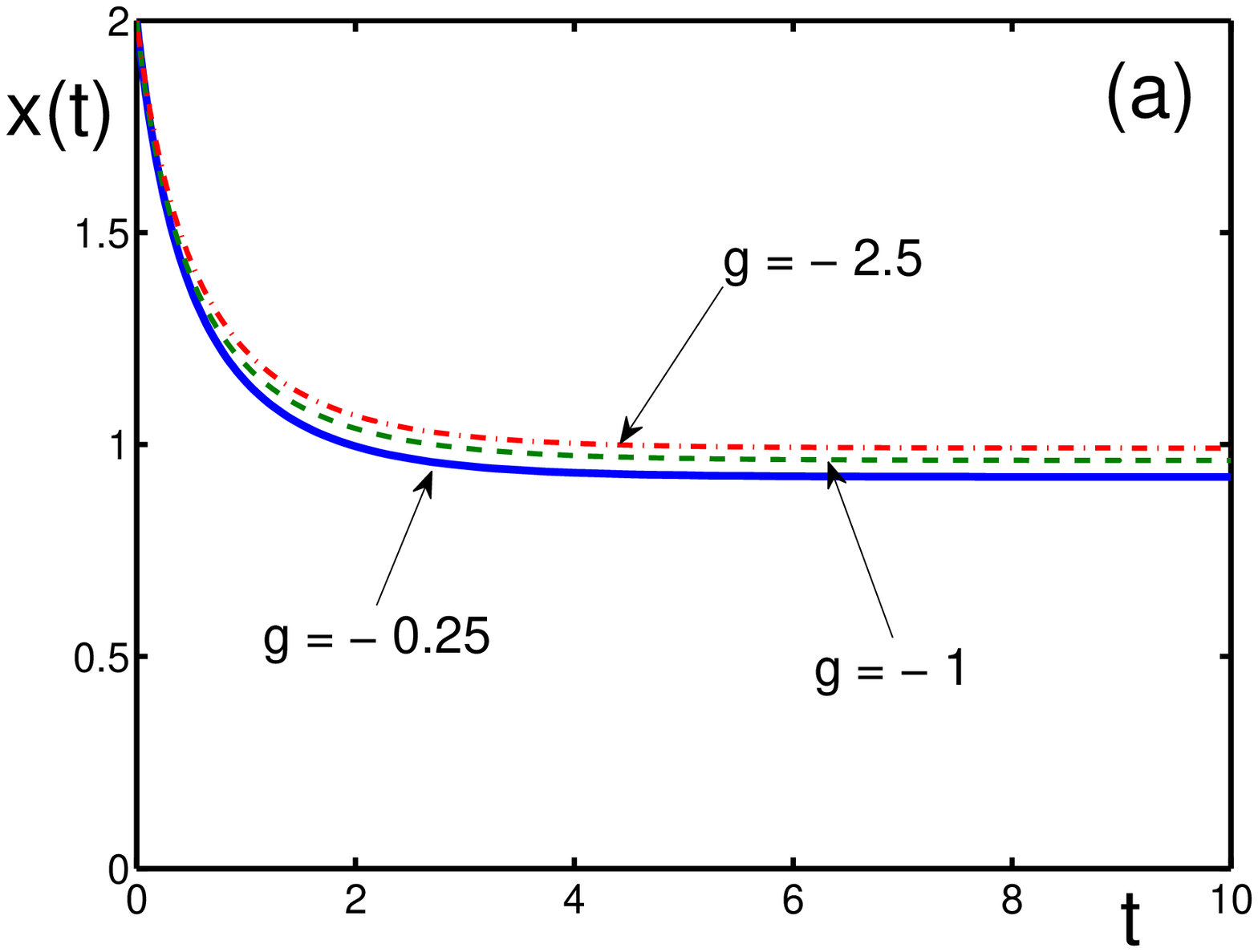,width=8cm}  &
\epsfig{file=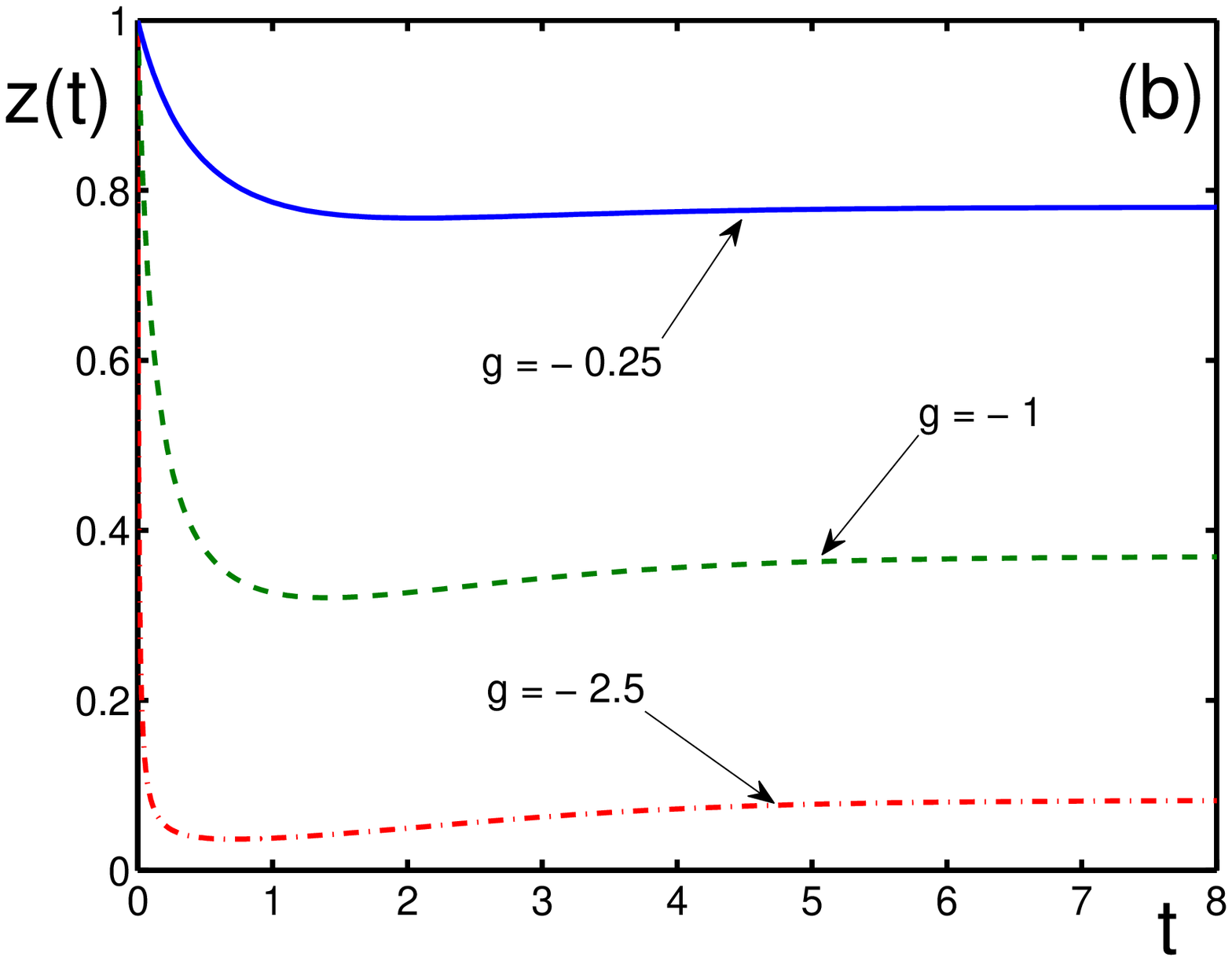,width=8cm} \\
\epsfig{file=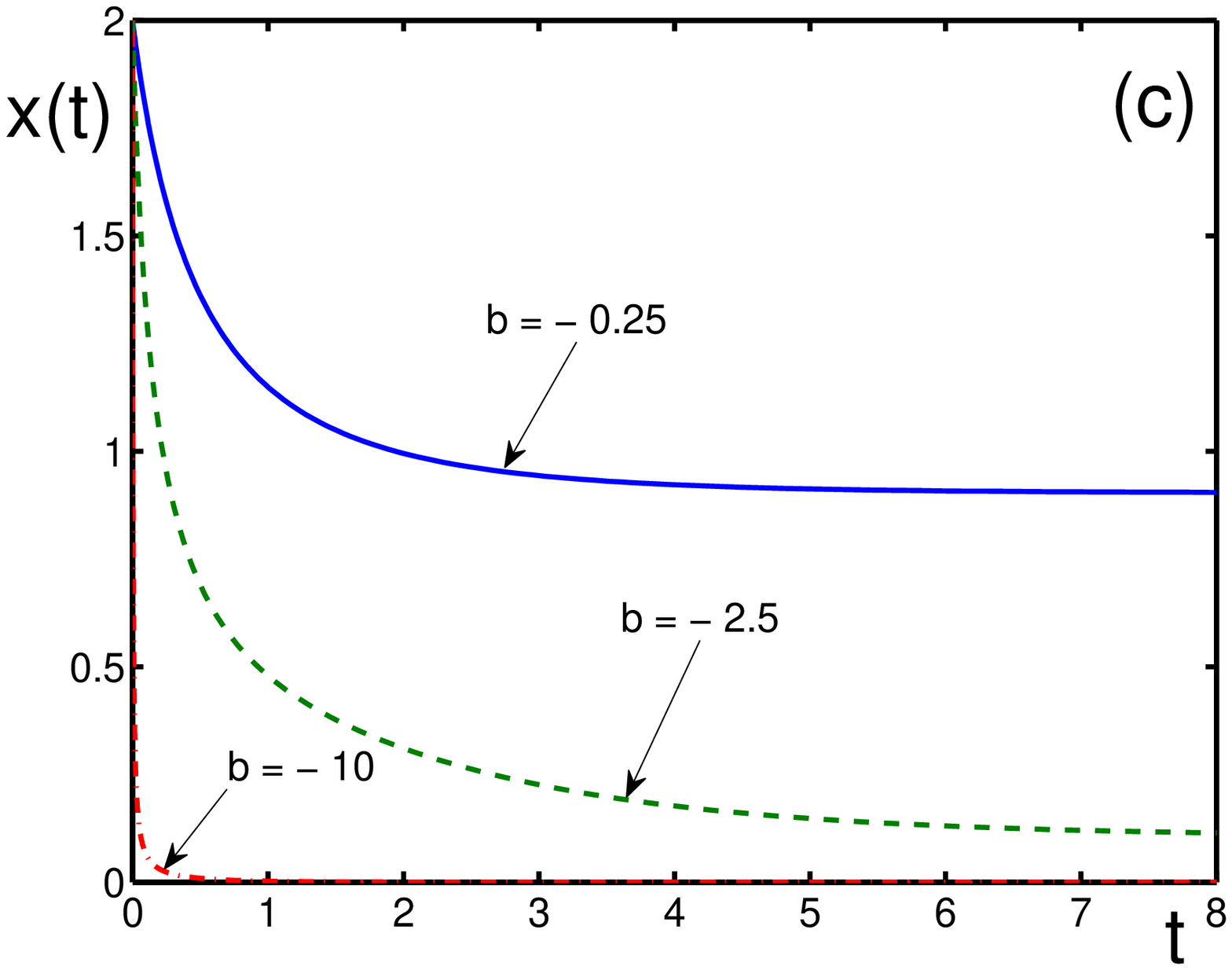,width=8cm} &
\epsfig{file=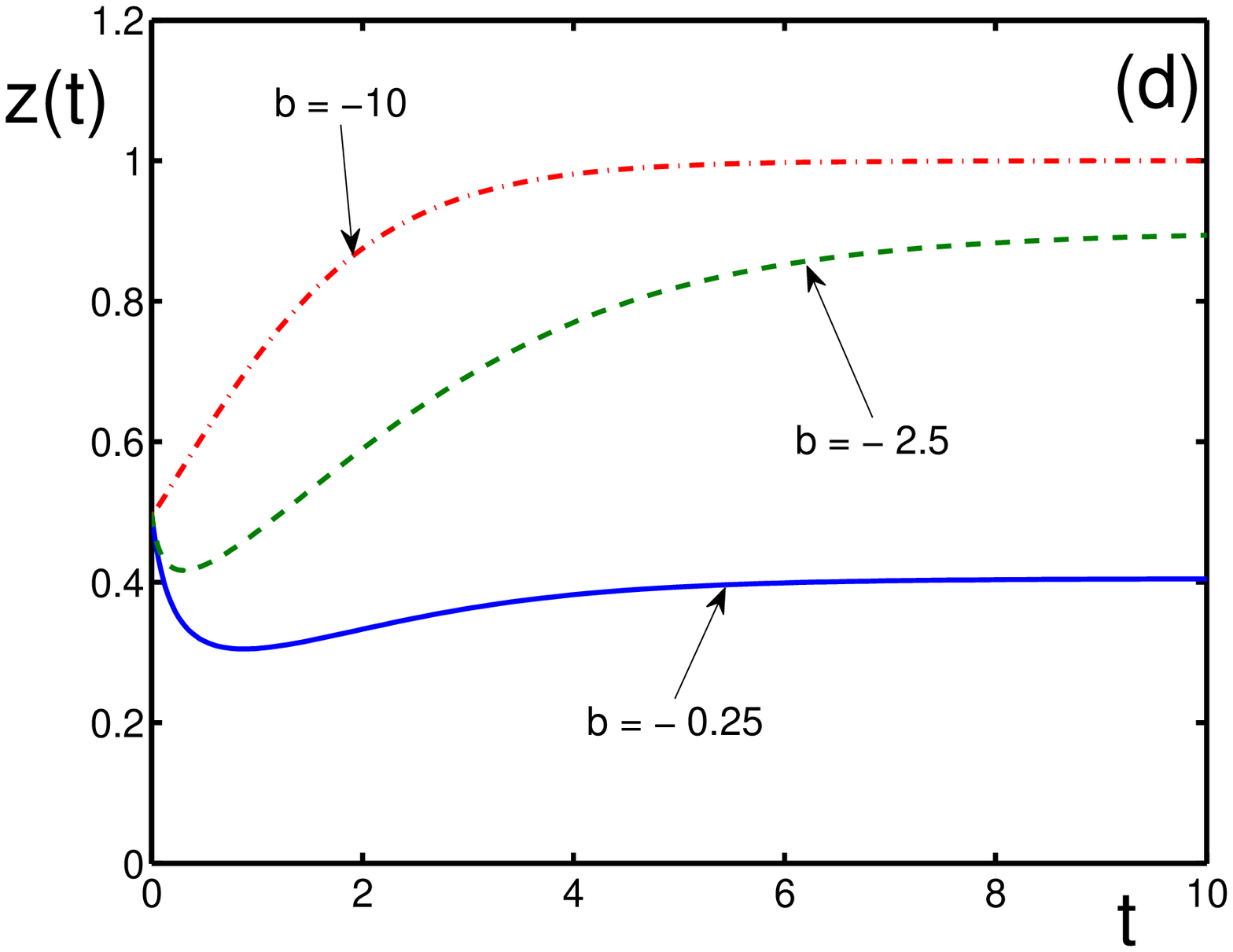,width=8cm}
\end{tabular}
\caption{Dynamics of populations under passive symbiosis with two parasitic
species. Influence of the symbiotic parameters $b < 0$ and $g < 0$ under the
fixed initial conditions $\{x_0 = 2, z_0 = 1\}$.
(a)
Dynamics of $x(t)$ for $b = -0.1$ and different parameters $g$. The
population $x(t)$ monotonically tends to its stationary state:
$\{x^* = 0.924, z^* = 0.794 \}$ for $g = -0.25$ (solid line);
$\{x^* = 0.963, z^* = 0.382 \}$ for $g = -1$ (dashed line); and
$\{x^* = 0.992, z^* = 0.084 \}$ for $g = -2.5$ (dashed-dotted line).
(b)
Dynamics of $z(t)$ for the same, as above, parameters $b = -0.1$ and
different $g$. Population $z(t)$ monotonically, or non-monotonically, from
above tends to the corresponding stationary states.
(c)
Dynamics of $x(t)$ for $g = -1$ and different parameters $b$. The
population $x(t)$ monotonically tends from above to its stationary state:
$\{x^* = 0.904, z^* = 0.405 \}$ for $b = -0.25$ (solid line);
$\{x^* = 0.105, z^* = 0.899 \}$ for $b = -2.5$ (dashed line); and
$\{x^* = 0.454 \times 10^{-4}, z^* = 0.99996\}$ for $b = -10$
(dashed-dotted line).
(d)
Dynamics of $z(t)$ for the same, as in Fig. 5c, parameters $g = -1$
and different $b$. Population $z(t)$ converges to the corresponding
stationary states, either monotonically or non-monotonically.
}
\label{fig:Fig.5}
\end{figure}

\newpage

%Figure 6
\begin{figure} [ht]
\centering
\begin{tabular}{lr}
\epsfig{file=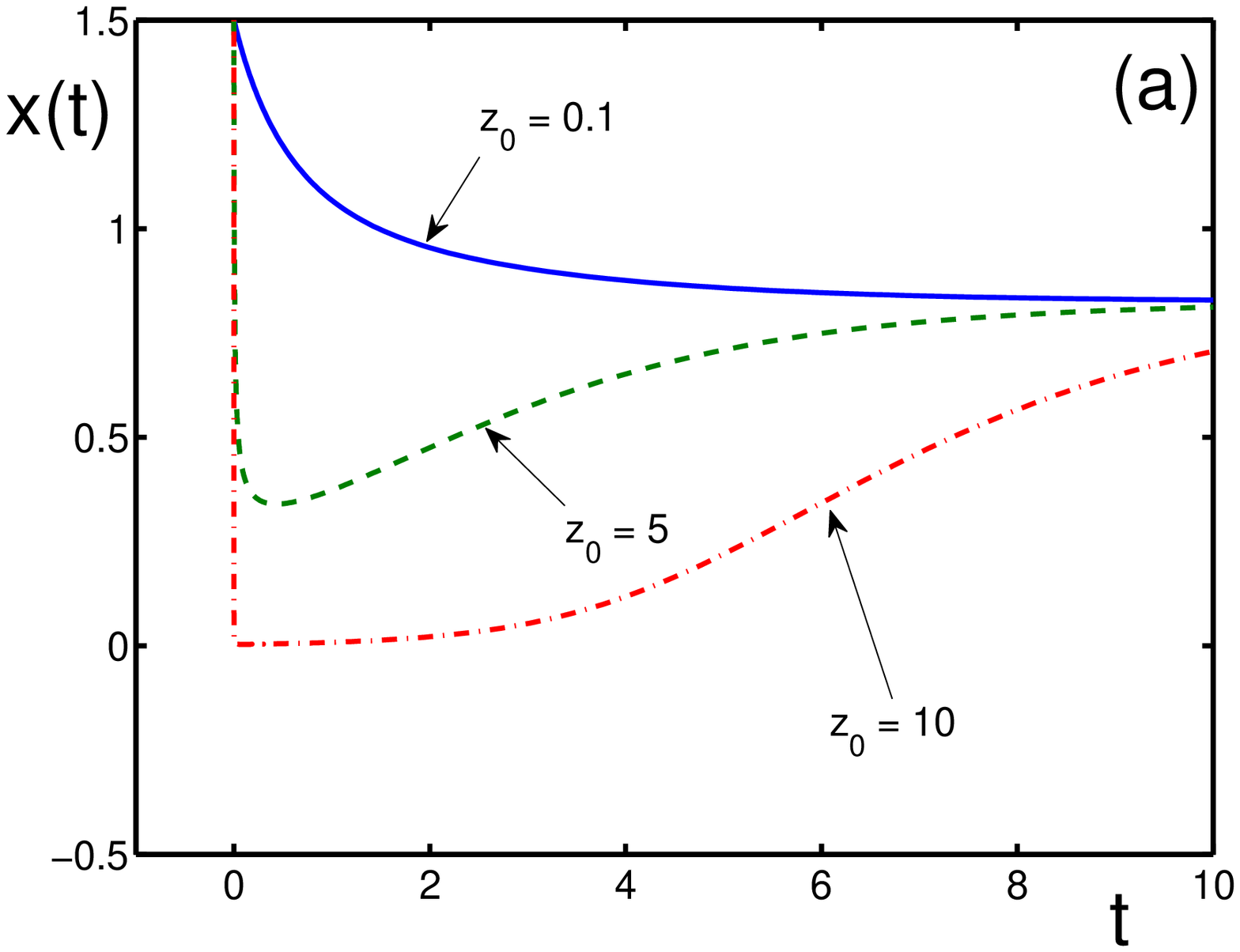,width=8cm}  &
\epsfig{file=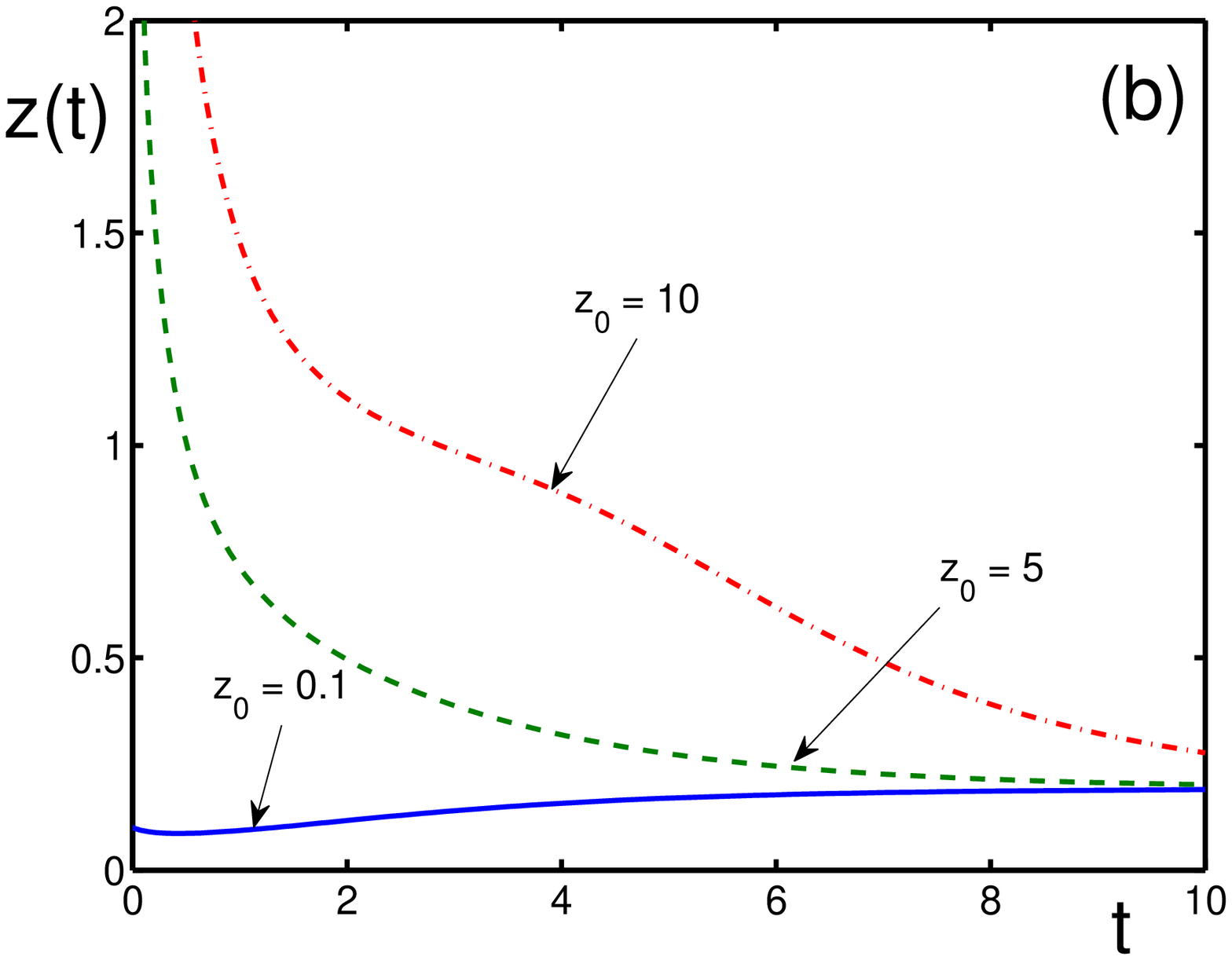,width=8cm} \\
\epsfig{file=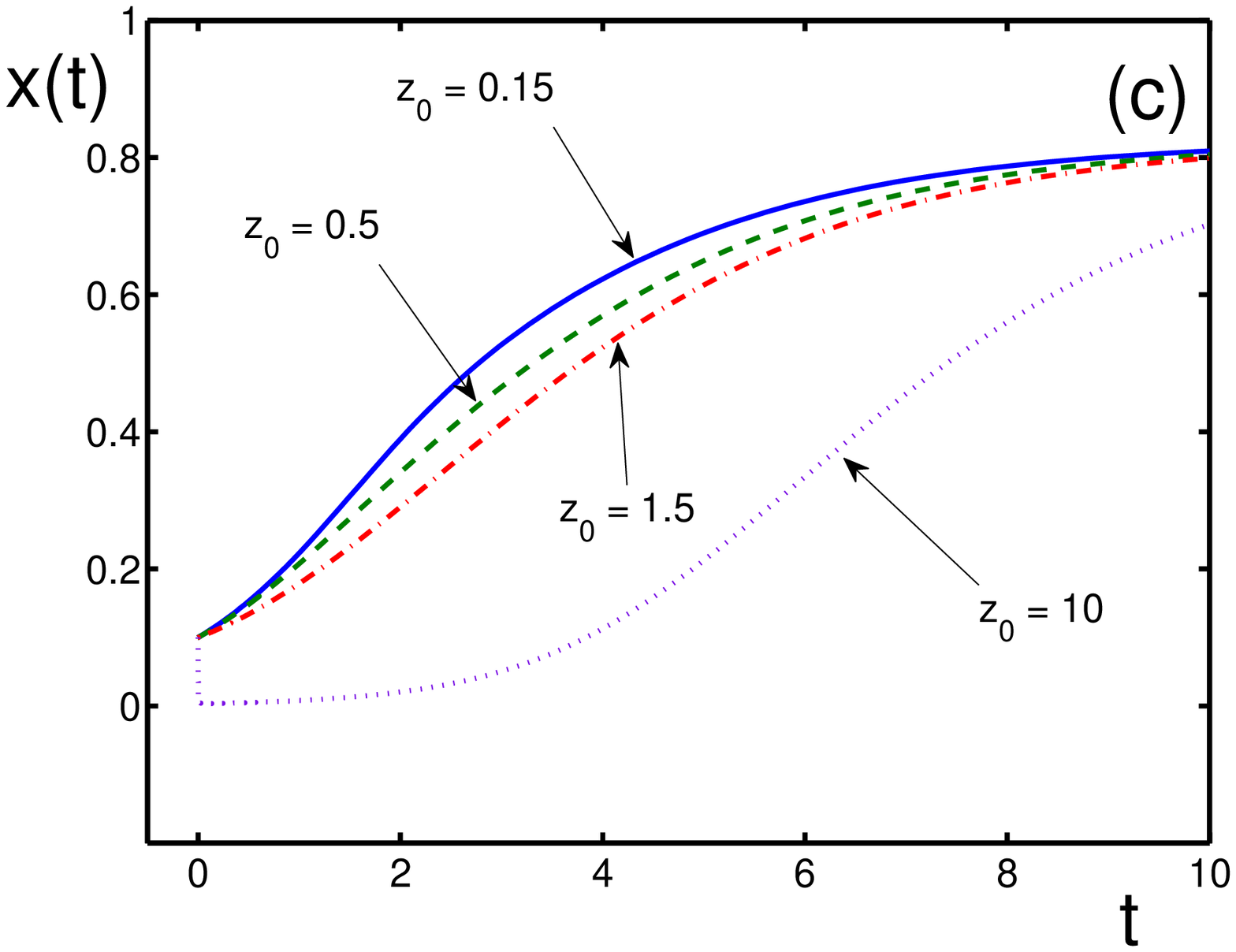,width=8cm} &
\epsfig{file=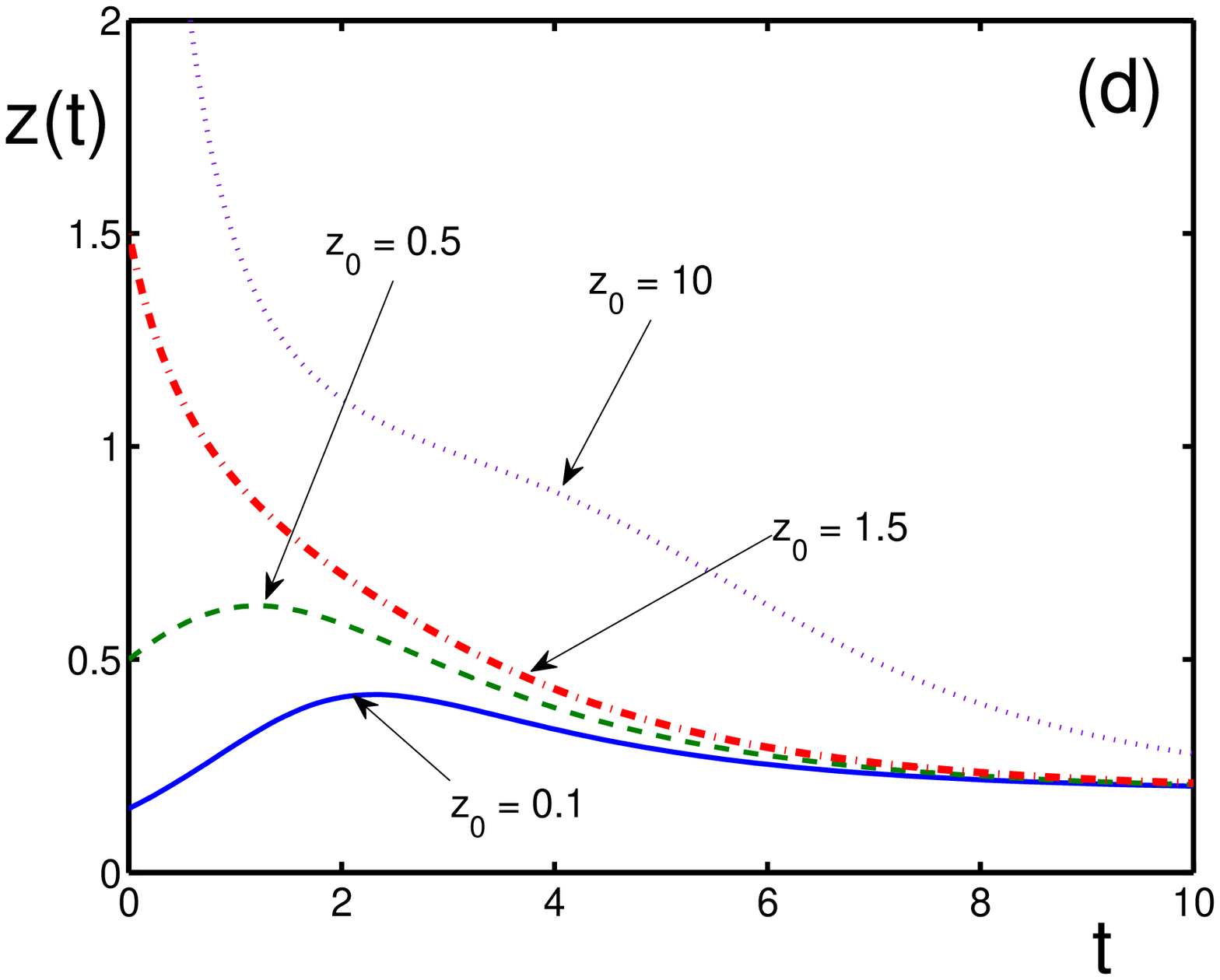,width=8cm}
\end{tabular}
\caption{Dynamics of populations under passive symbiosis with two parasitic
species. Influence of initial conditions $\{x_0, z_0\}$ on the behaviour of
the parasitic populations with $b = -1$ and $g = -2$. The corresponding
stable fixed point is $\{x^* = 0.825, z^* = 0.192\}$.
(a)
Dynamics of $x(t)$ for $x_0 = 1.5 > x^*$ and different $z_0$. Population
$x(t)$ converges to the stationary state $x^*$: monotonically for
$z_0 = 0.1 < z^*$ (solid line); non-monotonically for $z_0 = 5$
(dashed line); and non-monotonically for $z_0 = 10 > z^*$ (dashed-dotted line);
(b)
Dynamics of $z(t)$ for the same, as in Fig. 6a, initial condition $x_0 = 1.5$
and different $z_0$. Population $z(t)$ tends to the stationary state $z^*$:
non-monotonically for $z_0 = 0.1$ (solid line); monotonically for $z_0 = 5$
(dashed line); and monotonically for $z_0 = 10$ (dashed-dotted line).
(c)
Dynamics of $x(t)$ for $x_0 = 0.1 < x^*$ and different $z_0$. Population
$x(t)$ converges to the stationary state $x^*$: monotonically for $z_0 = 0.15$
(solid line); for $z_0 = 0.5$ (dashed line); and $z_0 = 1.5$ (dashed-dotted line);
while non-monotonically for $z_0 = 10$ (dotted line).
(d)
Dynamics of $z(t)$ for the same, as in Fig. 6c, initial condition $x_0 = 0.1$
and different $z_0$. Population $z(t)$ tends to the stationary state $z^*$:
non-monotonically for $z_0 = 0.1$ (solid line); and $z_0 = 0.5$ (dashed line);
but monotonically for $z_0 = 1.5$ (dashed-dotted line); and $z_0 = 10$ (dotted line).
}
\label{fig:Fig.6}
\end{figure}

\newpage

%Figure 7
\begin{figure} [ht]
\centering
\begin{tabular}{lr}
\epsfig{file=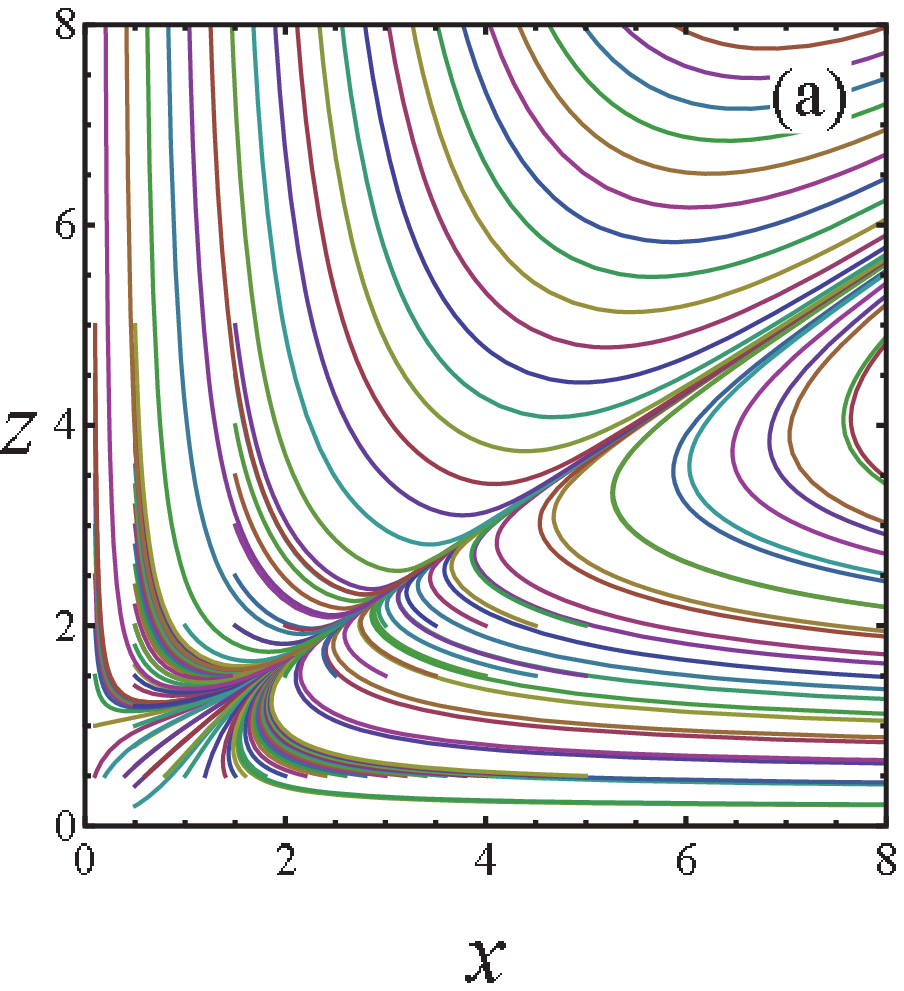,width=6.6cm}  & ~~~~
\epsfig{file=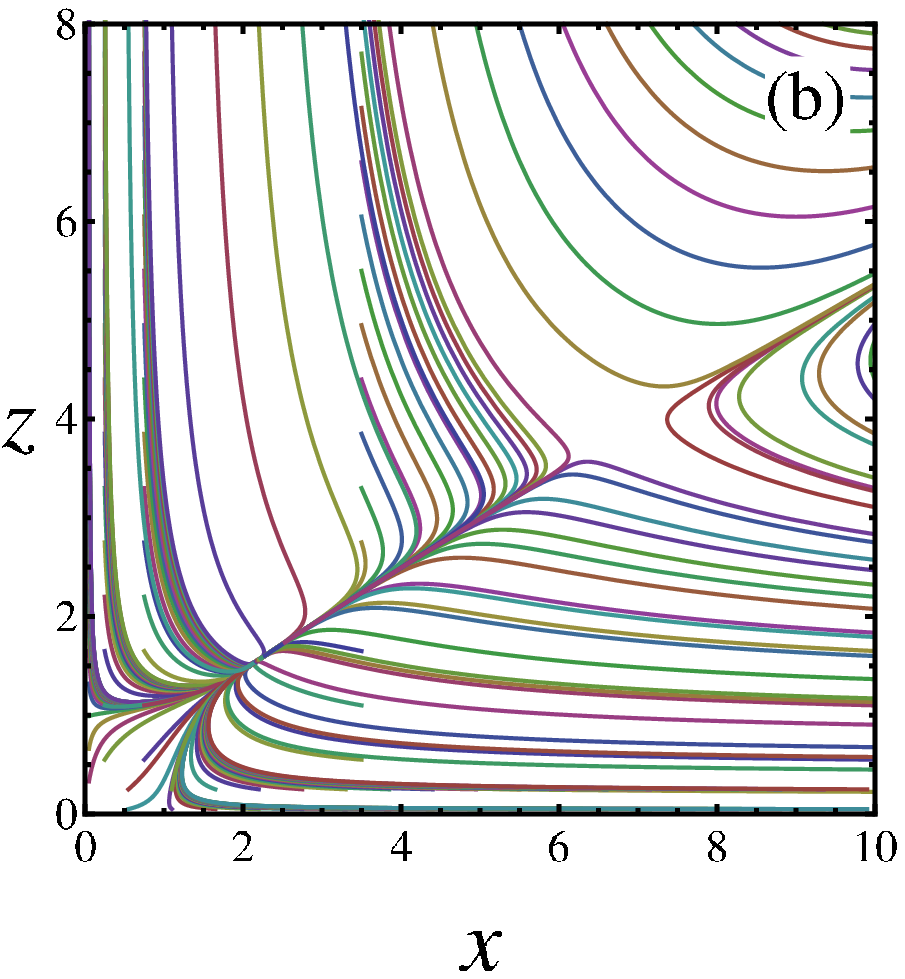,width=6.5cm} \\
&  \\
\epsfig{file=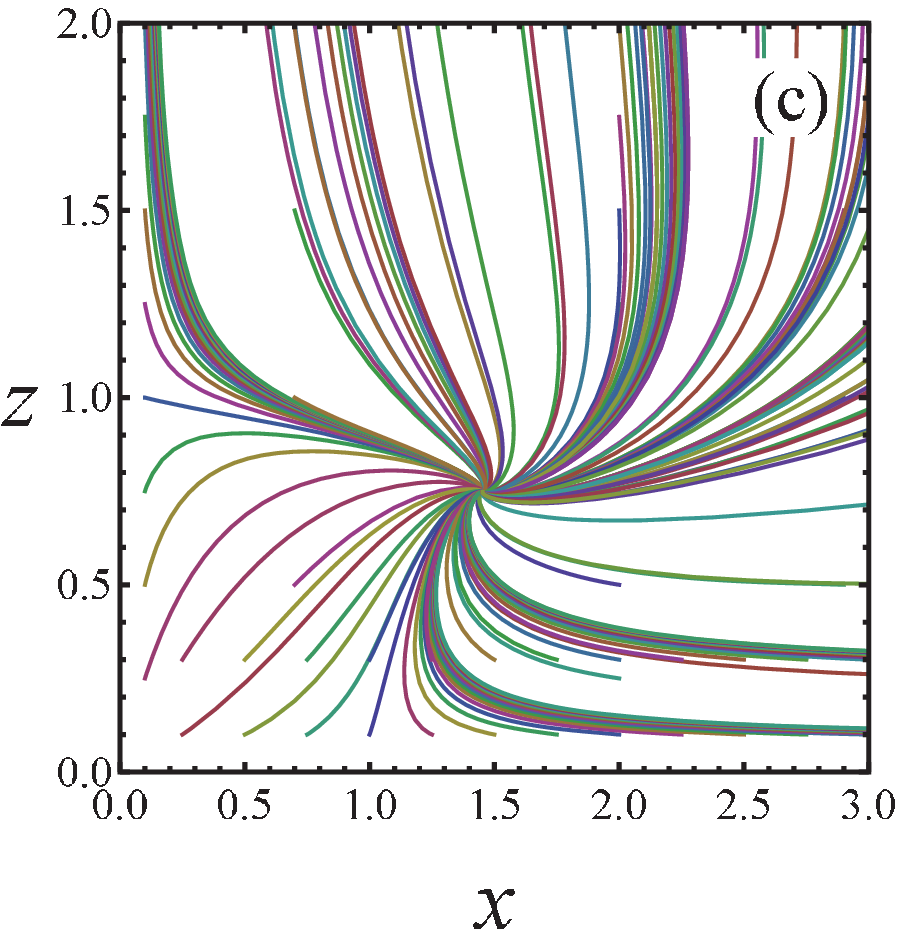,width=6.5cm} & ~~~~
\epsfig{file=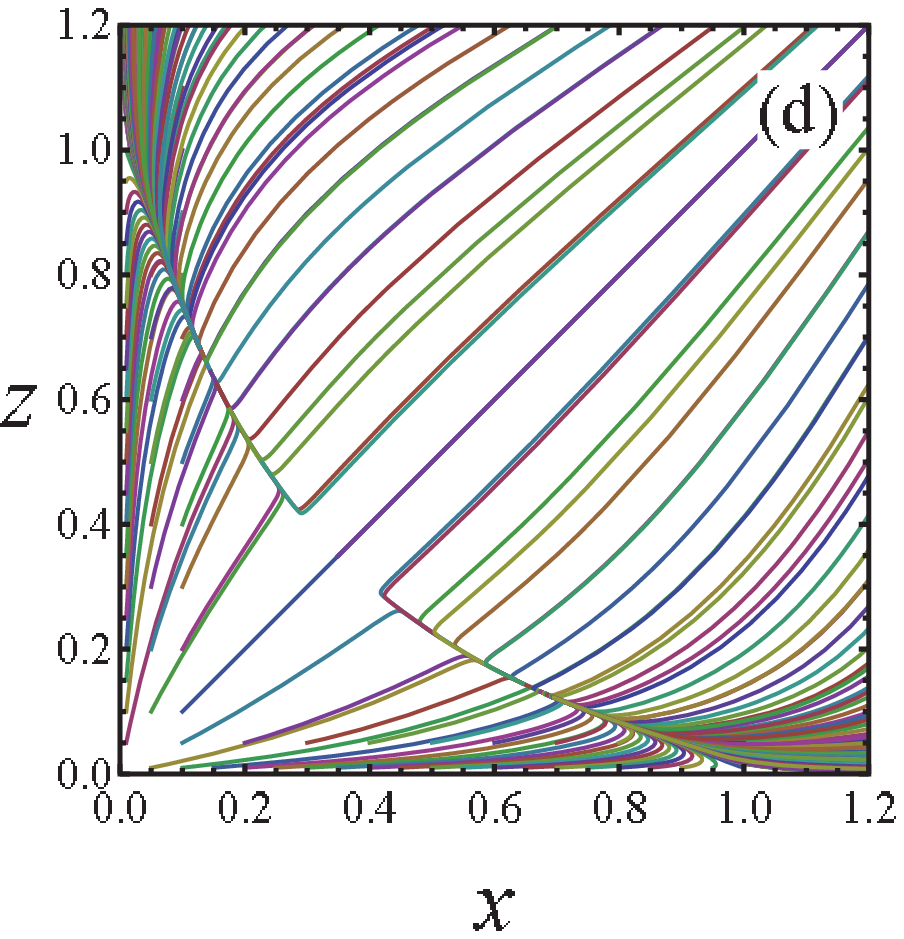,width=6.5cm}
\end{tabular}
\caption{
Phase portrait for passive symbiosis, with the symbiotic parameters $b$
and $g$ in qualitatively different regions:
(a)
Phase portrait for $b = 0.5$ and $g = 0.3$, where $g > g_c(b) \simeq 0.263$.
The parameters are in the region, where no fixed points exist for the system
of equations (\ref{eq1}).
(b)
Phase portrait for $b = 0.5$ and $0 < g = 0.2 < g_c(b) \simeq 0.263$. The
parameters are in the region of $b, g$, where two fixed points exist. The
first fixed point $\{ x_1^* = 2.16019, z_1^* = 1.54039 \}$ is a stable node,
with the Lyapunov exponents $\lbd_1 = -1.57685$, $\lbd_2 = -0.42315$, while
the second point $\{ x_2^* = 6.66791, z_2^* = 3.7946 \}$ is a saddle, with the
Lyapunov exponents $\lbd_1 = -2.59067$, $\lbd_2 = 0.59067$.
(c)
Phase portrait for $b = 0.5$ and $g = -0.2$. The parameters are in the region
of $b, g$, where only one fixed point exists, which is a stable focus
$\{ x^* = 1.45336, z_1^* = 0.74776 \}$, with the characteristic exponents
$\lbd_1 = -1 - 0.32966 i$, $\lbd_2 = -1 + 0.32966 i$.
(d)
Phase portrait for $b = g = -3$. The parameters are in the bistability region
of $b, g$, where three fixed points exist, two of them being stable and one,
unstable. Here, the first stable fixed point is the node
$\{x_1^* = 0.88474, z_1^* = 0.13612 \}$, with the Lyapunov exponents
$\lbd_1 = -1.90242$, $\lbd_2 = -0.0975818$. The second stable fixed point,
due to symmetry, is the node $\{ x_2^* = z_1^*, z_2^* = x_1^* \}$, with the same
Lyapunov exponents. And the third fixed point $\{ x_3^* = z_3^* = 0.34997 \}$
is the saddle, with the Lyapunov exponents $\lbd_1 = -2.04991$,
$\lbd_2 = 0.0499089$.
}
\label{fig:Fig.7}
\end{figure}

\newpage

%Figure 8
\begin{figure} [ht]
\centering
\epsfig{file=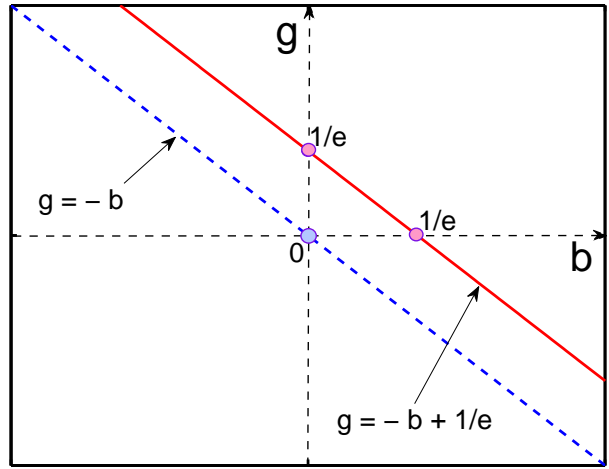,width=8cm}
\caption{
The regions of existence of the fixed-point for Eqs. (\ref{eqq1}) in the plane
$b-g$. When $b + g > 1/e$, then there are no fixed points. If $0 < b + g < 1/e$,
then there exist two fixed points, one of them being stable and another, unstable.
If $b + g \leq 0$, then there exists a single fixed point that is stable.
}
\label{fig:Fig.8}
\end{figure}

\newpage

%Figure 9
\begin{figure} [ht]
\centering
\begin{tabular}{lr}
\epsfig{file=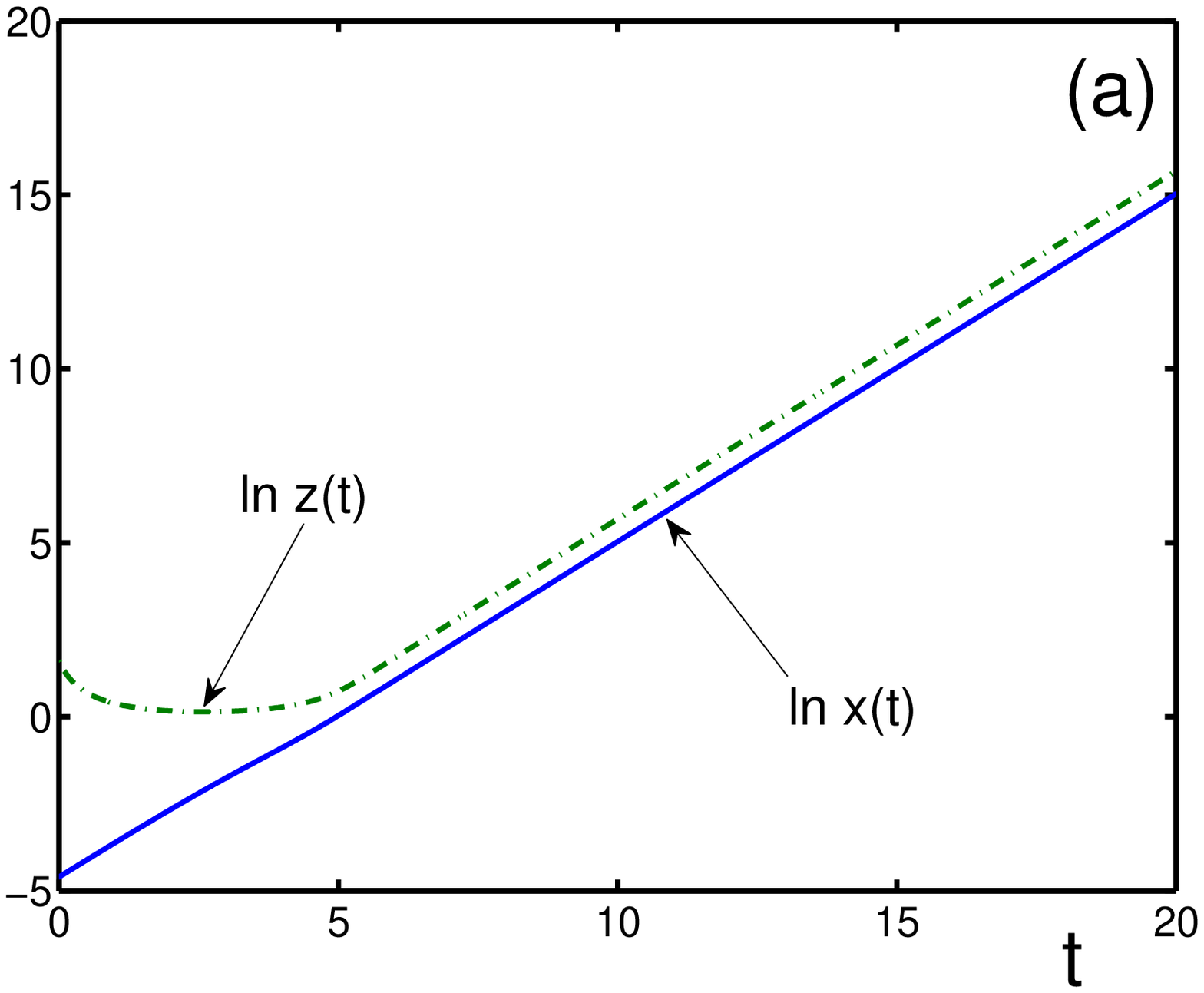,width=8cm}  &
\epsfig{file=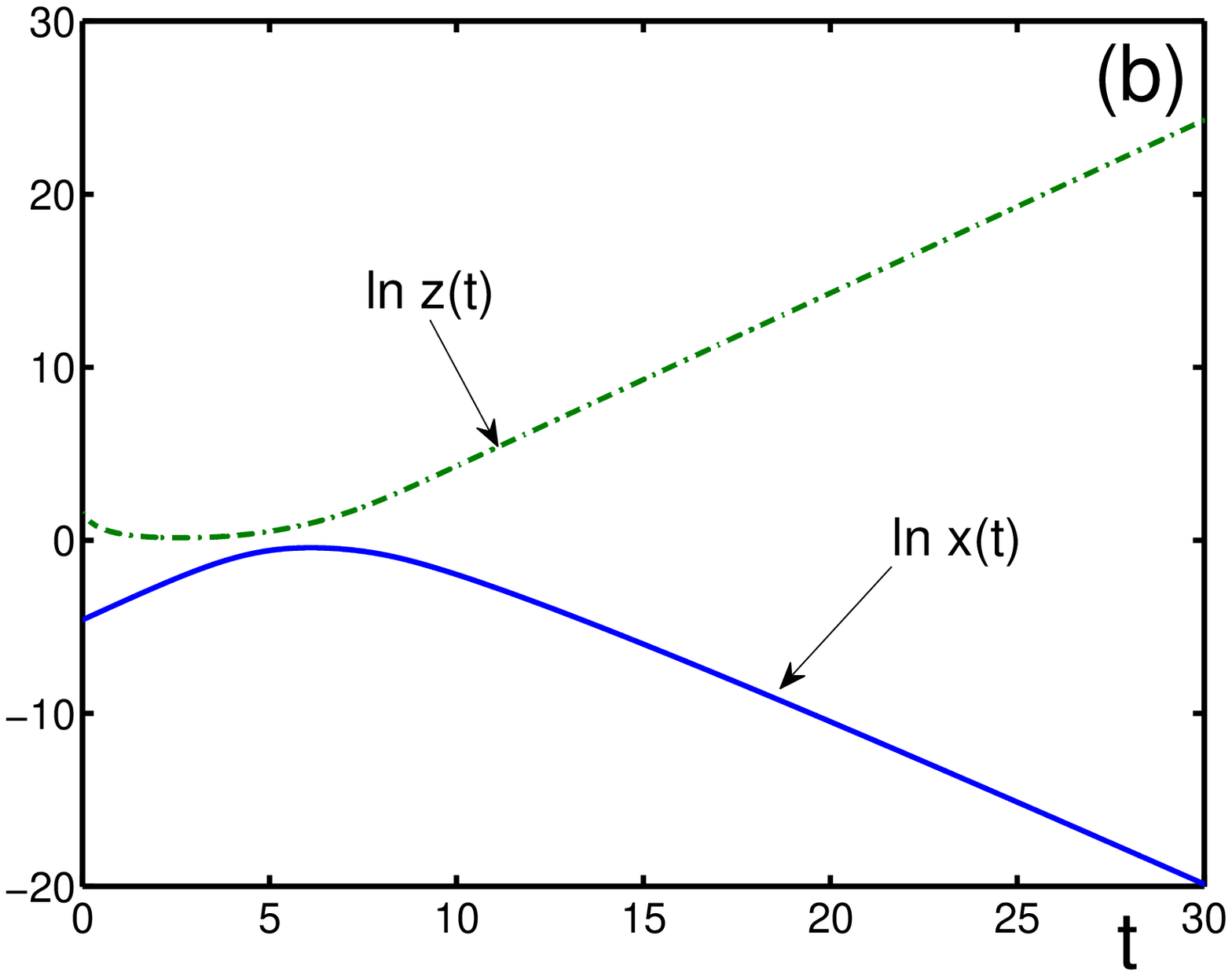,width=8cm} \\
\epsfig{file=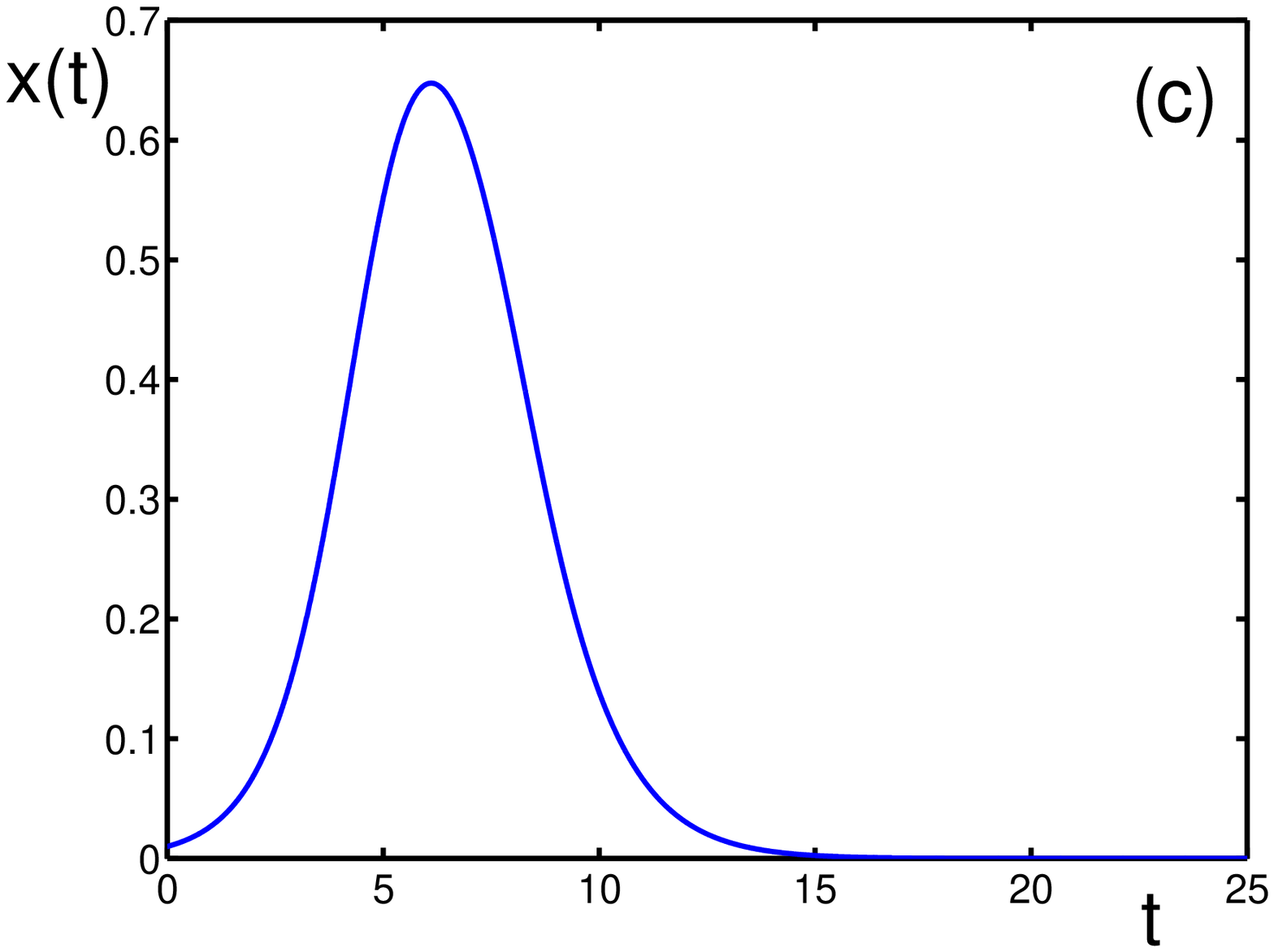,width=8cm}  &
\epsfig{file=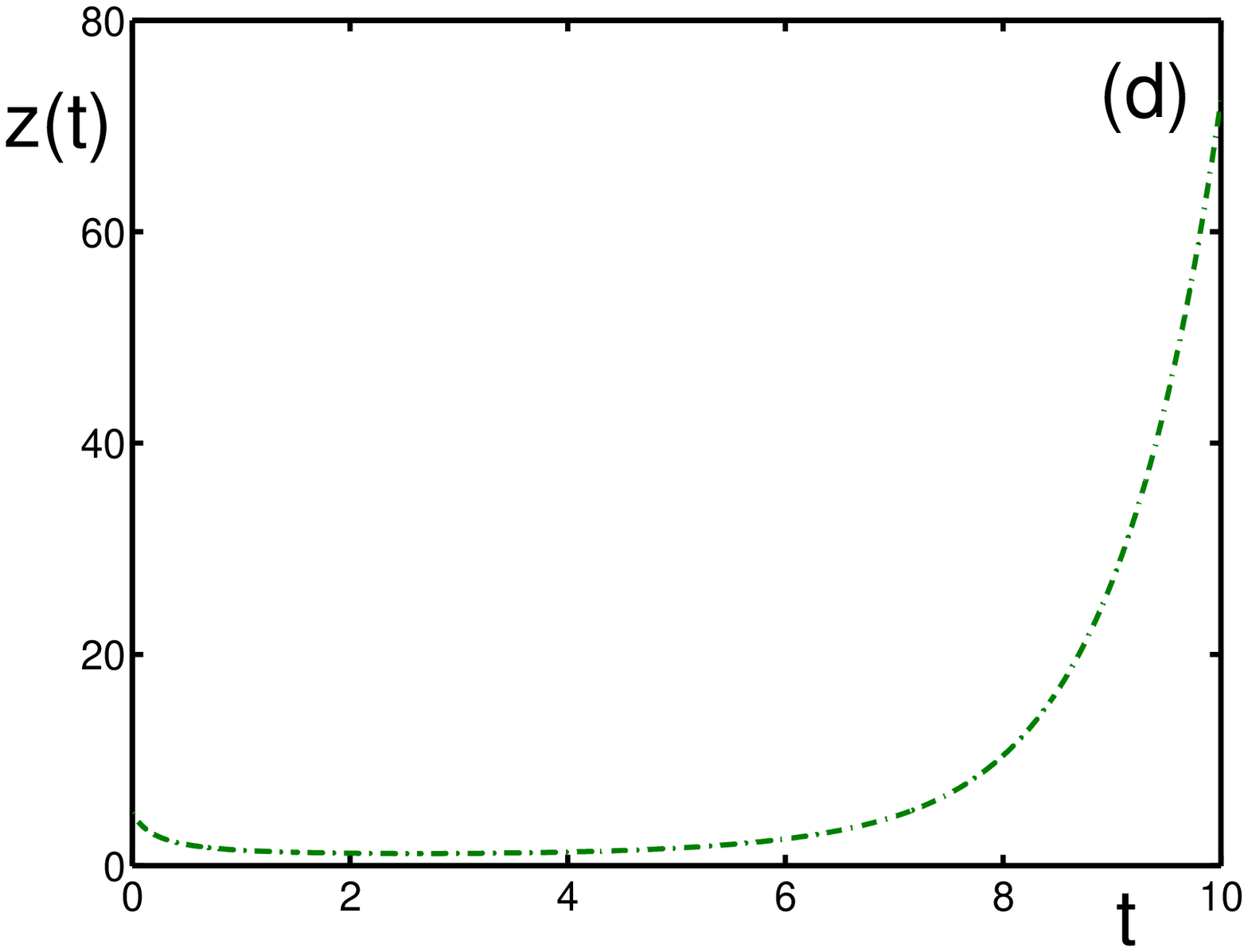,width=8cm}
\end{tabular}
\caption{
Dynamics of populations under $b + g > 1/e$, with the initial conditions
$x_0 = 0.01$ and $z_0 = 5$ for different symbiotic parameters:
(a)
Logarithmic behavior of mutualistic populations for $b = 2$ and $g = 1$.
Both populations grow, as $t \ra \infty$, with $\ln x(t) \ra \infty$ increasing
monotonically, while $\ln z(t) \ra \infty$, non-monotonically.
(b)
Logarithmic behavior of the populations, when one of them is parasitic,
for $b = -0.5$ and $g = 1$. At $t \ra \infty$, the host population dies out,
$\ln x(t) \ra -\infty$, that is, $x(t) \ra 0$, and the parasitic population
non-monotonically increases, $\ln z(t) \ra \infty$ .
(c)
Dynamics of the host population $x(t)$, for the same parameters $b = -0.5$
and $g = 1$, demonstrating that $x(t) \ra 0$ dies out, as $t \ra \infty$, in
a non-monotonic way.
(d)
Dynamics of the parasitic population $z(t)$, for the same parameters $b = -0.5$
and $g = 1$, showing the details of the non-monotonic increase of
$z(t) \ra \infty$, as $t \ra \infty$.
}
\label{fig:Fig.9}
\end{figure}

\newpage

%Figure 10
\begin{figure} [ht]
\centering
\begin{tabular}{lr}
\epsfig{file=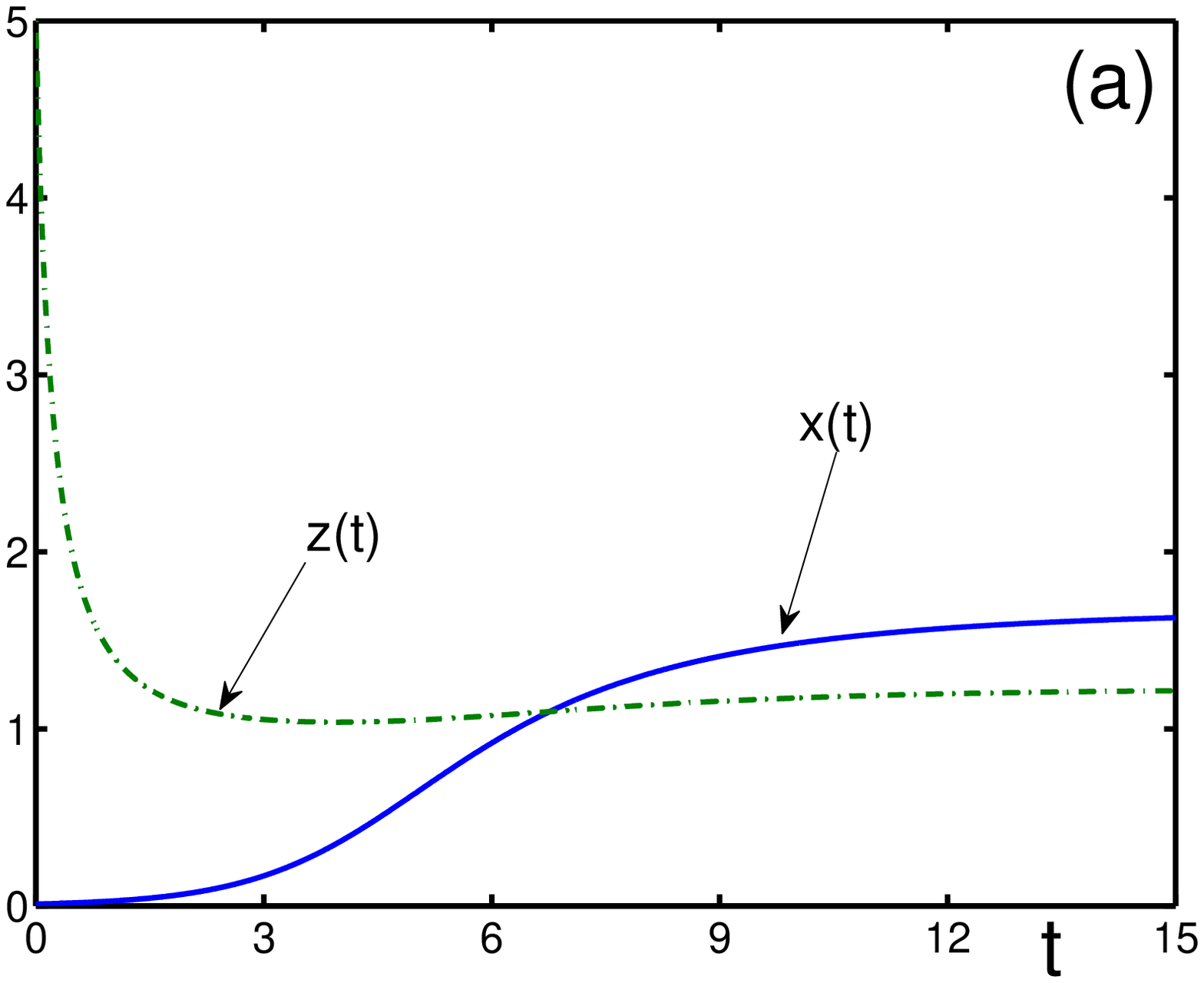,width=8cm}  &
\epsfig{file=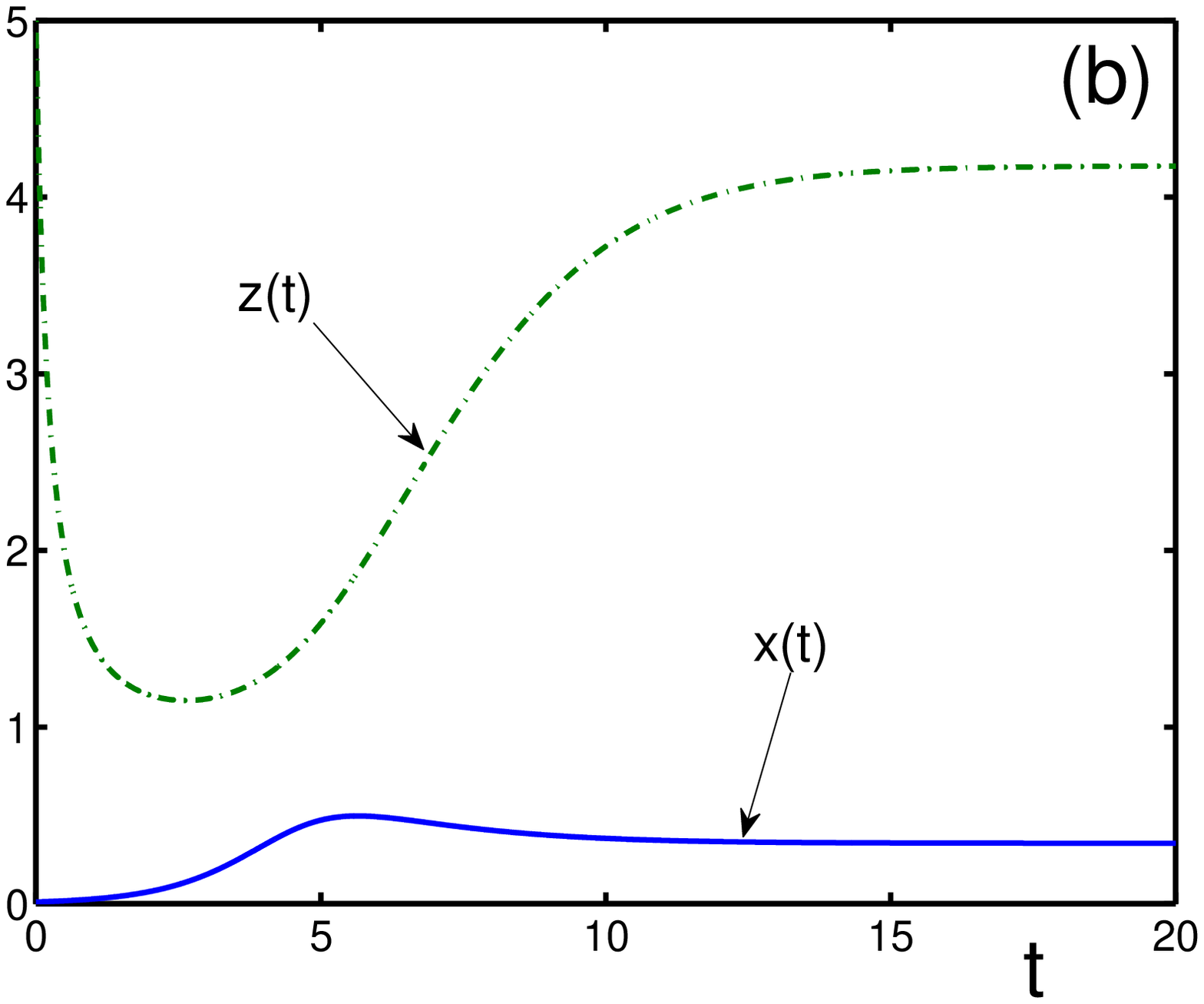,width=8cm} \\
\epsfig{file=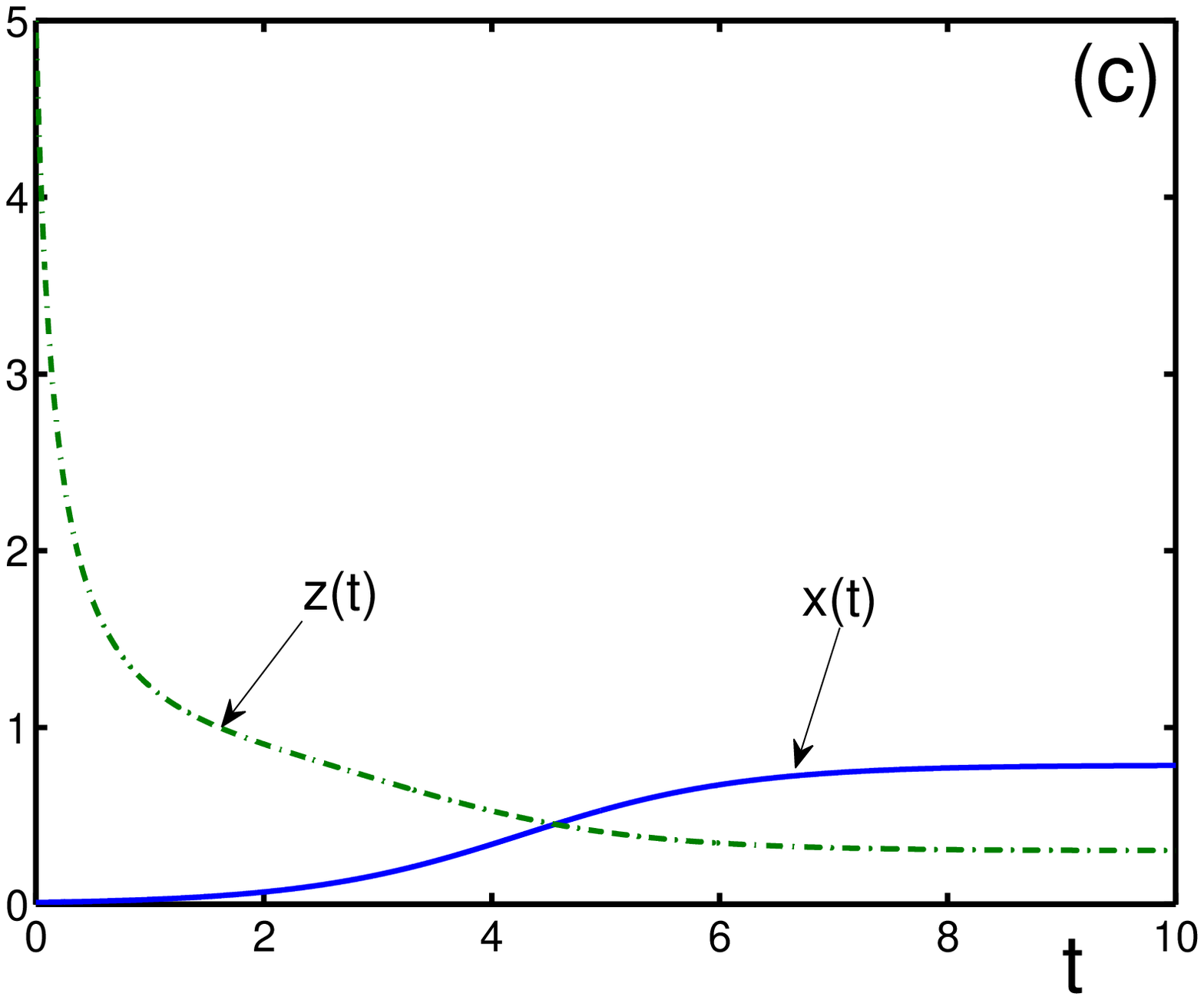,width=8cm}
\end{tabular}
\caption{
Population dynamics for $0 < b + g < 1/e$, with the initial conditions
$x_0 = 0.01$ and $z_0 = 5$.
(a)
For $b = 0.25$ and $g = 0.1$, the population $x(t) \ra x^* = 1.668$
monotonically from below, and population $z(t) \ra z^* = z^*=1.227$
non-monotonically from above.
(b)
For $b = -0.75$ and $g = 1$, the population $x(t) \ra x^* = 0.342$,
nonmonotonically from below, and population $z(t) \ra z^* = 4.177$,
non-monotonically from above.
(c)
For $b = -1$ and $g = -5$, the population $x(t) \ra x^* = 0.788$,
monotonically from below, and population $z(t) \ra z^* = 0.303$
monotonically from above.
}
\label{fig:Fig.10}
\end{figure}

\newpage

%Figure 11
\begin{figure} [ht]
\centering
\begin{tabular}{lr}
\epsfig{file=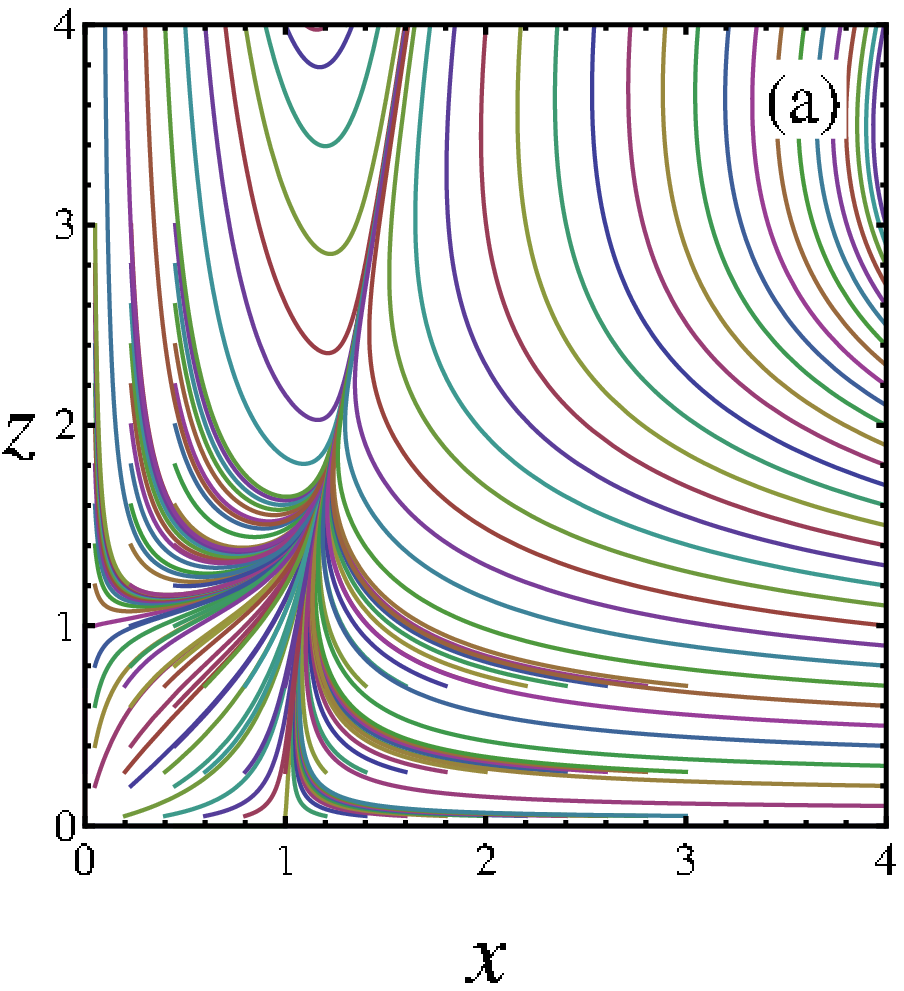,width=6.5cm}  &
\epsfig{file=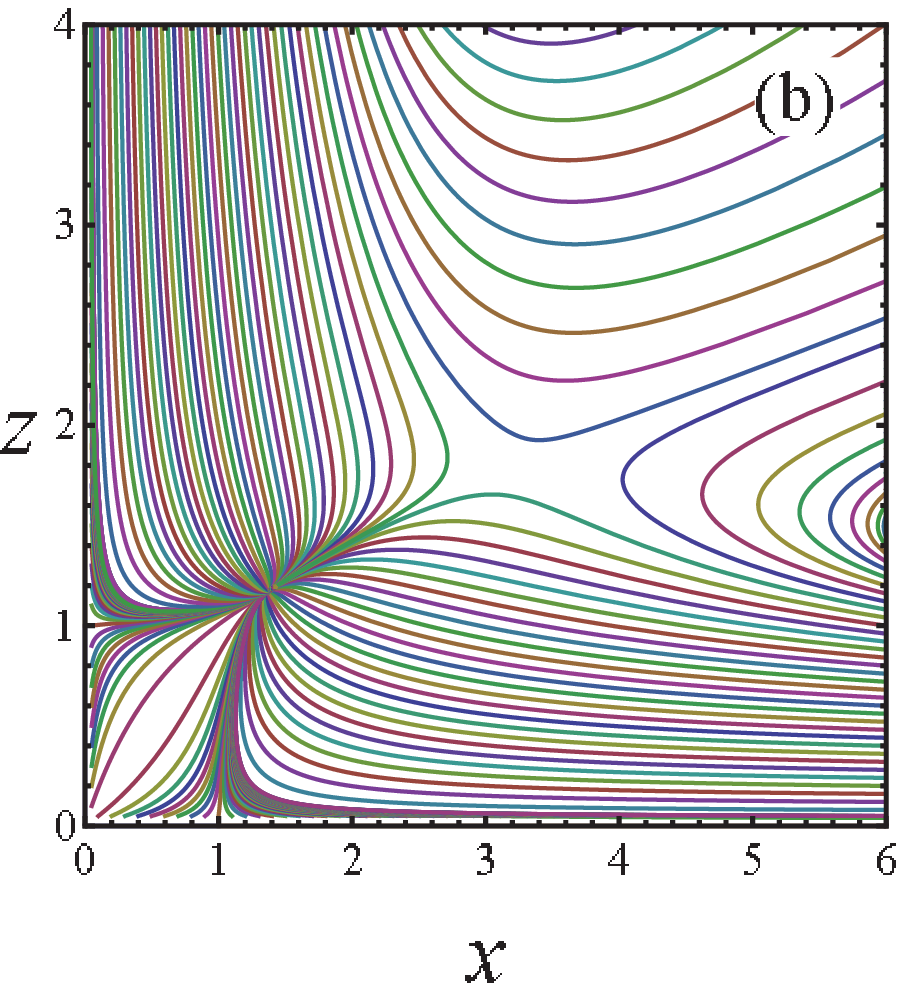,width=6.3cm} \\
&  \\
\epsfig{file=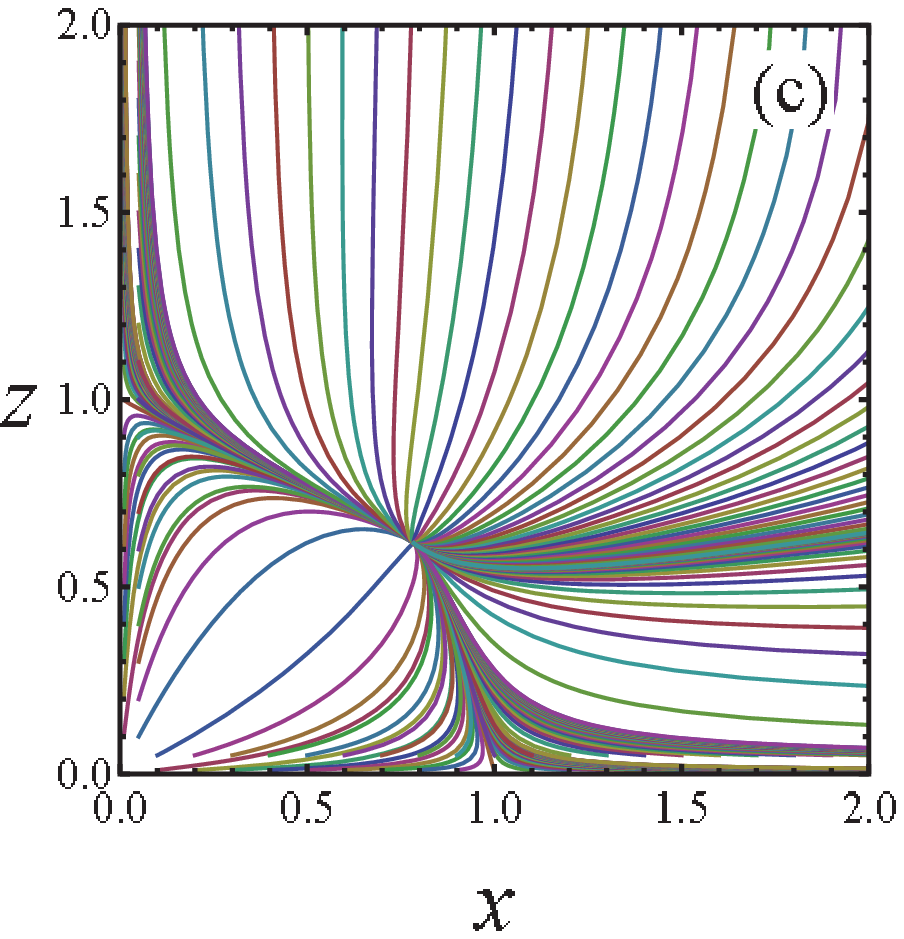,width=6.5cm}
\end{tabular}
\caption{
Phase portrait for active symbiosis, with the symbiosis parameters $b$
and $g$ from qualitatively different regions:
(a)
Phase portrait for $b = 0.1$ and $g = 0.3$, where $b + g > 1/e$. The
parameters are in the region, where there are no fixed points.
(b)
Phase portrait for $b = 0.2$ and $g = 0.1$, with $0 < b + g_c(b) < 1/e$.
The parameters are in the region of $b, g$, where two fixed points exist.
The first fixed point $\{ x_1^* = 1.38579, z_1^* = 1.17719 \}$ is the stable
node, with the Lyapunov exponents $\lbd_1 = -1$, $\lbd_2 = -1.72586$,
and the second point $\{x_2^* = 3.27906, z_2^* = 1.81082 \}$ is a saddle,
with the Lyapunov exponents $\lbd_1 = -1$, $\lbd_2 = 0.781337$.
(c)
Phase portrait for $b = -0.5$ and $g = -1$. The parameters are in the region
of $b, g$, where only one fixed point exists, which is the stable node
$\{ x^* = 0.78509, z^* = 0.61637 \}$, with the Lyapunov exponents
$\lbd_1 = -1$, $\lbd_2 = -1.72586$.
}
\label{fig:Fig.11}
\end{figure}

\clearpage

%Figure 12
\begin{figure} [ht]
\centering
\epsfig{file=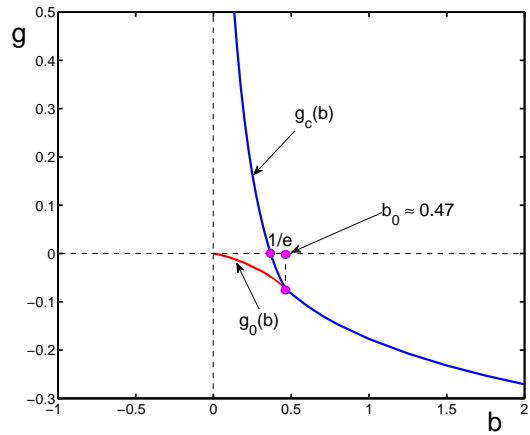,width=8cm}
\caption{Regions of existence of the fixed-point for mixed symbiosis. For
either $0 < b < 1/e$ and $g > g_c(b)$, or $b > 1/e$ and $g \geq 0$,
Eqs. (\ref{eqqq2}) do not have solutions, hence, there are no fixed points.
For $0 < b < 1/e$ and $0 < g < g_c(b)$, Eqs. (\ref{eqqq2}) have two solutions,
but only one fixed point is stable. For $b > 1/e$ and $g_c(b) < g <0$, there
exists one unstable fixed point. When either $0 < b < 1/e$ and $g_0(b) < g < 0$,
or $1/e < b < b_0 \approx 0.47$ and $g_0(b) < g < g_c(b)$, then there are
three fixed points, but only one of them is stable. When either $0 < b < b_0$
and $g < g_0(b)$, or $b \geq b_0$ and $g < g_c(b)$, then there exists a single
fixed point that is stable. When $b \leq 0$ and $g \in (-\infty, \infty)$, then
the unique solution to Eqs. (\ref{eqqq2}) is a stable fixed point.
}
\label{fig:Fig.12}
\end{figure}

\clearpage

%Figure 13
\begin{figure} [ht]
\centering
\begin{tabular}{lr}
\epsfig{file=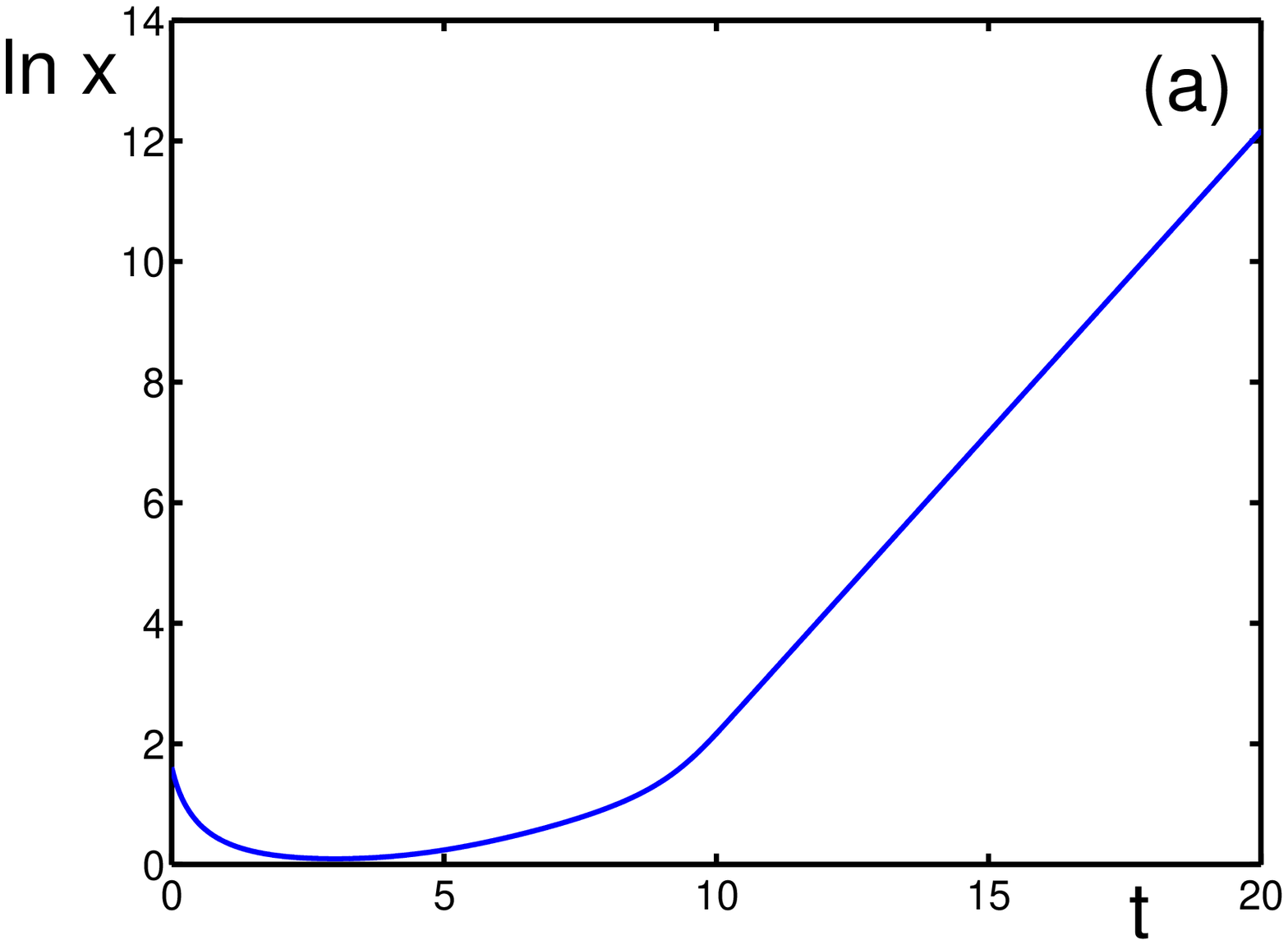,width=8cm}  &
\epsfig{file=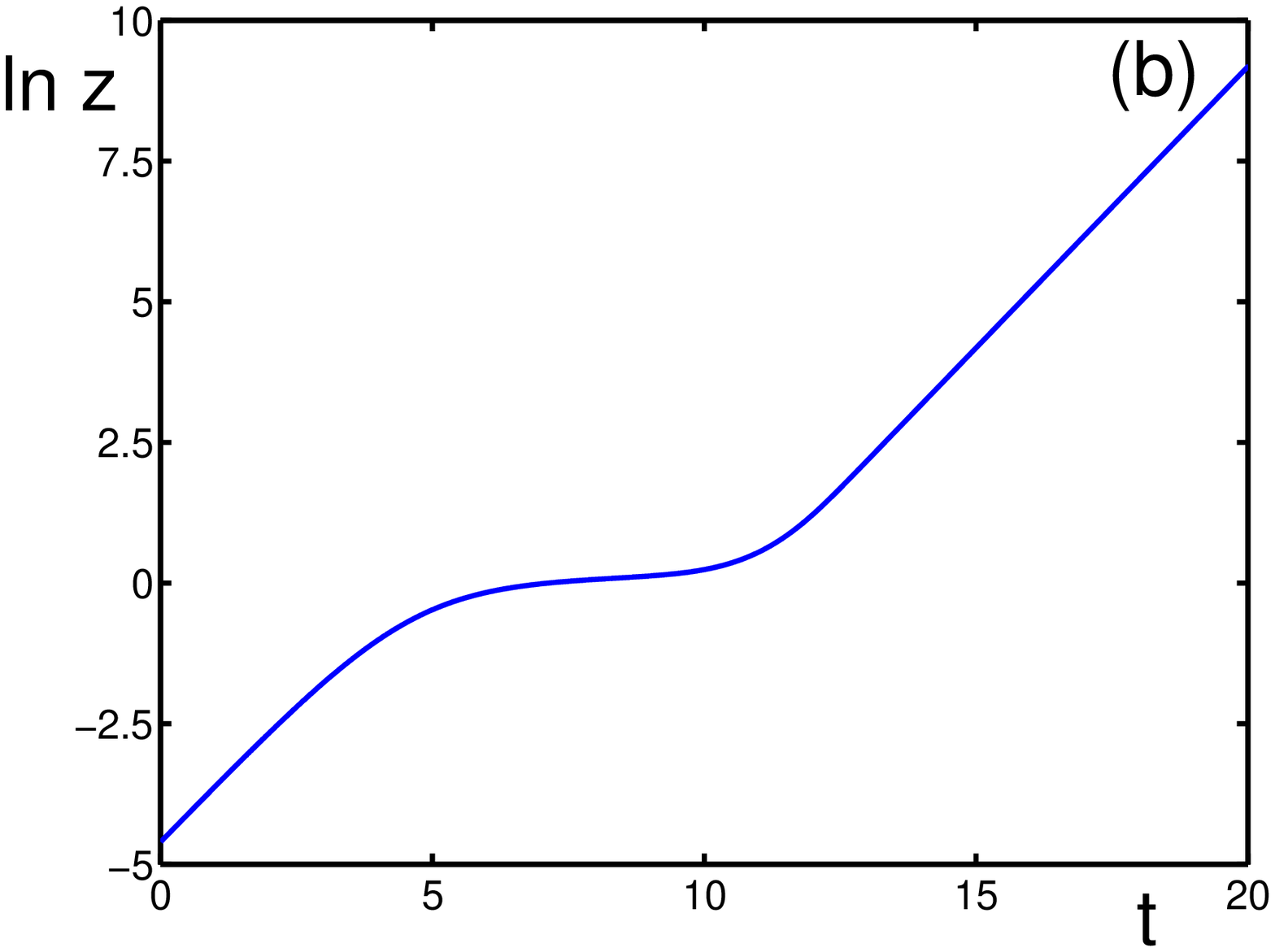,width=8cm} \\
\epsfig{file=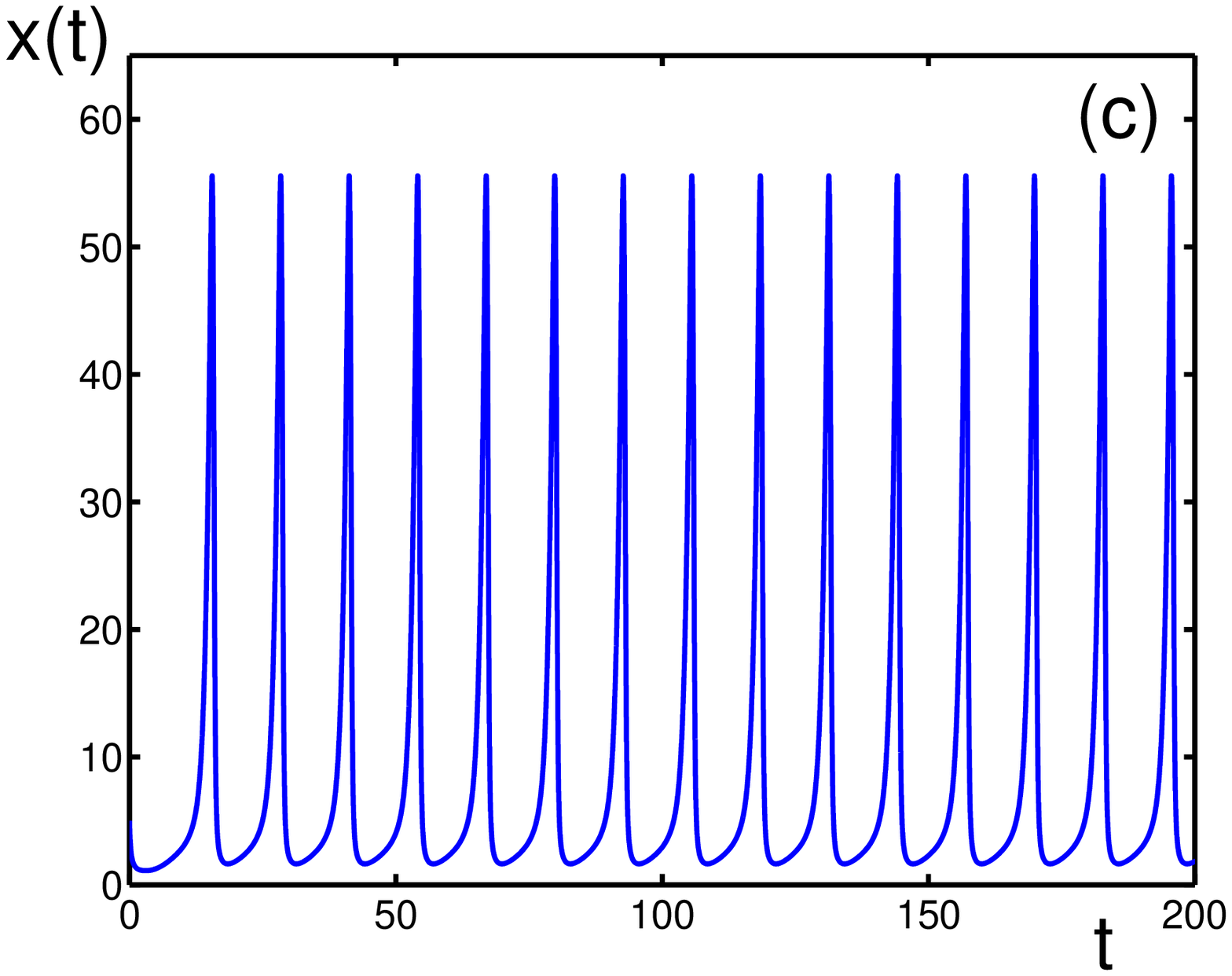,width=8cm} &
\epsfig{file=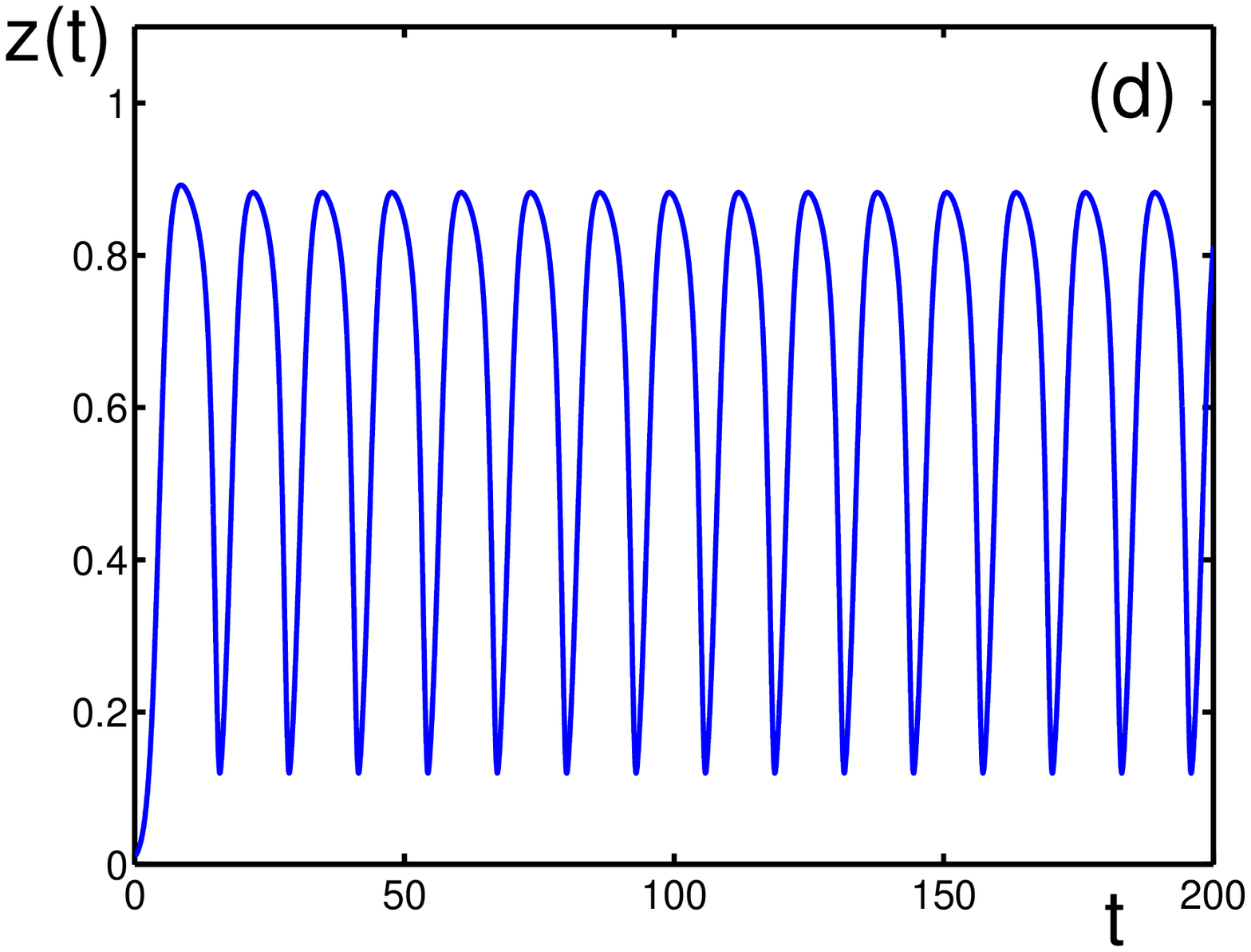,width=8cm}
\end{tabular}
\caption{
Dynamics of populations under mixed symbiosis for the symbiotic parameters,
when unbounded growth changes to everlasting oscillations. The symbiotic
parameter of the active species is fixed, $b = 0.5 > 1/e$, while that of
the passive species varies. For the given $b$, the change of the behavior
occurs on the line $g = 0$ that is higher than the Hopf bifurcation point
$g_c(b) = -0.083065$. The initial conditions are $\{x_0 = 5, z_0 = 0.01\}$.
(a)
Logarithmic behaviour of the active population $x(t)$ for $g = 0.05$, where
there are no fixed points.
(b)
Logarithmic behaviour of the passive population $z(t)$ for $g = 0.05$.
(c)
Active species population $x(t)$ oscillates without convergence for
$g = -0.05 > g_c(b) = -0.083065$.
(d)
Passive species population $z(t)$ oscillates without convergence for the same
$g = -0.05$.
}
\label{fig:Fig.13}
\end{figure}

\clearpage

%Figure 14
\begin{figure} [ht]
\centering
\begin{tabular}{lr}
\epsfig{file=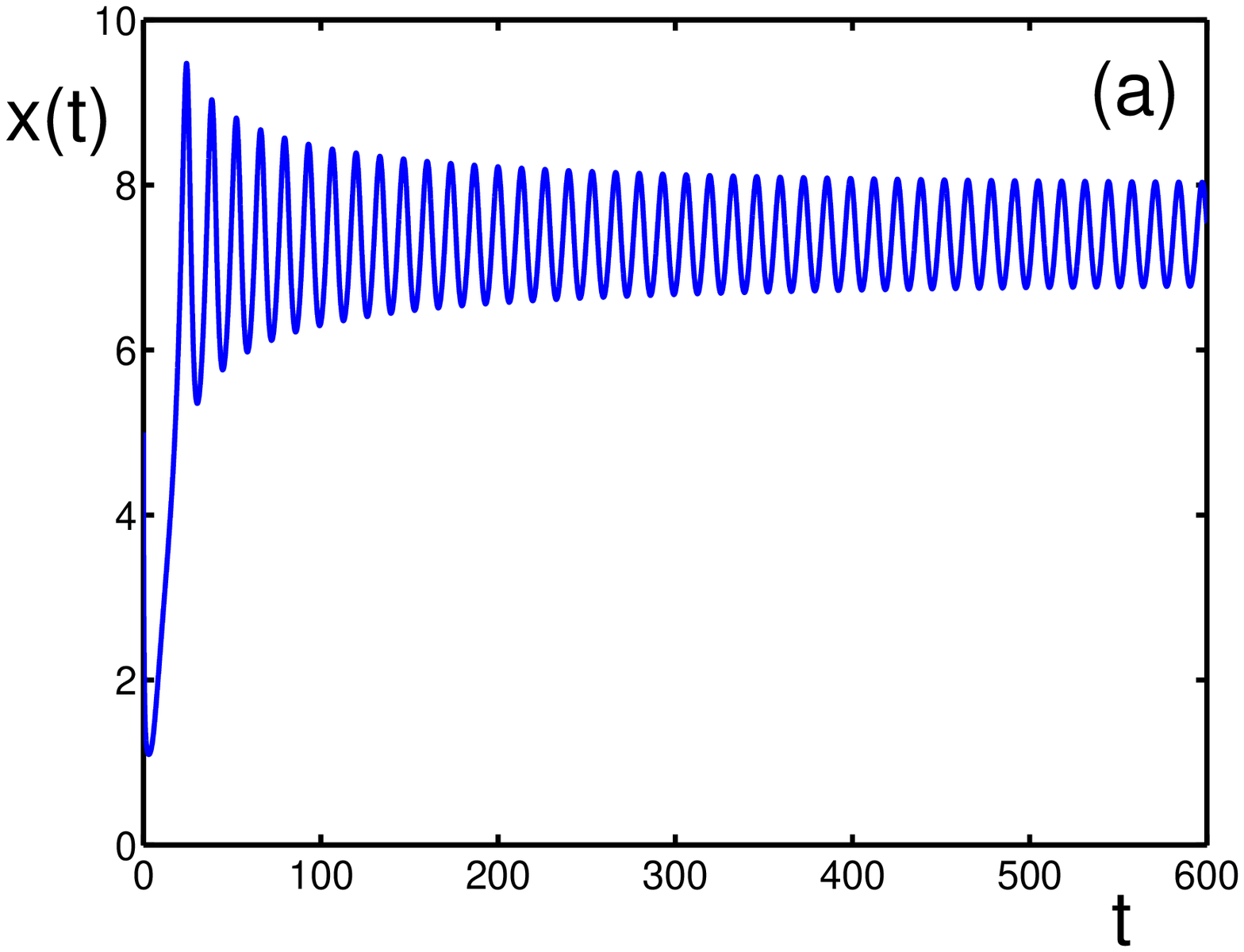,width=8cm}  &
\epsfig{file=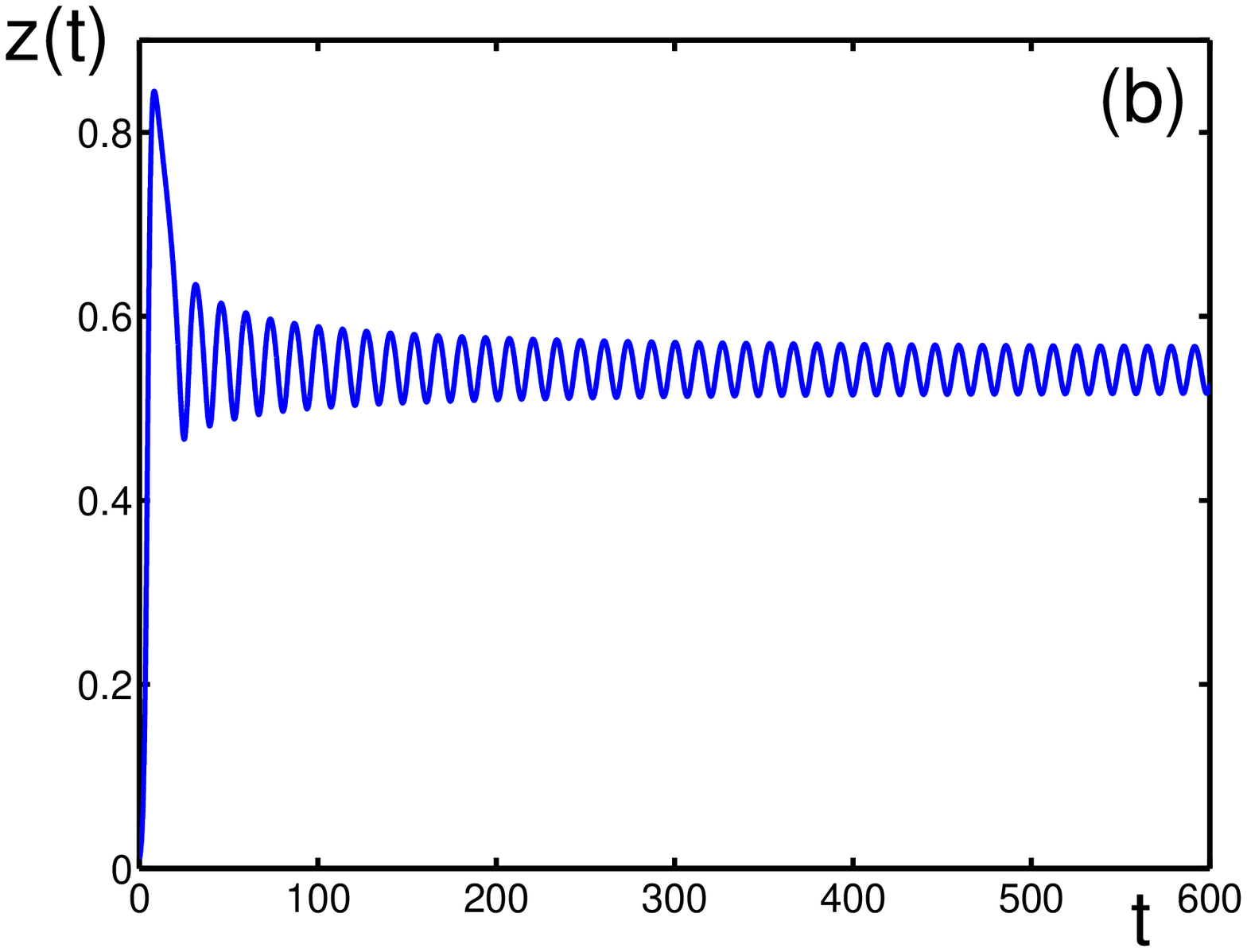,width=8cm} \\
\epsfig{file=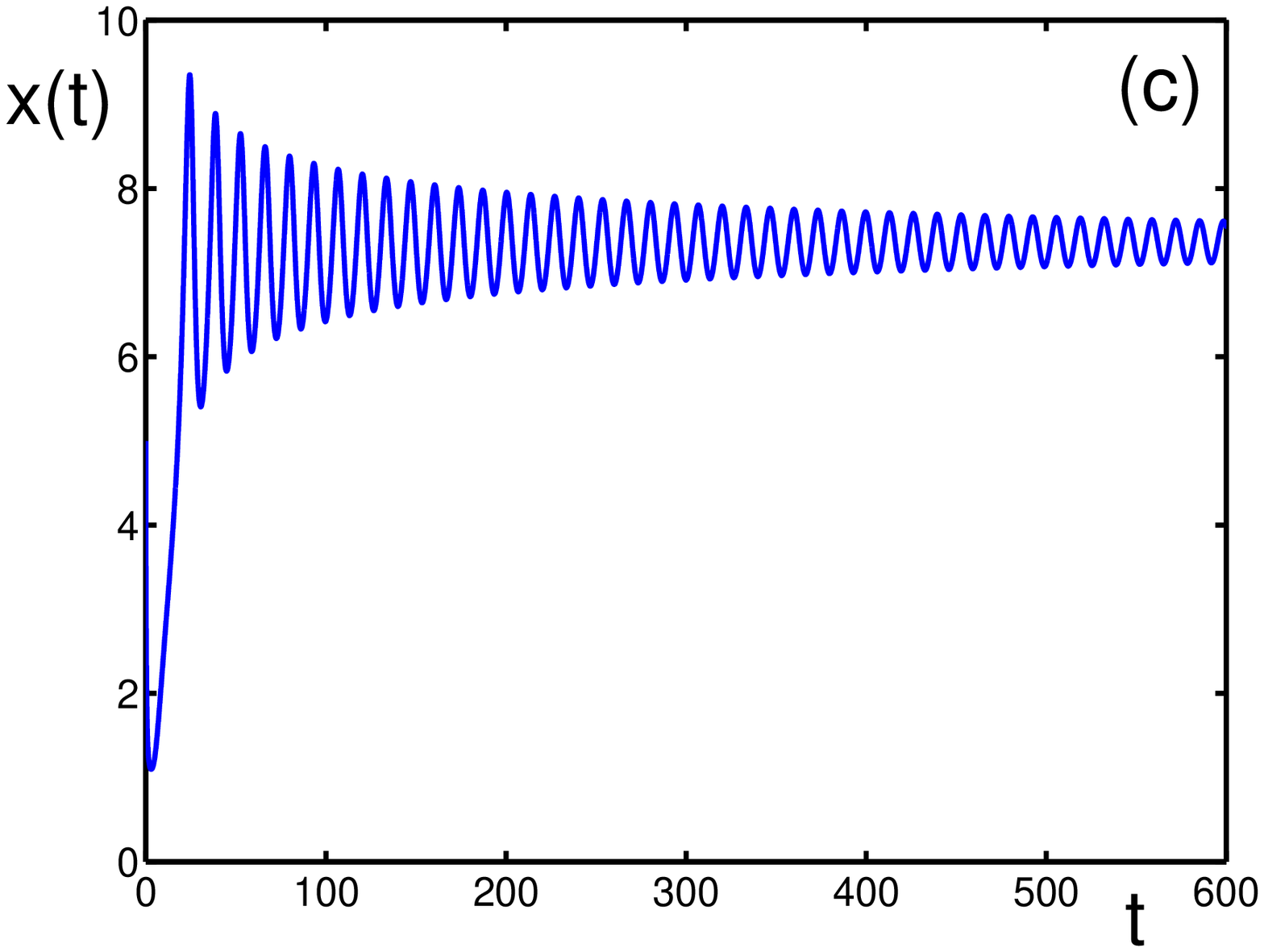,width=8.5cm} &
\epsfig{file=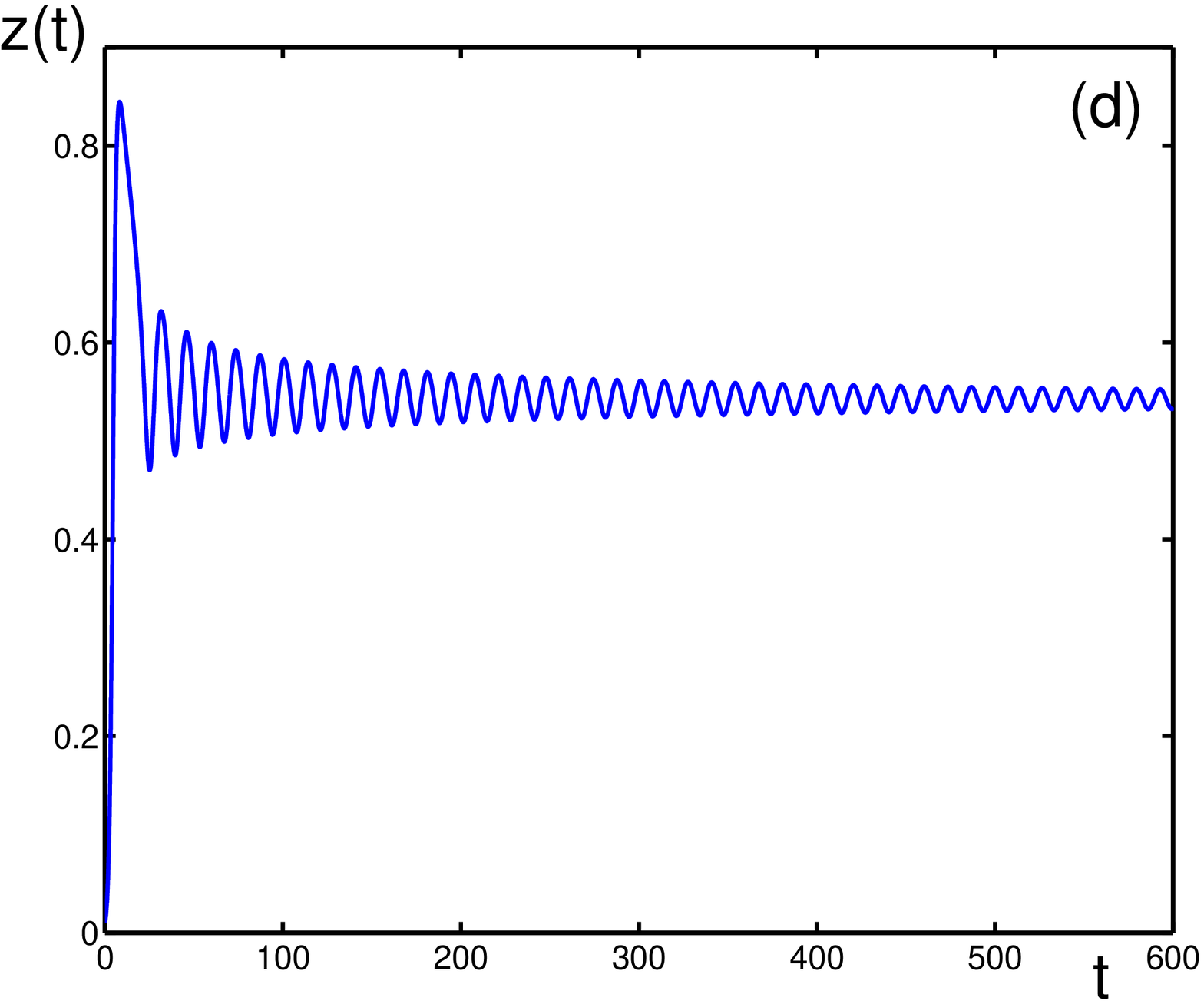,width=8cm}
\end{tabular}
\caption{
Dynamics of populations under mixed symbiosis for the symbiotic parameters,
when everlasting oscillations change to the oscillating convergence to a
a stable focus. The symbiotic parameter of the active species is fixed,
$b = 0.5 > 1/e$, with the Hopf bifurcation point $g_c(b) = -0.083065$. The
initial conditions are the same as in Fig. 13, $\{x_0 = 5, z_0 = 0.01\}$.
(a)
Active species population $x(t)$ oscillates with a small amplitude, as
$t \ra \infty$, when $g = -0.083 > g_c(b)$ is slightly larger than the Hopf
bifurcation point. The characteristic exponents for the unstable focus are
$\lbd_1 = 0.0018 + 0.479 i$ and $\lbd_2 = \lbd_1^*$.
(b)
Passive species population $z(t)$ oscillates, as $t \ra \infty$, when
$g = -0.083 > g_c(b)$ is slightly larger than the Hopf bifurcation point.
(c)
Active species population $x(t)$ converges with oscillations to its stable
stationary state $x^* = 7.368$, when $g = -0.0831 < g_c(b)$ is slightly
lower than the Hopf bifurcation point. The characteristic exponents for the
stable focus are $\lbd_1 = -0.0014 + 0.475 i$ and $\lbd_2 = \lbd_1^*$.
(d)
Passive species population $z(t)$ converges oscillating to its stable
stationary state $z^* = 0.542$, when $g = -0.0831 < g_c(b)$ is slightly
lower than the Hopf bifurcation point.
}
\label{fig:Fig.14}
\end{figure}

\clearpage

%Figure 15
\begin{figure} [ht]
\centering
\begin{tabular}{lr}
\epsfig{file=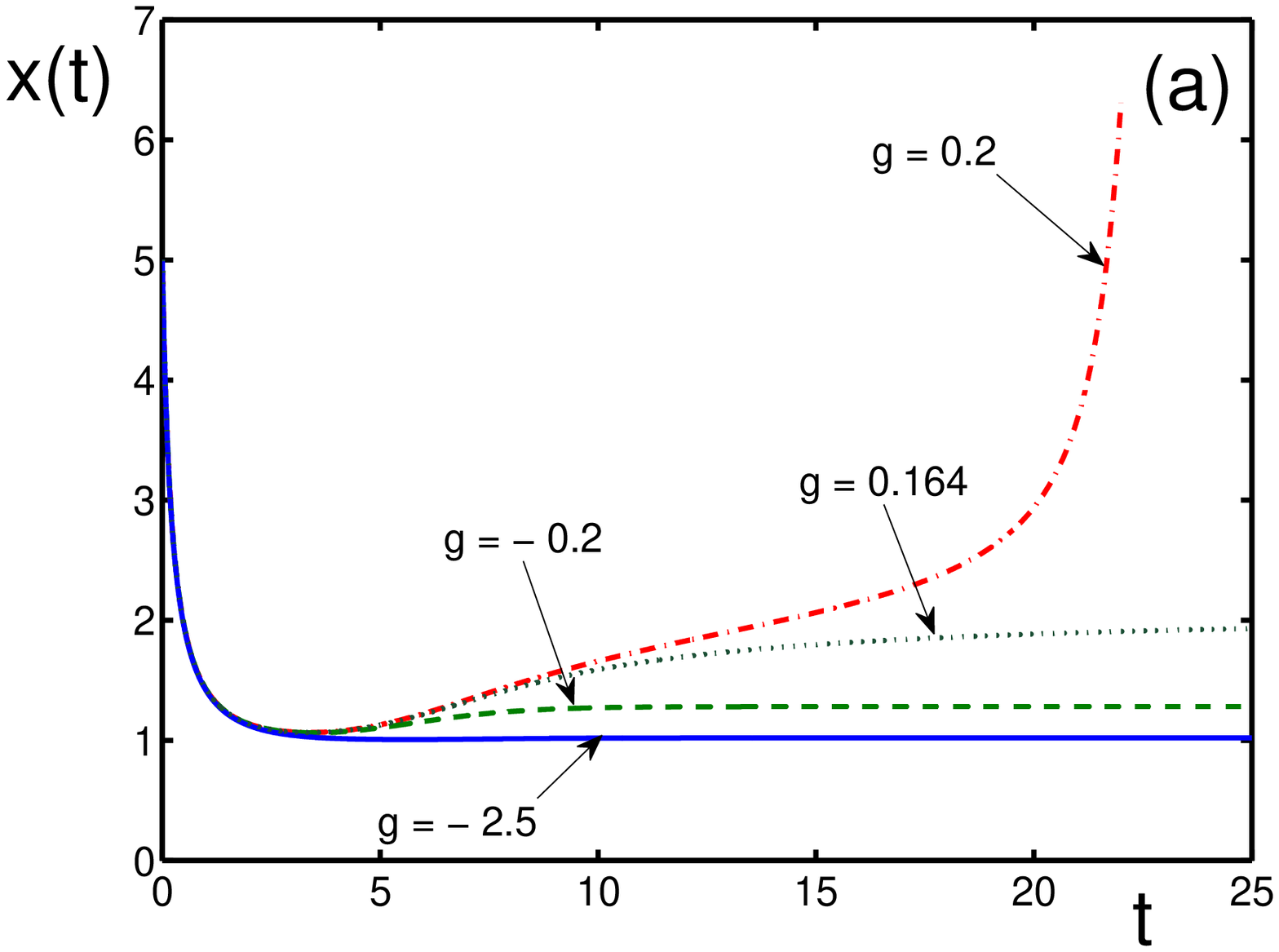,width=8cm}  &
\epsfig{file=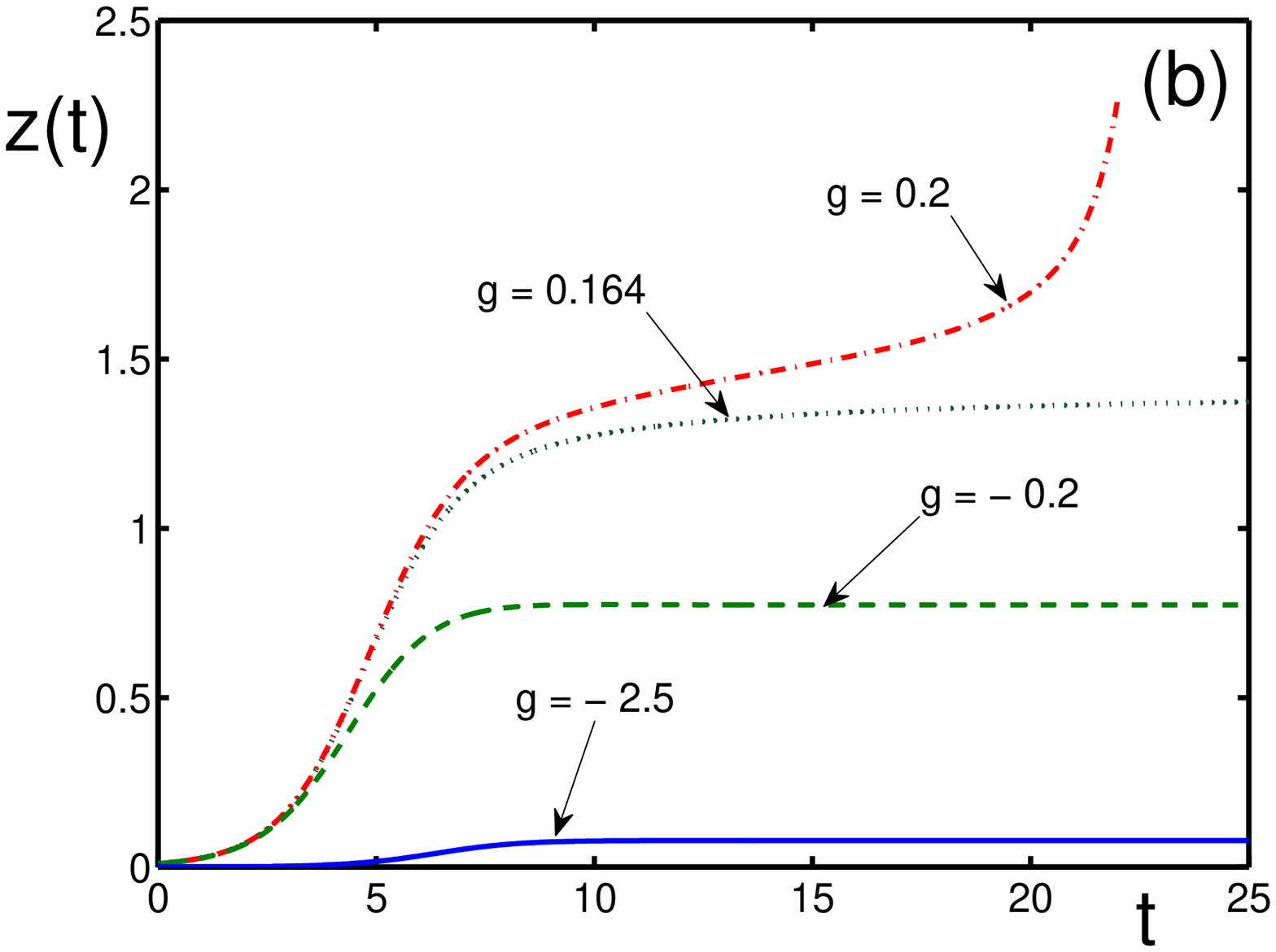,width=8cm} \\
\epsfig{file=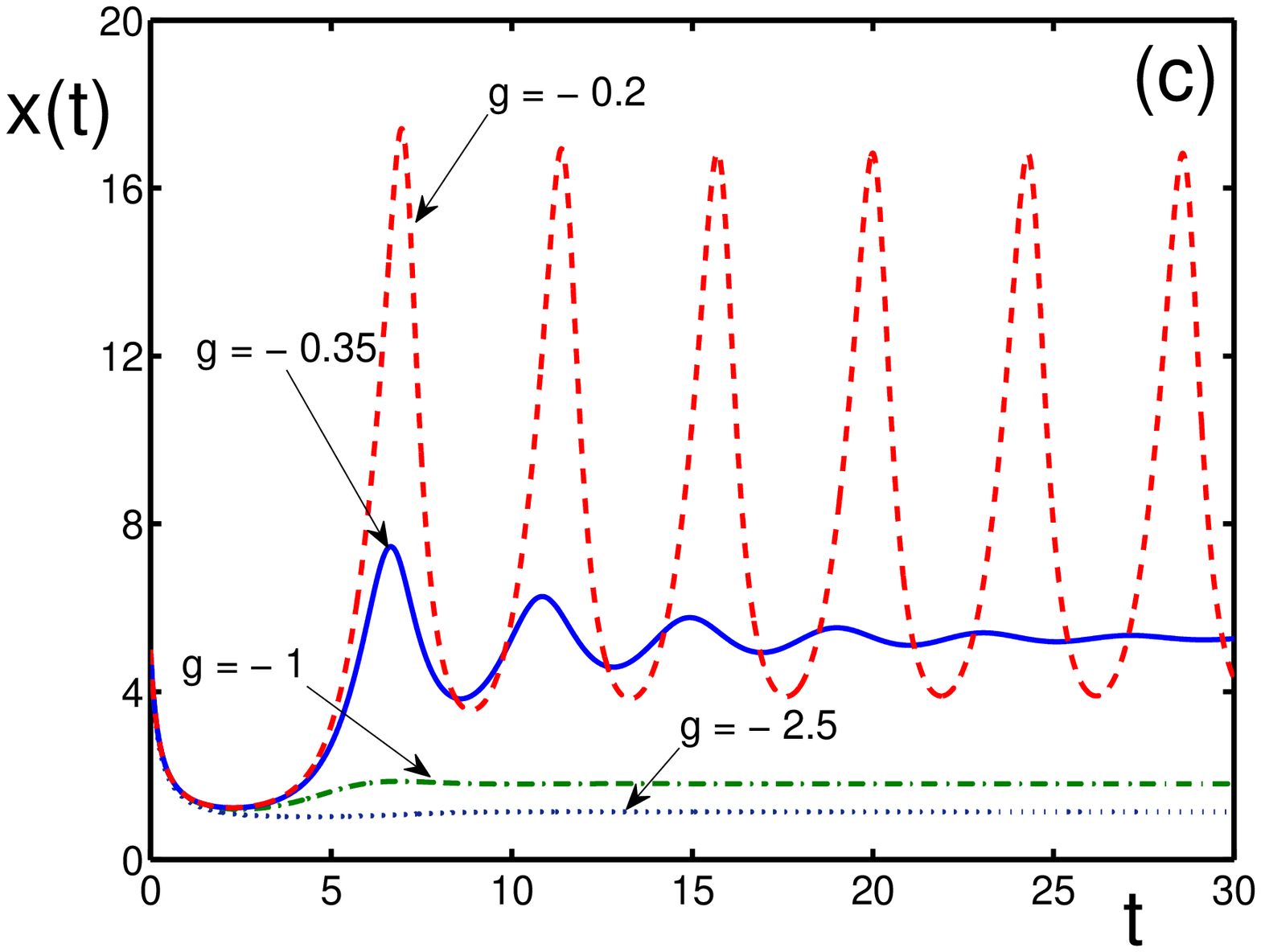,width=8cm} &
\epsfig{file=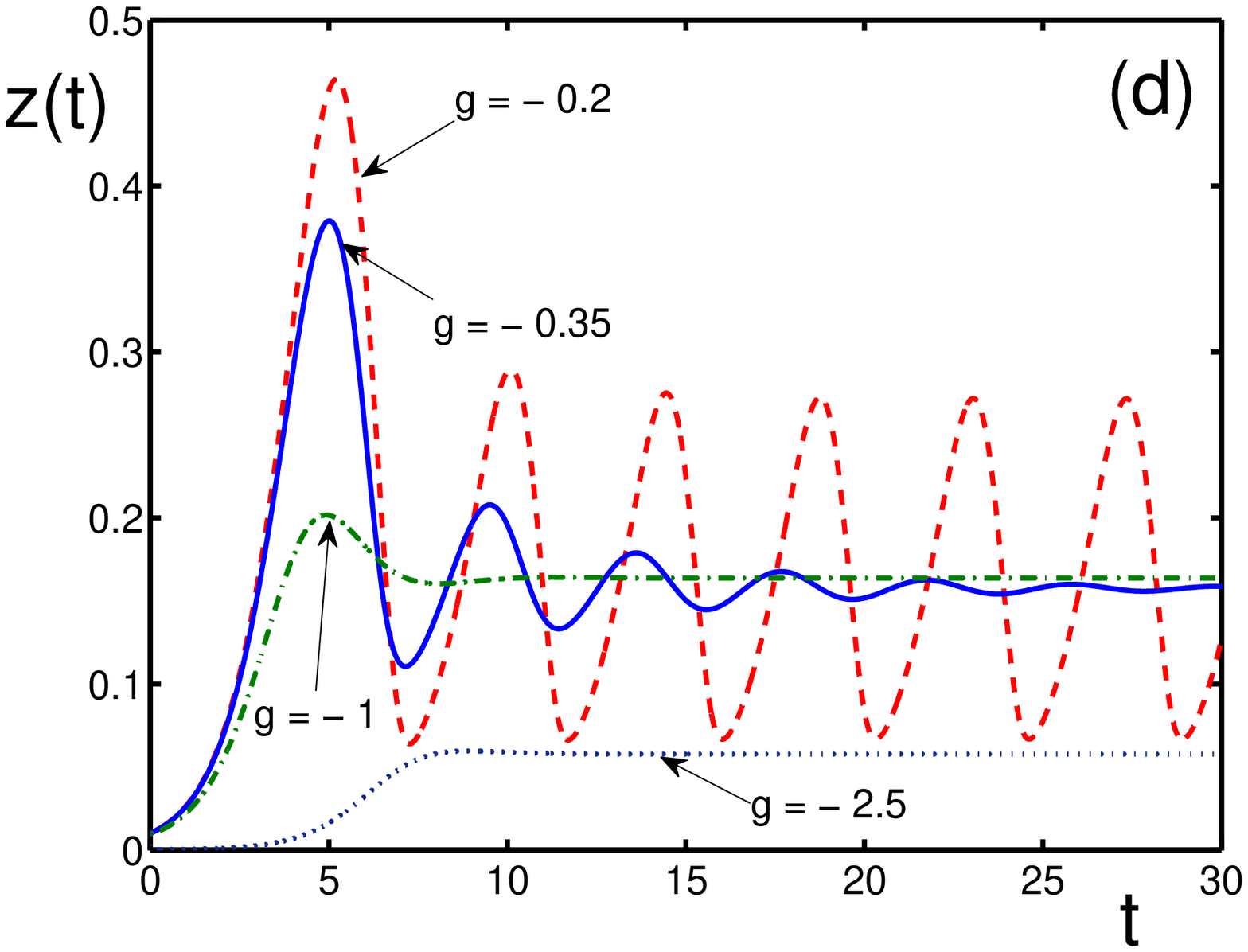,width=8cm}
\end{tabular}
\caption{
Population dynamics under mixed symbiosis for varying symbiotic parameters,
with the initial conditions $\{x_0 = 5, z_0 = 0.01\}$.
(a)
Active species population dynamics for the fixed $b = 0.25 < 1/e$, when
$g_c(b) = 0.1645$ and $g_0(b) = -0.026564$, for different passive species symbiotic
parameters: $g = -2.5 < g_0(b)$ (solid line), when the population
$x(t)$ tends to $x^* = 1.02$; $g = -0.2$  (dashed line), when the population
$x(t) \ra x^* = 1.28$; $0 < g = 0.164 < g_c(b)$ (dotted line), when the
population $x(t) \ra x^* = 2.03$;  and $g = 0.2 > g_c(b)$ (dashed-dotted line),
when $x(t) \ra \infty$, as $t \ra \infty$.
(b)
Passive species population dynamics for $b = 0.25$ and the same initial
conditions, with the varying symbiotic parameter $g$ of the passive species:
$g = -2.5$ (solid line), when $z(t) \ra z^* = 0.0781$; $g = -0.2$ (dashed line),
when $z(t) \ra z^* = 0.774$; $g = 0.164$ (dotted line), when $z(t) \ra z^* = 1.39$;
and $g = 0.2 > g_c(b)$ (dashed-dotted line), when $x(t) \ra \infty$, as
$t \ra \infty$.
(c)
Active species population dynamics under the fixed $b = 2 > b_0 \approx 0.47$,
with $g_c(b) = -0.271$, for different $g$: $g = -2.5$ (dotted line), when
$x(t) \ra x^* = 1.14$; $g = -1$ (dashed-dotted line), with $x(t) \ra x^* = 1.81$;
$g = -0.35 < g_c(b)$ (solid line), when $x(t) \ra x^* = 5.28$ with a few
oscillations; and $g = -0.2 > g_c(b)$ (dashed line), when $x(t)$ oscillates
without convergence, as $t\ra\infty$.
(d)
Passive species population dynamics for $b = 2$ and the same initial
conditions, with the varying symbiotic parameter $g$: $g = -2.5$ (dotted line),
when $z(t) \ra z^* = 0.0577$; $g = -1$ (dashed-dotted line), with
$z(t) \ra z^* = 0.164$; $g = -0.35 < g_c(b)$ (solid line), when
$z(t) \ra z^* = 0.157$ with a few oscillations, and $g = -0.2 > g_c(b)$
(dashed line), with $z(t)$ oscillating without convergence, as $t \ra \infty$.
}
\label{fig:Fig.15}
\end{figure}

\clearpage

%Figure 16
\begin{figure} [ht]
\centering
\begin{tabular}{lr}
\epsfig{file=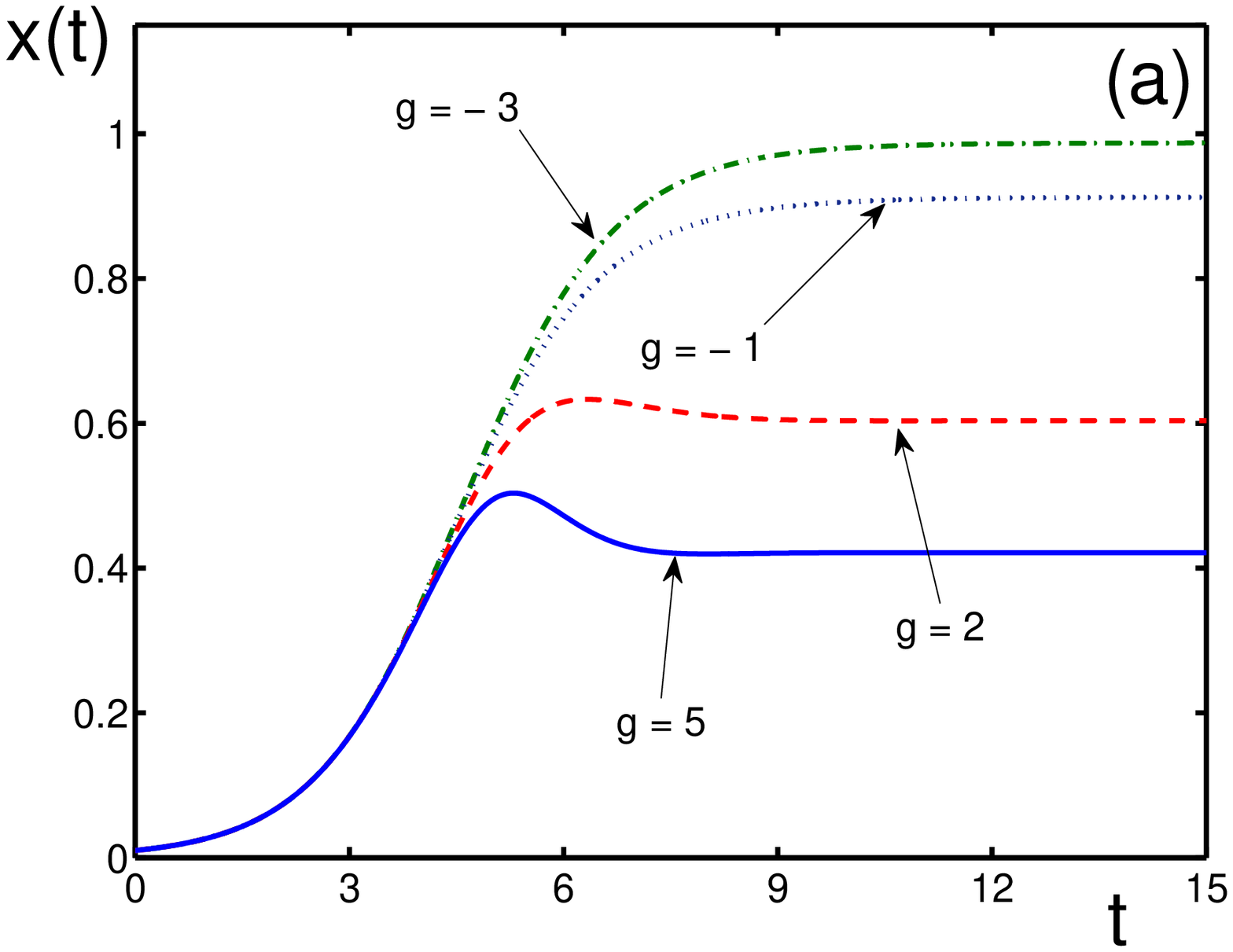,width=8cm}  &
\epsfig{file=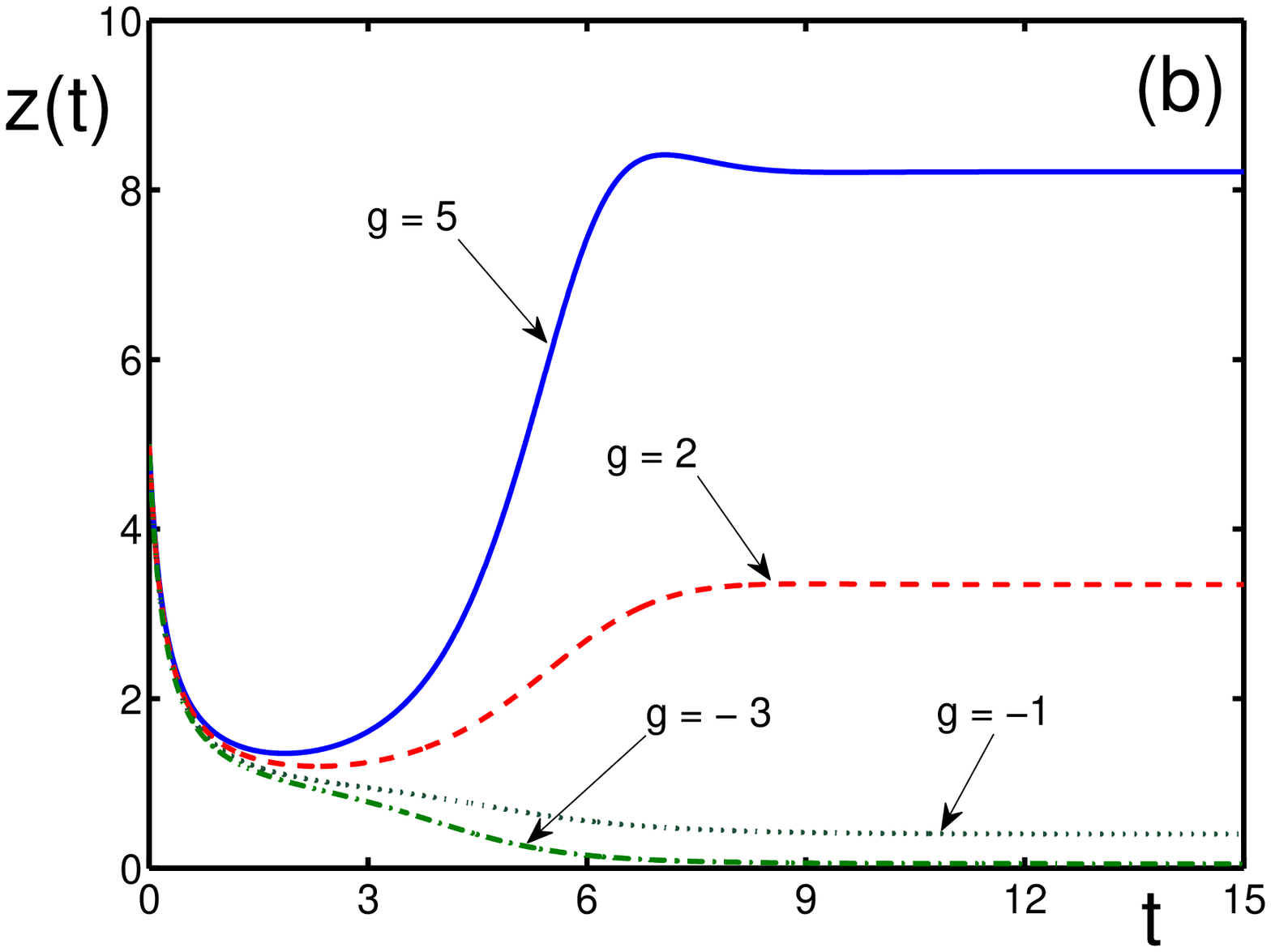,width=8cm} \\
\epsfig{file=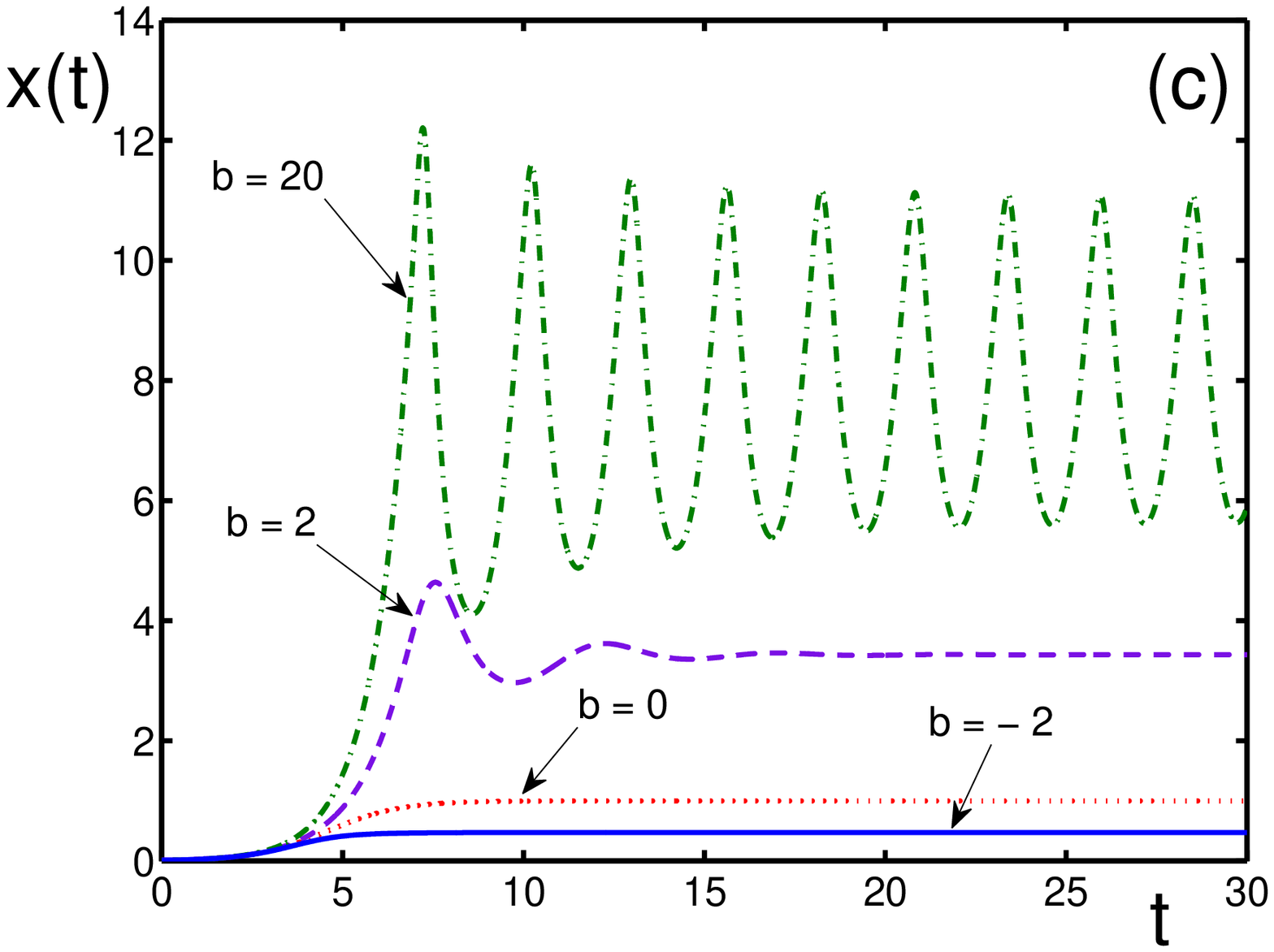,width=8cm} &
\epsfig{file=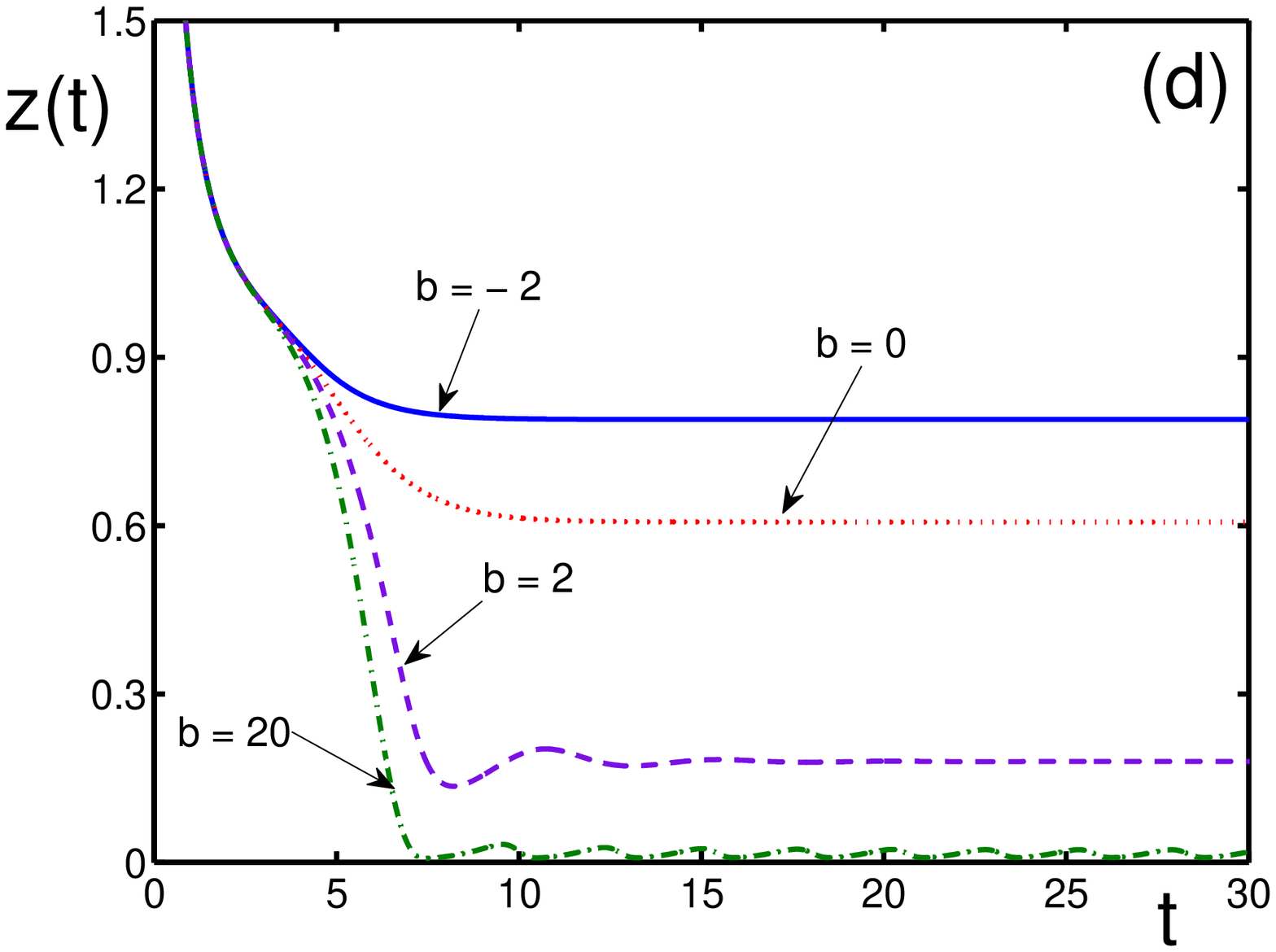,width=8cm}
\end{tabular}
\caption{
Population dynamics under mixed symbiosis for varying symbiotic parameters,
with the initial conditions $\{x_0 = 0.01, z_0 = 5 \}$.
(a)
Active species population dynamics under the fixed $b = -0.25$ and different
$g$: $g = 5$ (solid line), when $x(t) \ra x^* = 0.421$; $g = 2$ (dashed line),
with $x(t) \ra x^* = 0.604$; $g = -1$ (dotted line), when $x(t) \ra x^* = 0.912$;
and $g = -2$ (dashed-dotted line), with $x(t) \ra x^* = 0.987$.
(b)
Passive species population dynamics under the fixed $b = -0.25$ and different
$g$: $g = 5$ (solid line), with $z(t) \ra z^* = 8.21$ non-monotonically from
below; $g = 2$ (dashed line), when $z(t) \ra z^* = 3.34$ non-monotonically from
above; $g = -1$ (dotted line), when $z(t)\ra z^* = 0.402$; and $g = -2$
(dashed-dotted line ), when $z(t) \ra z^* = 0.0517$ monotonically from above.
(c)
Active species population dynamics under the fixed $g = -0.5$ and different $b$:
$b = -2$ (solid line), when $x(t)\ra x^*=0.474$; $b = 0$ (dotted line), with
$x(t) \ra x^* = 1$ monotonically from below; $b = 2$ (dashed line), when
$g_c(b) = -0.271 > g$, and $x(t) \ra x^* = 3.43$ with a few oscillations;
$b = 20$ (dashed-dotted line), when $g_c(b) = -0.5823 < g < 0$, and solution
$x(t)$ oscillates without convergence, as $t \ra \infty$.
(d)
Passive species population dynamics under the fixed $g = -0.5$ and different
$b$: $b = -2$ (solid line), when $z(t)\ra z^* = 0.789$; $b = 0$ (dotted line),
with $z(t) \ra z^* = 0.607$ monotonically from above; $b = 2$ (dashed line),
when $z(t) \ra z^* = 0.1796$ with a few oscillations; and $b = 20$
(dashed-dotted line), when $z(t)$ oscillates without convergence, as $t \ra \infty$.
}
\label{fig:Fig.16}
\end{figure}

\newpage

%Figure 17
\begin{figure} [ht]
\centering
\begin{tabular}{lr}
\epsfig{file=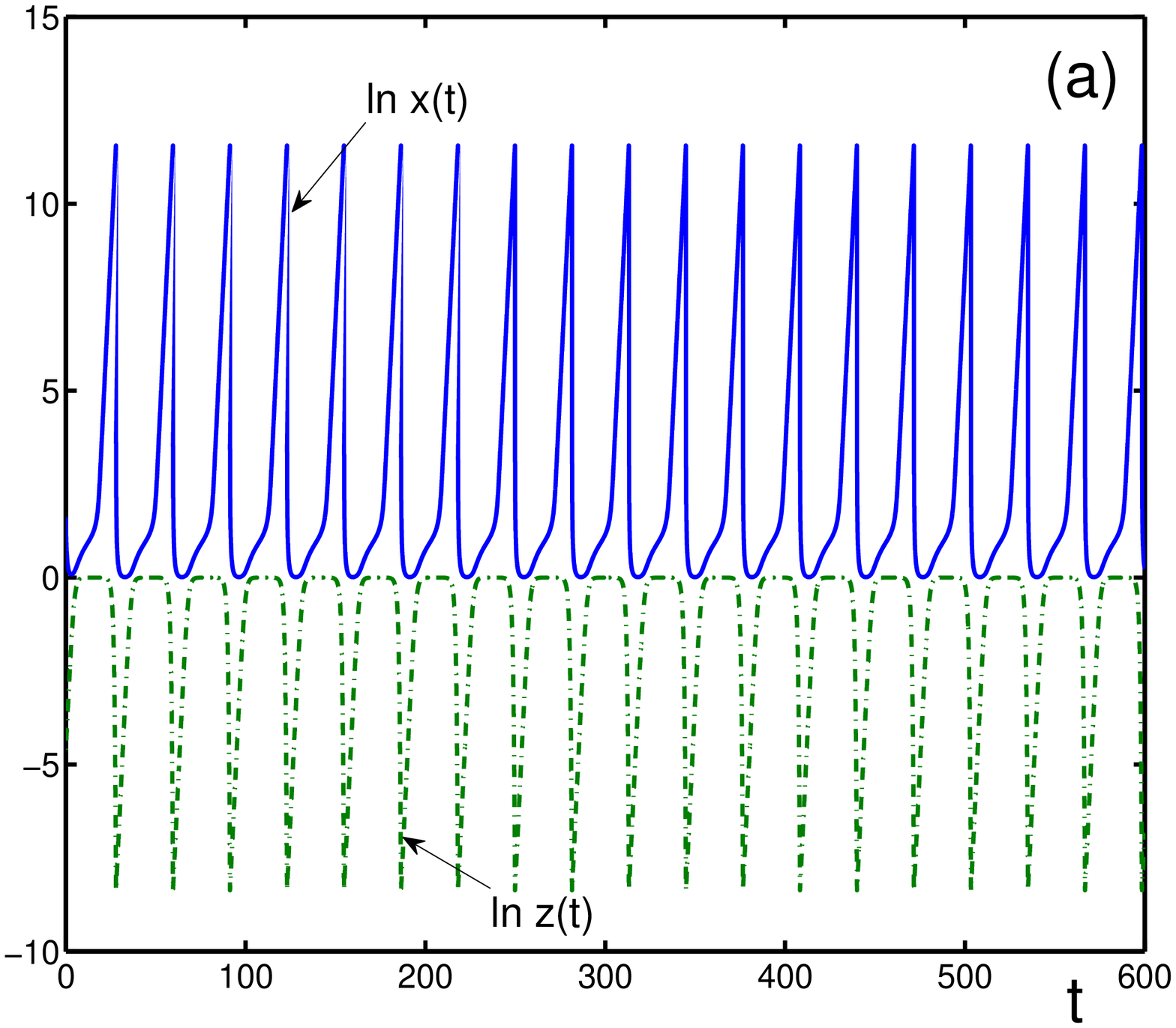,width=8cm}  &
\epsfig{file=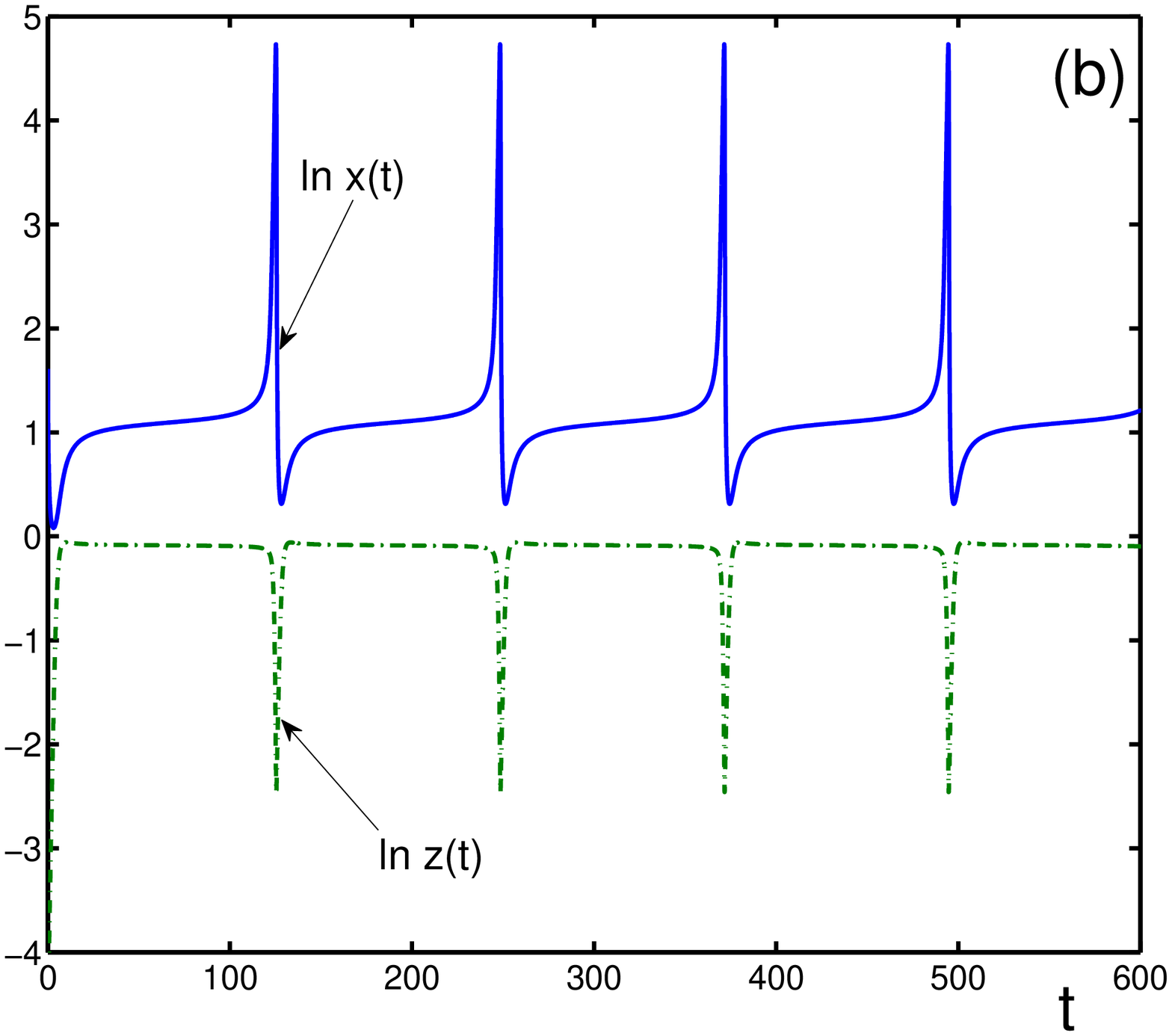,width=8cm}
\end{tabular}
\caption{Logarithmic behaviour of populations $x(t)$ (solid line)
and $z(t)$ (dashed-dotted line) for $b = 0.4$, and the initial conditions
$\{x_0 = 5, z_0 = 0.01 \}$, when $g_c(b) \approx -0.02942408$ and
$g_0(b) \approx -0.05524965$.
(a)
Everlasting oscillations of populations for $g_c(b) < g = -0.0001 < 0$.
(b)
Everlasting oscillations of populations for $g_0(b) < g = -0.029 < g_c(b)$.
}
\label{fig:Fig.17}
\end{figure}

\clearpage

%Figure 18
\begin{figure} [ht]
\centering
\begin{tabular}{lr}
\epsfig{file=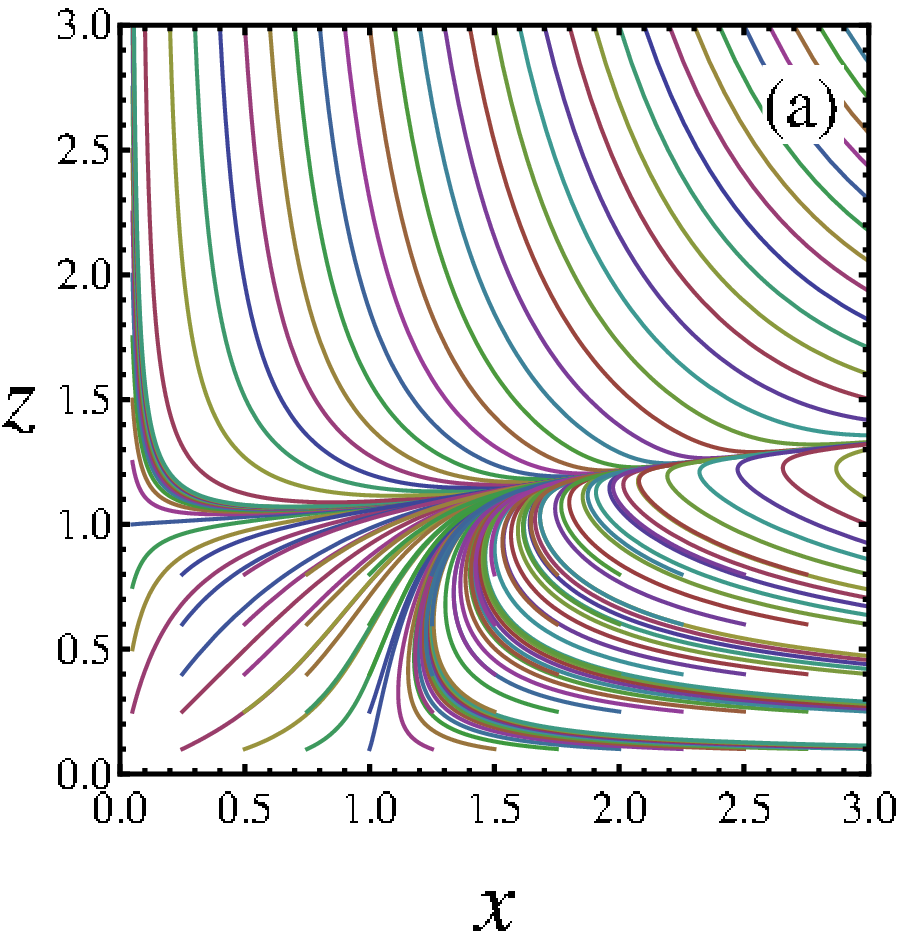,width=6.5cm}  & ~~~~
\epsfig{file=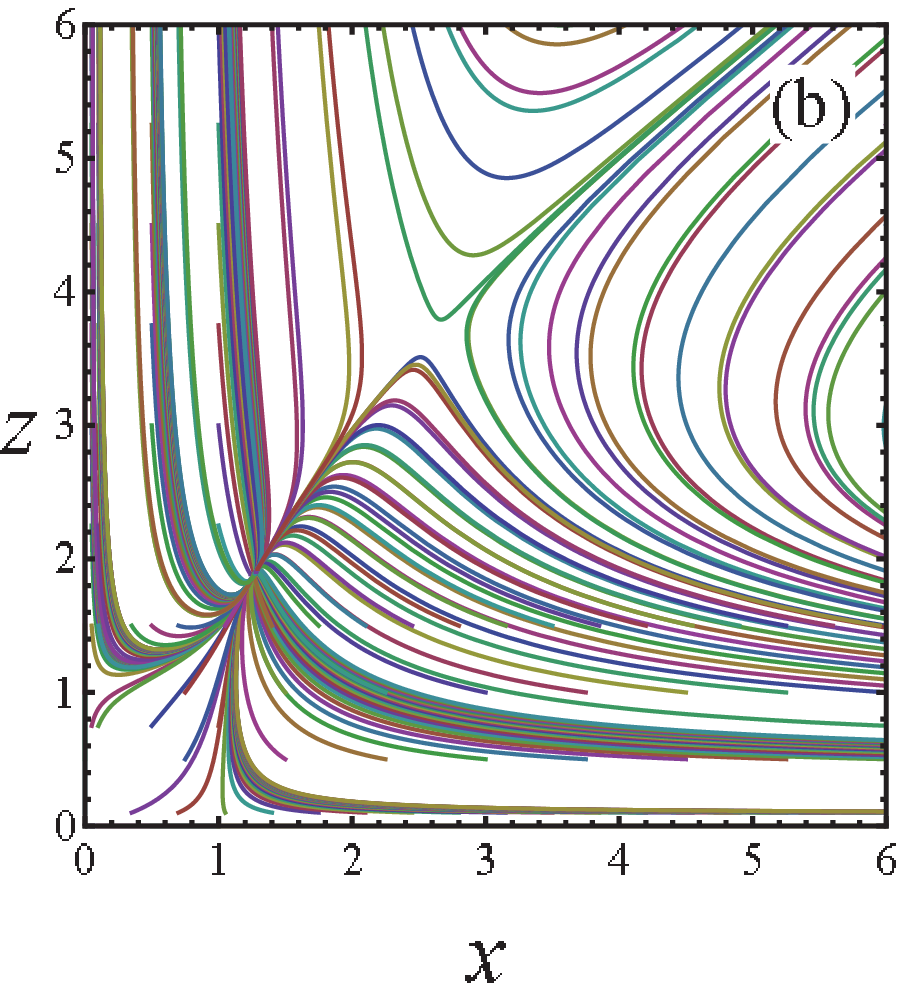,width=6.2cm} \\
&  \\
\epsfig{file=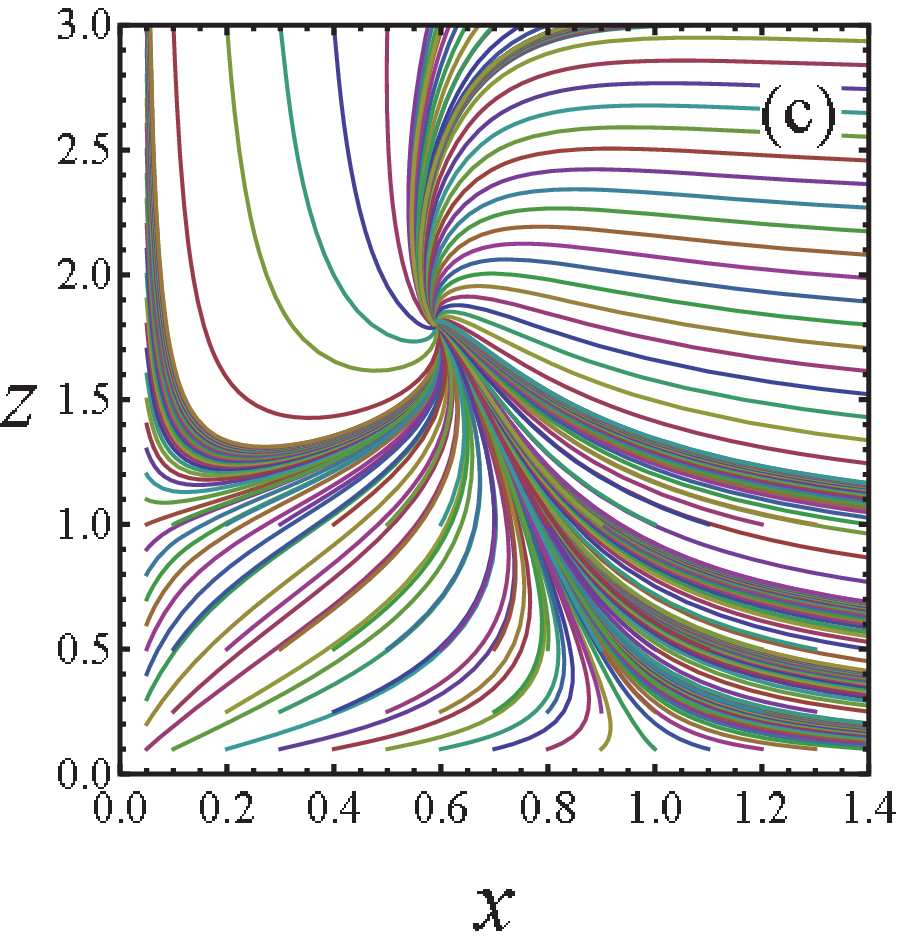,width=6.5cm}
\end{tabular}
\caption{
Phase portrait for mixed symbiosis, with the symbiotic parameters $b$
and $g$ from qualitatively different regions:
(a)
Phase portrait for the symbiotic parameters, where there are no fixed points.
The parameters are $b = 0.3$ and $g = 0.1 > g_c(b)$. Here $g_c(b)\approx 0.0817$.
(b)
Phase portrait for the parameters $b = 0.1$ and $g = 0.5 < g_c(b)$, with
$g_c(b)\approx 0.673$, where there are two fixed points. The first fixed point
$\{ x_1^* = 1.27133, z_1^* = 1.8882\}$ is stable, with the Lyapunov exponents
$\lbd_1 = -1.28863$ and $\lbd_2 = -0.471306$. The second fixed point
$\{x_2^* = 2.60324, z_2^* = 3.67525\}$ is a saddle, with the Lyapunov exponents
$\lbd_1 = -1.73578$ and $\lbd_2 = 0.69254$.
(c)
Phase portrait for the parameters $b = -0.5$ and $g = 1$, where there is a
single fixed point $\{ x^* = 0.588548, z^* = 1.80137 \}$ that is a stable focus,
with the characteristic exponents $\lbd_1 = -1.26505 - 0.49167 i$ and
$\lbd_2 = \lbd_1^*$).
}
\label{fig:Fig.18}
\end{figure}

\clearpage

%Figure 19
\begin{figure} [ht]
\centering
\begin{tabular}{lr}
\epsfig{file=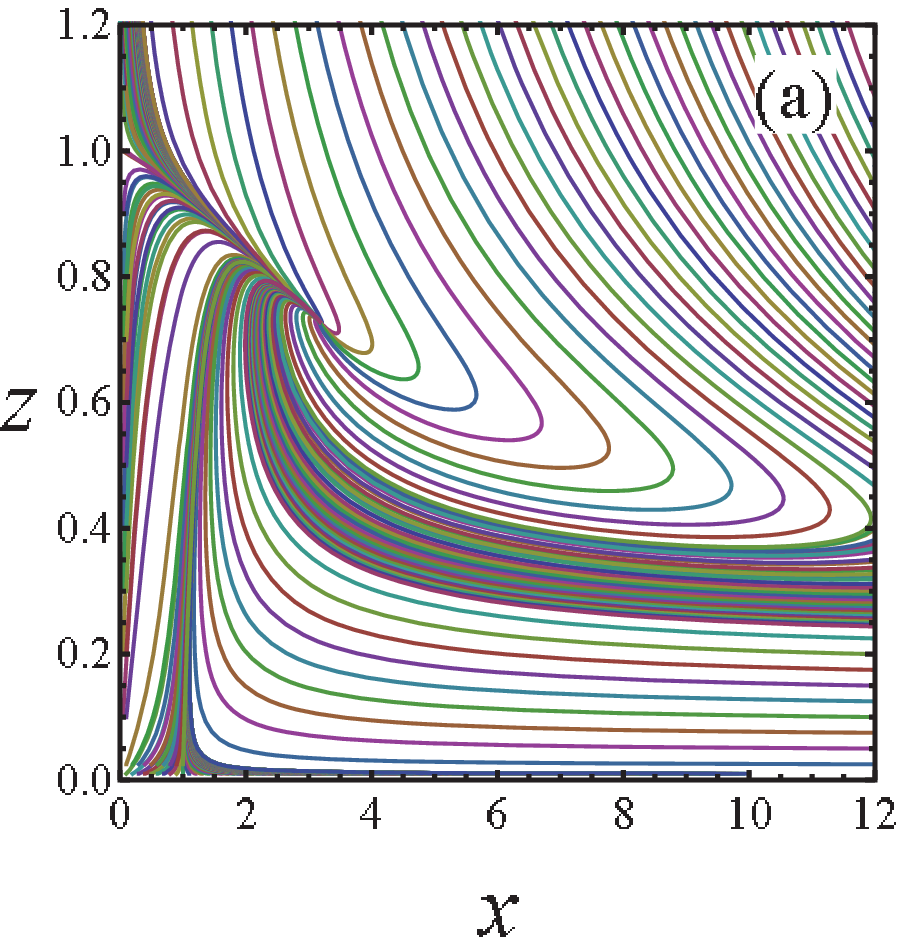,width=6.5cm}  & ~~~~
\epsfig{file=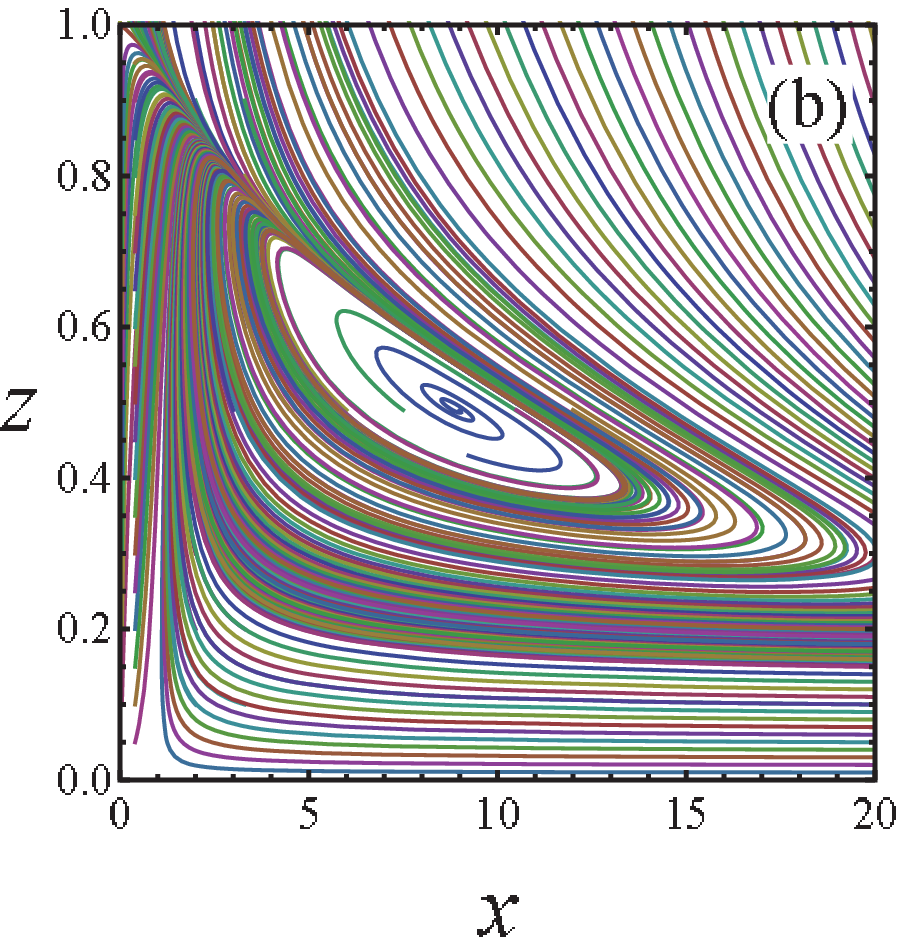,width=6.5cm} \\
&  \\
\epsfig{file=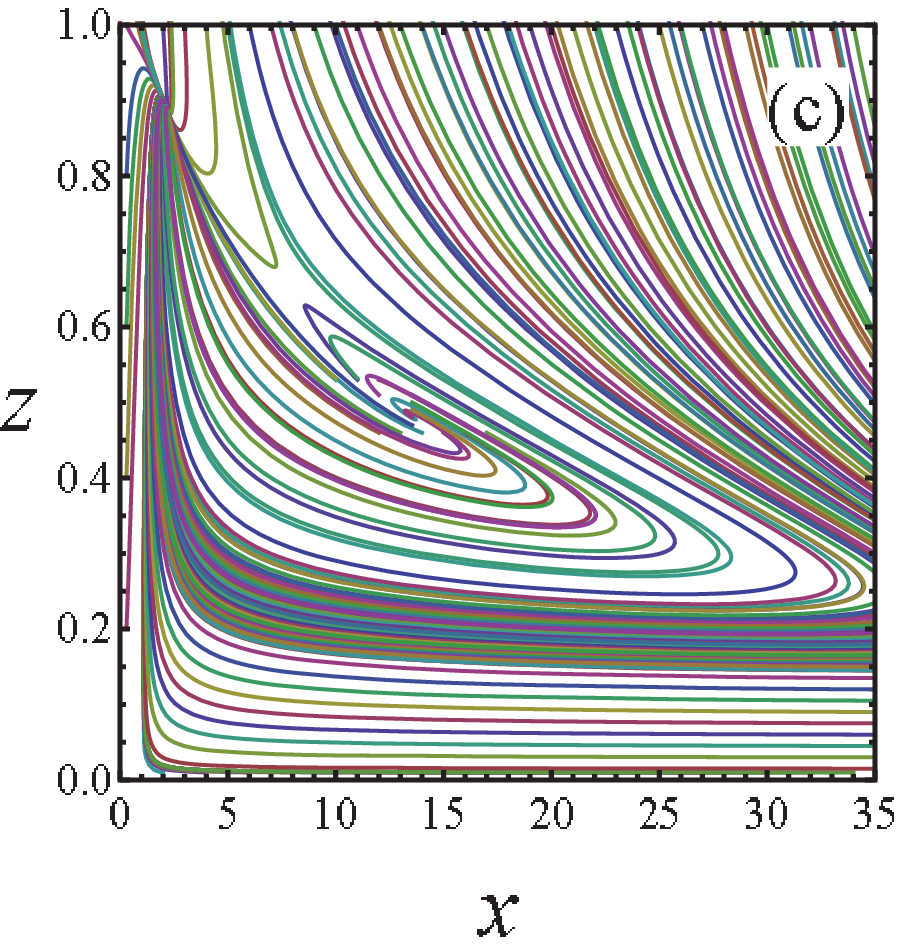,width=6.5cm}
\end{tabular}
\caption{
Phase portrait for mixed symbiosis, with the symbiotic parameters $b > 0$
and $g < 0$ from qualitatively different regions.
(a)
Phase portrait for $b = 0.5$ and $g = -0.1 < g_c(b) < 0$, with
$g_c(b) \approx -0.083056$, in the region, where there is a single fixed
point $\{x^* = 3.1801, z^* = 0.7276\}$ that is a stable focus, with the
characteristic exponents $\lbd_1 = -0.4215 - 0.1825 i$ and $\lbd_2 = \lbd_1^*$.
(b)
Phase portrait for the same $b = 0.5$, but with $g_c(b) < g = -0.083 < 0$. In
this region, there exist a single fixed point $\{ x^*=8.8369, z^*=0.49315\}$,
which is an unstable focus, with the characteristic exponents
$\lbd_1=0.08947-0.5945i$, $\lbd_2 = \lbd_1^*$, and a limit cycle.
(c)
Phase portrait for $b = 0.4$ and $g = -0.054$, when $g_c(b) \approx -0.02943$
and $g_0(b) \approx -0.05524965$, so that $g_0(b) < g < g_c(b) < 0$. The
parameters are in the region of $b > 0, g < 0$, where three fixed points exist.
One of them, $\{x_1^* = 2.15176, z_1^* = 0.890302\}$ is a stable node, with the
Lyapunov exponents $\lbd_1 = -0.85719$ and $\lbd_2 = -0.37654$. The two other
fixed points are unstable. The point $\{x_2^* = 8.08875, z_2^* = 0.646106\}$
is a saddle, with the Lyapunov exponents $\lbd_1 = -0.37834$ and
$\lbd_2 = 0.468814$. And $\{x_3^* = 13.6002, z_3^* = 0.479787\}$ is an unstable
focus, with the characteristic exponents $\lbd_1 = 0.305044 - 0.462324 i$ and
$\lbd_2 = \lbd_1^*$.
}
\label{fig:Fig.19}
\end{figure}

\clearpage

%Figure 20
\begin{figure} [ht]
\centering
\begin{tabular}{lr}
\epsfig{file=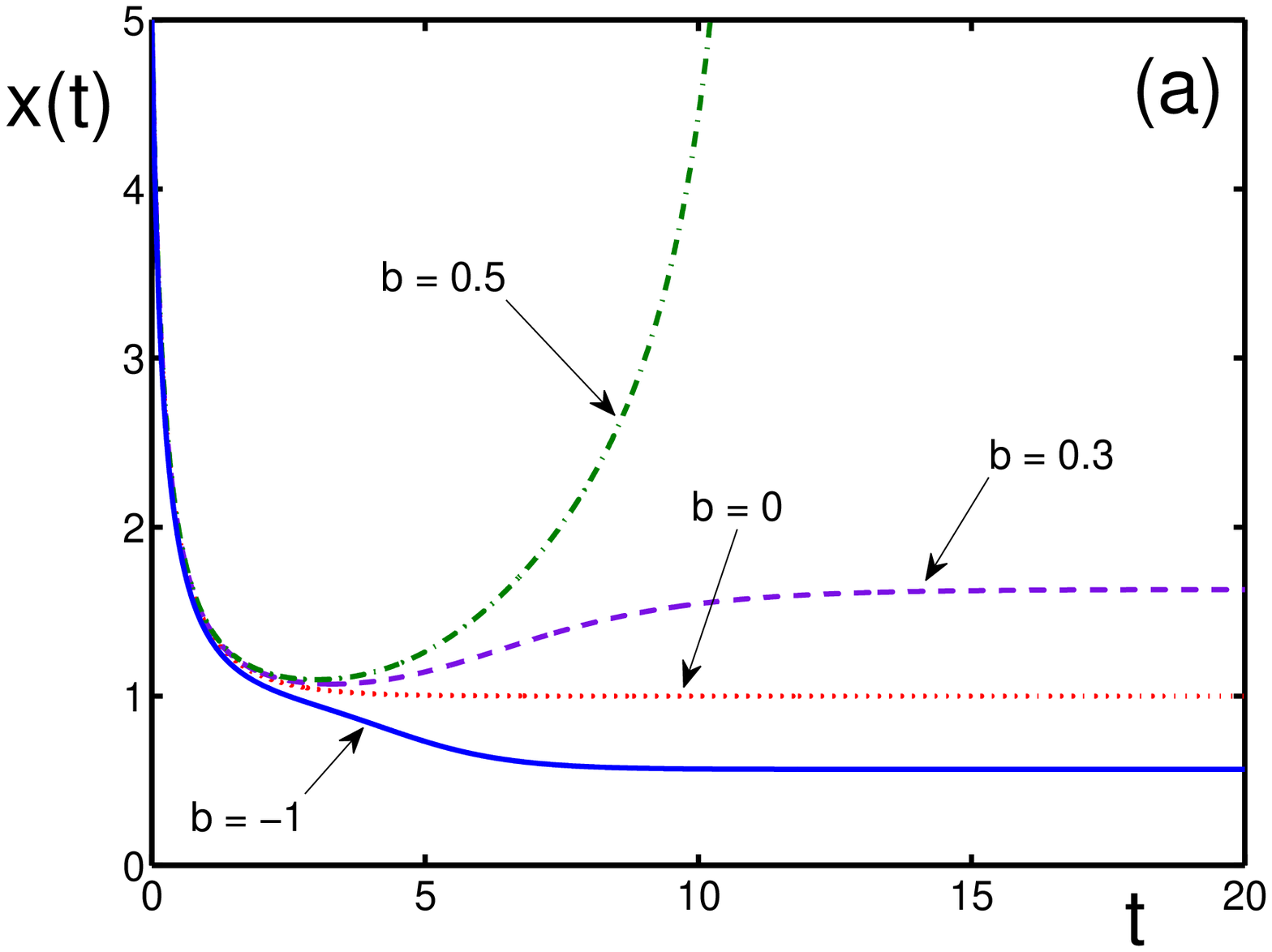,width=8cm}  &
\epsfig{file=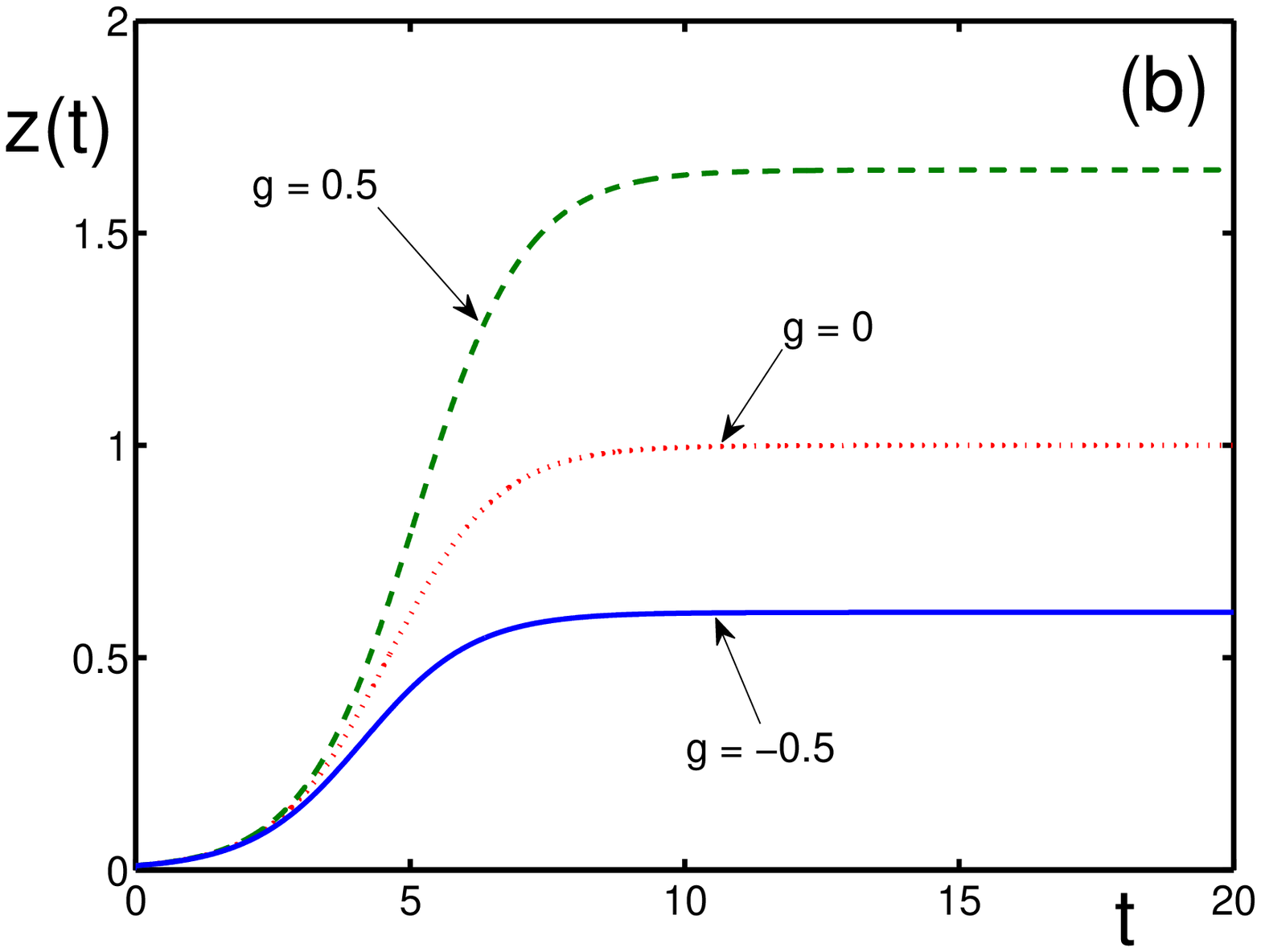,width=8cm}
\end{tabular}
\caption{Dynamics of populations under commensalism, characterized by the
degenerate systems of equations (\ref{d}), (\ref{b}), and (\ref{c}). Initial
conditions are $\{x_0 = 5, z_0 = 0.01\}$.
(a)
Dynamics of the population $x(t)$ for different $b$ and $g = 0$: For $b = 0$
(dotted line), $x(t) \ra x^* = 1$; for $b = 0.5 > 1/e$ (dashed-dotted line),
$x(t) \ra \infty$; for $b = 0.3 < 1/e$ (dashed line), $x(t) \ra x^* = 1.63$;
and for $b = -1$ (solid line), $x(t) \ra x^* = 0.567$, as $t \ra \infty$.
(b)
Dynamics of the population $z(t)$ for different $g$ and $b = 0$: For $g = 0$
(dotted line), $z(t) \ra z^* = 1$; for $g = 0.5$ (dashed-dotted line),
$z(t) \ra z^* = e^g = 1.65$; and for $g = -0.5$ (solid line),
$z(t) \ra z^* = 0.607$, as $t \ra \infty$.
}
\label{fig:Fig.23}
\end{figure}


\newpage


\begin{thebibliography}{9}


\bibitem[Ahmadjian \& Paracer(2000)]{Ahmadjian_4}
Ahmadjian, V. \& Paracer, S. [2000]
{\it An Introduction to Bilogical Associations}
(Oxford University, Oxford).

\bibitem[Boucher(1988)]{Boucher_1}
Boucher, D. [1988]
{\it The Biology of Mutualism: Ecology and Evolution}
(Oxford University, New York).

\bibitem[Chen et al.(2003)]{Chen_33}
Chen, G., Hill, D.J. \& Yu, X.H. [2003]
{\it Bifurcation Control: Theory and Applications} (Springer, Berlin).

\bibitem[Cobiaga \& Reartes(2013)]{Cobiaga_35}
Cobiaga, R, \$ Reartes, W. [2013]
``A new approach in the search for periodic orbits",
{\it Int. J. Bifur. Chaos} {\bf 23}, 1350186.

\bibitem[Desroches et al.(2012)]{Desroches_36}
Desroches, M., Guckenheimer, J., Krauskopf, B., Kuehn, C., Osinga, H. M.
\& Wechselberger, M. [2012]
``Mixed-mode oscillations with multiple time scales",
{\it SIAM Review} {\bf 54}, 211--288.

\bibitem[Douglas(1994)]{Douglas_2}
Douglas, A. E. [1994]
{\it Symbiotic Interactions}
(Oxford University, Oxford).

\bibitem[Gluzman \& Yukalov(1997)]{Gluzman_24}
Gluzman, S. \& Yukalov, V. I. [1997]
``Algebraic self-similar renormalization in the theory of critical phenomena",
{\it Phys. Rev. E} {\bf 55}, 3983--3999.

\bibitem[Gluzman \& Yukalov(1998a)]{Gluzman_26}
Gluzman, S. \& Yukalov, V. I. [1998a]
``Resummation method for analyzing time series",
{\it Mod. Phys. Lett. B} {\bf 12}, 61--74.

\bibitem[Gluzman \& Yukalov(1998b)]{Gluzman_27}
Gluzman, S. \& Yukalov, V. I. [1998b]
``Renormalization group analysis of October market crashes",
{\it Mod. Phys. Lett. B} {\bf 12}, 75--84.

\bibitem[Gluzman \& Yukalov(1998c)]{Gluzman_28}
Gluzman, S. \& Yukalov, V. I. [1998c]
``Booms and crashes in self-similar markets",
{\it Mod. Phys. Lett. B} {\bf 12}, 575--587.

\bibitem[Gluzman et al.(2003)]{Gluzman_30}
Gluzman, S., Sornette, D. \& Yukalov, V. I. [2003]
``Reconstructing generalized exponential laws by self-similar exponential approximants",
{\it Int. J. Mod. Phys. C} {\bf 14}, 509--527.

\bibitem[Graedel \& Allenby(2003)]{Graedel_10}
Graedel, T. \& Allenby, B. [2003]
{\it Industrial Ecology}
(Prentice Hall, Englewood Cliffs).

\bibitem[Grossman \& Helpman(1991)]{Grossman_8}
Grossman, G. M. \& Helpman, E. [1991]
{\it Innovation and Growth in the Global Economy}
(MIT, Massachusetts).

\bibitem[Kuznetsov(1995)]{Kuznetsov_32}
Kuznetsov, Y. A. [1995]
{\it Elements of Applied Bifurcation Theory}
(Springer, Berlin).

\bibitem[Leonov \& Kuznetsov(2007)]{Leonov_37}
Leonov, G. A. \& Kuznetsov, N. V. [2007]
``Time-varying linearization and the Perron effect",
{\it Int. J. Bifur. Chaos} {\bf 17}, 1079--1107.

\bibitem[Leonov \& Kuznetsov(2013)]{Leonov_34}
Leonov, G. A. \& Kuznetsov, N. V. [2013]
``Hidden attractors in dynamical systems",
{\it Int. J. Bifur. Chaos} {\bf 23}, 1330002.

\bibitem[Lotka(1925)]{Lotka_11}
Lotka, A. J. [1925]
{\it Elements of Physical Biology}
(Williams and Wilkins, Baltimore).

\bibitem[Perc \& Szolnoki(2010)]{Perc_13}
Perc, M. \& Szolnoki, A. [2010]
``Coevolutionary games - a mini review",
{\it Biosystems} {\bf 99}, 109--125.

\bibitem[Perc \& Szolnoki(2012)]{Perc_14}
Perc, M. \& Szolnoki, A. [2012]
``Self-organization of punishment in structured populations",
{\it New J, Phys.} {\bf 14}, 043013.

\bibitem[Perc et al.(2013)]{Perc_17}
Perc, M., Gomez-Gardenes, Szolnoki, A., Floria, L. M. \& Moreno, J. [2013]
``Evolutionary dynamics of group interactions on structural populations: a review"
{\it J. Roy. Soc. Interface} {\bf 10}, 20120997.

\bibitem[Pongvuthithum \& Likasiri(2010)]{Pong_38}
Pongvuthithum, R. \& Likasiri, C. [2010]
``Analytical discussions on species extinction in competitive communities
due to habitat destruction",
{\it Ecol. Model.} {\bf 221}, 2634--2641.

\bibitem[Richard(1993)]{Richard_9}
Richard, R. R. [1993]
{\it National Innovation Systems: A Comparative Analysis}
(Oxford University, Oxford).

\bibitem[Szolnoki \& Perc(2013a)]{Szolnoki_15}
Szolnoki, A. \& Perc, M. [2013a]
``Effectiveness of conditional punishment for the evolution of public cooperation",
{\it J. Theor. Biol.} {\bf 325}, 34--41.

\bibitem[Szolnoki \& Perc(2013b)]{Szolnoki_16}
Szolnoki, A. \& Perc, M. [2013b]
``Information sharing promotes prosocial behavior",
{\it New J. Phys.} {\bf 15}, 053010.

\bibitem[Sapp(1994)]{Sapp_3}
Sapp, J. [1994]
{\it Evolution by Association: A History of Symbiosis}
(Oxford University, Oxford).

\bibitem[Thompson et al.(1994)]{Thompson_31}
Thompson, J. M., Stewart, H. B. \& Ueda, Y. [1994]
``Safe, explosive, and dangerous bifurcations in dissipative dynamical systems",
{\it Phys. Rev. E} {\bf 49}, 1019--1027.

\bibitem[Townsend et al.(2002)]{Townsend_5}
Townsend, C. R., Begon, M. \& Harper, J. D. [2002]
{\it Ecology: Individuals, Populations and Communities}
(Blackwell Science, Oxford).

\bibitem[Von Hippel(1988)]{Hippel_7}
Von Hippel, E. [1988]
{\it The Sources of Innovation}
(Oxford University, Oxford).

\bibitem[Yukalov(1990)]{Yukalov_20}
Yukalov, V. I. [1990]
``Self-similar approximations for strongly interacting systems",
{\it Physica A} {\bf 167},  833--860.

\bibitem[Yukalov(1991)]{Yukalov_21}
Yukalov, V. I. [1991]
``Method of self-similar approximations",
{\it J. Math. Phys.} {\bf 32}, 1235--1239.

\bibitem[Yukalov(1992)]{Yukalov_22}
Yukalov, V. I. [1992]
``Stability conditions for method of self-similar approximations",
{\it J. Math. Phys.} {\bf 33},  3994--4001.

\bibitem[Yukalov \& Yukalova(1996)]{Yukalov_23}
Yukalov, V. I. \&  Yukalova, E.P. [1996]
``Temporal dynamics in perturbation theory",
{\it Physica A} {\bf 225}, 336--362.

\bibitem[Yukalov \& Gluzman(1998)]{Yukalov_25}
Yukalov, V. I. \& Gluzman, S. [1998]
``Self-similar exponential approximants",
{\it Phys. Rev. E} {\bf 58}, 1359--1382.

\bibitem[Yukalov \& Gluzman(1999)]{Yukalov_29}
Yukalov, V. I. \& Gluzman, S. [1999]
``Weighted fixed points in self-similar analysis of time series",
{\it Int. J. Mod. Phys. B} {\bf 13}, 1463--1476.

\bibitem[Yukalov et al.(2012a)]{Yukalov_6}
Yukalov, V. I., Yukalova, E. P. \& Sornette, D. [2012a]
``Modeling symbiosis by interactions through species carrying capacities",
{\it Physica D} {\bf 241}, 1270--1289.

\bibitem[Yukalov et al.(2012b)]{Yukalov_12}
Yukalov, V. I., Yukalova, E. P. \&  Sornette, D. [2012b]
``Extreme events in population dynamics with functional carrying capacity",
{\it Eur. Phys. J. Spec. Top.} {\bf 205}, 313--354.

\bibitem[Yukalov et al.(2013)]{Yukalov_18}
Yukalov, V. I., Yukalova, E. P. \&  Sornette, D. [2013]
``Utility rate equations of group population dynamics in biological and social systems",
{\it PLOS One} {\bf 8}, e83225.

\bibitem[Yukalov et al.(2014)]{Yukalov_19}
Yukalov, V. I., Yukalova, E. P. \&  Sornette, D. [2014]
``Population dynamics with nonlinear delayed carrying capacity",
{\it Int. J. Bifur. Chaos} {\bf 24}, 1450021.

\end{thebibliography}
\end{document}